\documentclass[%
reprint,
nobibnotes,
 amsmath,amssymb,
 prx,
floatfix,
longbibliography
]{revtex4-2}

\usepackage{xcolor}
\usepackage{graphicx}% Include figure files
\usepackage{dcolumn}% Align table columns on decimal point
\usepackage{bm}% bold math
\usepackage[mathlines]{lineno}
\usepackage{bm}

\newcommand{\tavg}[1]{\left\langle #1 \right\rangle}

\newcommand{\aavg}[1]{\left\langle #1 \right\rangle}
\newcommand{\tbtmat}[4]{\left(\begin{array}{cc} #1 & #2 \\ #3 & #4 \end{array}\right)}
\newcommand{\tbomat}[2]{\left(\begin{array}{c} #1 \\ #2 \end{array}\right)}
\newcommand{\obtmat}[2]{\left(\begin{array}{cc} #1 & #2 \end{array}\right)}
\newcommand{\pderiv}[2]{\frac{\partial #1}{\partial #2}}
\newcommand{\diag}[1]{\text{diag}\hspace{-0.2em}\left(#1\right)}
\newcommand{\ca}{{\sim}}
\newcommand{\blue}[1]{{#1}}

\begin{document}
\title{Theory of Coupled Neuronal-Synaptic Dynamics}
\author{David G. Clark} \email{dgc2138@cumc.columbia.edu, he/him}
\author{L.F. Abbott} \email{lfa2103@columbia.edu}
\affiliation{{Zuckerman Institute, Department of Neuroscience, Columbia University, New York, New York 10027, USA}}
\date{\today}

\begin{abstract}

In neural circuits, synaptic strengths influence neuronal activity by shaping network dynamics, and neuronal activity influences synaptic strengths through activity-dependent plasticity. Motivated by this fact, we study a recurrent-network model in which neuronal units and synaptic couplings are interacting dynamic variables, with couplings subject to Hebbian modification with decay around quenched random strengths. Rather than assigning a specific role to the plasticity, we use dynamical mean-field theory and other techniques to systematically characterize the neuronal-synaptic dynamics, revealing a rich phase diagram. Adding Hebbian plasticity slows activity in already chaotic networks and can induce chaos in otherwise quiescent networks. Anti-Hebbian plasticity quickens activity and produces an oscillatory component. Analysis of the Jacobian shows that Hebbian and anti-Hebbian plasticity push locally unstable modes toward the real and imaginary axes, respectively, explaining these behaviors. Both random-matrix and Lyapunov analysis show that strong Hebbian plasticity segregates network timescales into two bands, with a slow, synapse-dominated band driving the dynamics, suggesting a flipped view of the network as synapses connected by neurons. For increasing strength, Hebbian plasticity initially raises the complexity of the dynamics, measured by the maximum Lyapunov exponent and attractor dimension, but then decreases these metrics, likely due to the proliferation of stable fixed points. We compute the marginally stable spectra of such fixed points as well as their number, showing exponential growth with network size. Finally, in chaotic states with strong Hebbian plasticity, a stable fixed point of neuronal dynamics is destabilized by synaptic dynamics, allowing any neuronal state to be stored as a stable fixed point by halting the plasticity. This phase of freezable chaos offers a new mechanism for working memory. 

% Overall, we provide a theoretical basis for understanding computation via coupled neuronal-synaptic dynamics.

\end{abstract}

\maketitle

\section{Introduction}
\label{sec:introduction}

Computations in neural circuits are commonly thought to be implemented through the coordinated dynamics of neurons \cite{gao2015simplicity, pandarinath2018inferring, vyas2020computation}. Under this view, the role of synaptic connectivity is to sculpt neuronal dynamics to implement computations. In actuality, synapses undergo plasticity over diverse timescales in response to neuronal activity and thus constitute dynamic degrees of freedom in their own right \cite{benna2016computational}. A more accurate picture of computation in neural circuits should involve the coupled dynamics of neurons and synapses. Indeed, it is possible that a network is better described by the states of its synapses than of its neurons \cite{abbott2004synaptic}. Here, we study the consequences of treating neurons and synapses as mutually coupled dynamic variables on equal footing.

Synaptic dynamics are often divided into short-term plasticity, which operates on short timescales and depends on presynaptic activity \cite{abbott1997synaptic, maass1997dynamic, barak2007persistent, mongillo2008synaptic, buonomano2009state, mi2017synaptic}; and long-term plasticity, which acts on much longer timescales and depends on both pre- and postsynaptic activity \cite{penney1993coupled, penney1994slow, roberts2000dynamics, ocker2015self}. However, short-term forms of Hebbian plasticity exist, suggesting that the timescale distinction is little more than a convention \cite{gustafsson1989onset, malenka1991postsynaptic, malenka1993nmda, volianskis2003transient, erickson2010single, volianskis2013different, driesen2013impact, park2014nmda, lisman2017glutamatergic} (see \cite{lansner2023hebbian} for a review).  Hebbian plasticity is more powerful than the presynaptic variety due to its ability to create attractor states of neuronal dynamics, the basis of Hopfield networks \cite{hopfield1982neural}.  We are therefore motivated to introduce ongoing Hebbian plasticity in a recurrent network, without necessarily imposing a separation of timescales between neuronal and synaptic dynamics. This has unexpected, computationally useful consequences, a key example being \textit{freezable chaos}, a phase in which a stable fixed point of neuronal dynamics is destabilized through synaptic dynamics. By contrast, introducing presynaptic plasticity to this model simply adds an effective constant input to each neuron (Appendix~\ref{sec:extensions-and-limits}). Our work thus provides a theoretical impetus for further experimental investigation of ongoing Hebbian plasticity mechanisms.

The view that synapses serve solely to sculpt neuronal dynamics is mirrored in machine learning. In artificial neural networks, weights are trained via gradient descent, then fixed. However, allowing weights to be modulated by the activity of the units has been shown to confer computational advantages \cite{miconi2018backpropamine, miconi2018differentiable, hinton1987using, miconi2022learning, tyulmankov2022meta}, particularly in tasks requiring short-term memory storage \cite{masse2019circuit}. For example, \citet{ba2016using} showed that recurrent networks benefit from a combination of ``slow weights'' trained via backpropagation and ``fast weights" that undergo activity-dependent updates (the model we study is essentially the continuous-time counterpart of this proposal). In practice, recurrent networks have been superseded by transformers \cite{vaswani2017attention}. While these models were not neuroscientifically motivated, \citet{schlag2021linear} showed that linearized transformers \cite{katharopoulos2020transformers} are equivalent to fast weight programmers \cite{schmidhuber1992learning}, a recurrent network with activity-dependent weight updates
\footnote{This equivalence arises due to the fact that \textit{attention}, the core component of transformers, is based on forming linear combinations of neuronal states weighted by inner products between pairs of neuronal states; this can be mimicked by multiplying a neuronal state by a coupling matrix that has been shaped through activity-dependent plasticity to encode outer products of neuronal states.}.

A major impediment to studying neuronal-synaptic dynamics, in both neuroscience and machine learning, is that the analytical methods developed for nonplastic networks often do not translate to plastic networks, particularly when the neuronal and synaptic timescales are not well separated. For example, a common simplification is to study nonplastic networks with linear neuronal dynamics; however, such networks become highly nonlinear when synaptic plasticity is introduced \cite{magnasco2009self}. Moreover, synaptic degrees of freedom increase the dimension of phase space from $\mathcal{O}(N)$ to $\mathcal{O}(N^2)$. In studying and training nonplastic networks, the theory of random networks has played a crucial role \cite{sussillo2009generating}. In seminal work, \citet{sompolinsky1988chaos} showed that random recurrent networks exhibit a phase transition to high-dimensional chaotic activity at a critical coupling variance \cite{clark2023dimension}. While this analysis was in firing-rate (nonspiking) networks with fully unstructured couplings, key phenomena, such as the transition to chaos, generalize to spiking networks with more realistic distributions over couplings \cite{kadmon2015transition}. Here, we extend this approach to plastic networks---that is, develop a theory for such a model in the thermodynamic limit, compute its phase diagram, and characterize its dynamics---thereby providing a foundation for understanding how coupled neuronal-synaptic dynamics could underlie computation.

\section{Model}
\label{sec:model}

\begin{figure*}
    \centering
    \includegraphics[width=\textwidth]{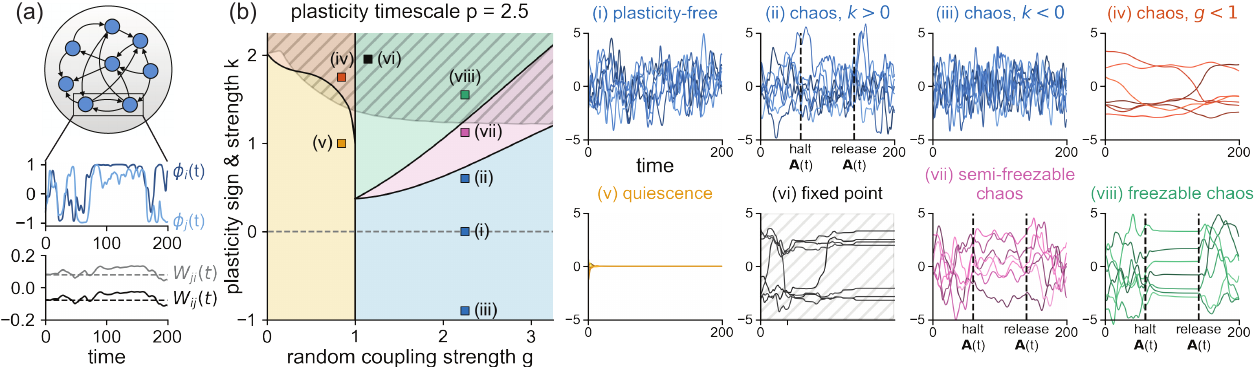}
    \caption{(a) Dynamics of a pair of neurons (top panel) and of the synapses through which they are reciprocally coupled (bottom panel). Synapses fluctuate about quenched random strengths (dashed lines) in response to pre- and postsynaptic activity according to a Hebbian rule. (b) Left: phase diagram of the plastic network for $p = 2.5$. Right: example neuronal traces $x_i(t)$ from simulations of each phase-diagram region, with parameters given by the location of the associated square marker.}
    \label{fig:phase-diagram}
\end{figure*}

We augment the random-network model of \citet{sompolinsky1988chaos} with dynamic couplings. There are $N$ neuronal units with pre-activations $x_i(t)$ and activations ${\phi_i(t) = \phi(x_i(t))}$, where ${\phi(\cdot)}$ is a nonlinearity. Throughout, we use ${\phi(\cdot) = \tanh(\cdot)}$. Neurons interact through all-to-all time-dependent couplings ${W}_{ij}(t)$ according to the neuronal dynamics
\begin{equation}
\left(1 + \partial_t\right){x}_i(t) = \sum_jW_{ij}(t){\phi_j}(t).
\label{eq:neuronal-dynamics}
\end{equation}
Concurrently, ${W}_{ij}(t)$ displays synaptic dynamics. We first express these couplings as a sum of quenched and fluctuating terms,
\begin{equation}
    {W}_{ij}(t) = {J}_{ij} + {A}_{ij}(t),
    \label{eq:form-of-W}
\end{equation}
where ${J_{ij} \sim \mathcal{N}\left(0, {g^2}/{N} \right)}$ provides quenched disorder. The fluctuating term ${A}_{ij}(t)$ follows a local plasticity rule,
\begin{equation}
    \left(1 + p \partial_t\right) {A}_{ij}(t) = \frac{k}{N}{\phi}_i(t) {\phi}_j(t),
\label{eq:syn-dynamics}
\end{equation}
where $k$ is the sign and strength of the plasticity, which is Hebbian or anti-Hebbian for ${k > 0}$ or ${k < 0}$, respectively; and $p$ is the synaptic decay timescale in units in which the neuronal decay timescale is unity. We do not require $p \gg 1$, though a reasonable constraint from biology is $p > 1$ since the synaptic timescale is unlikely to be shorter than the neuronal timescale. \blue{The couplings $\bm{W}(t)$ include self-connections (on-diagonals) with the same dynamics as non-self-connections (off-diagonals). However, the effect of these self-connections on each neuron is $\sim 1/\sqrt{N}$, and thus is negligible as $N\rightarrow \infty$.} The full set of dynamic variables comprises the $N$ neurons, ${x}_i(t)$, and $N^2$ fluctuating couplings, ${A}_{ij}(t)$. We study their collective dynamics as a function of $g$, $k$, and $p$ as $N \rightarrow \infty$ [Fig.~\ref{fig:phase-diagram}(a)].

% Because the synapses have intensive decay timescale $p$, $\bm{A}(t)$ is an average over outer products of at most $\ca p$ decorrelated neuronal states. The approximate rank of $\bm{A}(t)$ is therefore intensive.

\blue{Because $\bm{A}(t)$ is an average over outer products for each order-one timestep and decays with timescale $p$, it has rank of order $p$ or smaller. Given that $p$ is order-one (not order-$N$), the approximate rank of $\bm{A}(t)$ is intensive.} Note that ${J_{ij} \sim {1}/{\sqrt{N}}}$ and ${A_{ij}(t) \sim {1}/{N}}$, so plasticity is vanishingly weak at the single-synapse level as ${N\rightarrow\infty}$. Nevertheless, because $\bm{A}(t)$ is approximately intensive-rank, the random and fluctuating couplings both have order-one macroscopic effects. This is because when neuronal activity exhibits alignment to the states encoded in $\bm{A}(t)$, the $\bm{J}$ and $\bm{A}(t)$ terms each contribute an order-one input to neurons, as made clear by the mean-field analysis in Sec.~\ref{sec:dynamical-mean-field-theory}. Further intuition for this scaling can be obtained from the spectra of $\bm{J}$ and $\bm{A}(t)$; the chosen scaling implies that the eigenvalues of both matrices are order-one, allowing the modes they drive in the network to compete on equal footing. This scaling of low-rank structure relative to randomness at individual synapses is generic in models in which the couplings are the sum of random and low-rank terms \cite{mastrogiuseppe2018linking, schuessler2020dynamics, schuessler2020interplay} (but see \cite{landau2018coherent} who used a $\ca 1/\sqrt{N}$ rank-one term given by an outer product of orthogonal vectors). The scaling also appears in spiked matrix or tensor models in statistical physics \cite{baik2005phase}. In experiments that measure changes in synaptic strengths, plasticity on this weak scale could go unnoticed, but nevertheless exert dramatic influence at the network level.

% \textbf{Freezable chaos:} The dashed vertical lines in {(ii)}, {(vii)}, and {(viii)} pertain to freezable chaos. In these traces, synaptic dynamics are halted at the first line and released at the second. In {(ii)}, chaos is nonfreezable: upon halting synaptic dynamics, neuronal dynamics remain chaotic, with no trace of the halt-time neuronal state present in neuronal activity. \textbf{(vii)} For larger $g$, chaos is semifreezable: upon halting synaptic dynamics, neuronal dynamics remain chaotic, with fluctuations occurring about a nonzero baseline state near the halt-time neuronal state. \textbf{(viii)} For sufficiently large $k$, chaos is freezable: halting synaptic dynamics leaves a stable fixed point of neuronal dynamics near the halt-time neuronal state. Simulations are of size $N = 400$ for (vi) and $N = 2000$ otherwise.}

\section{Phase diagram}
\label{sec:phase-diagram}

We summarize the behavior of the model with a phase diagram in $(g, k)$ parameter space for $p = 2.5$ [Fig.~\ref{fig:phase-diagram}(b)], noting that constant-$p$ slices of the full $(g, k, p)$ diagram are similar for order-one values of $p$. We refer to this phase diagram for the rest of this section.

For $k = 0$, the model reduces to that of \citet{sompolinsky1988chaos} (dashed horizontal line). For ${g > 1}$, this nonplastic network is chaotic (i). For ${g < 1}$, the trivial neuronal fixed point of the nonplastic network, $\bm{x}(t) = \bm{0}_N$, is globally stable and the network is quiescent. In analogy with the nonplastic network, the plastic network 
\blue{can produce chaotic activity} for ${g > 1}$, and the activity is further shaped by synaptic plasticity. Hebbian plasticity, ${k > 0}$, slows activity (ii), while anti-Hebbian plasticity, ${k < 0}$, quickens activity and generates an oscillatory component (iii).

For ${g < 1}$, the trivial neuronal-synaptic fixed point, ${(\bm{x}(t), \bm{A}(t)) = (\bm{0}_N, \bm{0}_{N\times N})}$, is stable. However, in contrast to the nonplastic network, this fixed point is not necessarily globally stable, but coexists with chaotic states for large $k$ (iv). Thus, Hebbian plasticity can induce \blue{dynamic activity} in an otherwise quiescent network. For ${g < 1}$, if $k$ is not large enough to induce \blue{dynamic activity}, the network is globally quiescent (v).

In dynamic states, network activity is shaped by a proliferation of stable fixed points throughout phase space. In particular, if Hebbian plasticity is strong (hatched region), there exist stable fixed points to which finite-size networks settle following transient chaos, accompanied by the rank of $\bm{A}(t)$ collapsing to unity (vi). Such fixed points are stable with respect to the combined neuronal-synaptic dynamics. Their number, which we compute, is exponential in $N$.

Two scenarios can lead to transient chaotic states. First, when $g < 1$ and $k$ is large enough to induce chaotic activity, finite-size networks may collapse to the trivial fixed point. Additionally, when $(g, k)$ is in the hatched region, finite-size networks may collapse to nonzero fixed points.

% On the other hand, for networks of infinite size, as well as for finite-size networks that are not in the above-mentioned regions of the phase diagram, chaos is persistent.

Finally, we describe {freezable chaos}. Consider a chaotic state with Hebbian plasticity. Suppose we abruptly halt synaptic dynamics. If $k$ is small, the halted-synapse neuronal dynamics are chaotic, with no trace of the halt-time neuronal state in the activity (ii). If $k$ is larger, the halted-synapse neuronal dynamics are chaotic, but neurons fluctuate around the halt-time state, so a memory of this state is retained (vii). If $k$ is sufficiently large, the halted-synapse neuronal dynamics are nonchaotic; neurons flow to a stable fixed point near the halt-time state (viii). In all cases, releasing the synapses returns the network to neuronal-synaptic chaos. As ${N \rightarrow \infty}$, these three cases, which we label nonfreezable, semifreezable, and freezable chaos, respectively, are distinct phases. In freezable chaos, there is, at any instant, a stable fixed point of neuronal dynamics that is destabilized by synaptic dynamics. Thus, a stable fixed point can be created near any neuronal state by halting synaptic plasticity, enabling a new form of short-term memory storage.

\section{Dynamical Mean-Field Theory}
\label{sec:dynamical-mean-field-theory}

The temporal structure of network activity is described in the limit $N \rightarrow \infty$ by a dynamical mean-field theory (DMFT) whose main order parameter is the single-unit autocovariance (two-point) function,
\begin{equation}
    C(\tau) = \tavg{\phi_i(t) \phi_i(t +\tau)}_{\bm{J}},
\end{equation}
where we assume statistical stationarity in time. Integrating the synaptic dynamics, Eq.~(\ref{eq:syn-dynamics}), and inserting this into the neuronal dynamics, Eq.~(\ref{eq:neuronal-dynamics}), gives
\begin{multline}
(1 + \partial_t) x_i(t) = \sum_j J_{ij} \phi_j(t) \\
+ \frac{k}{p} \int^t_{-\infty} {dt'} e^{-({t-t'})/{p}} \left(\frac{1}{N}\sum_j \phi_j(t)\phi_j(t') \right) \phi_i(t'),
\label{eq:pure-neuronal-dynamics}
\end{multline}
where we have separated terms arising from $\bm{J}$ (first term on the rhs) and $\bm{A}(t)$ (second term on the rhs). Taking the limit ${N \rightarrow \infty}$ yields the single-site picture
\begin{multline}
\left(1  + \partial_t\right)x(t) =
\eta(t) \\
+ \frac{k}{p} \int^t_{-\infty} dt' e^{-({t-t'})/{p}}C(t - t')\phi(t'),
\label{eq:single-site-picture}
\end{multline}
where $\eta(t)$ is an effective Gaussian field with zero mean and second-order statistics
\begin{equation}
    \tavg{\eta(t) \eta(t + \tau)}_{\eta} = g^2 C(\tau).
\label{eq:eta-distr}
\end{equation}
The DMFT is closed by the self-consistency condition
\begin{equation}
    C(\tau) = \tavg{\phi(t) \phi(t +\tau)  }_{\eta}.
\end{equation}
In the single-site dynamics of Eq.~(\ref{eq:single-site-picture}), synaptic plasticity introduces a convolutional self-coupling with a kernel that depends self-consistently on $C(\tau)$. For $k = 0$, the self-coupling vanishes and the DMFT reduces to that of \cite{sompolinsky1988chaos}, which can be solved analytically. For $k \neq 0$, the nonlinearity of the self-coupling induces a non-Gaussian distribution over $x(t)$---in particular, a distribution that becomes increasingly bimodal with larger $k$ due to saturation of the tanh function---so we solve the DMFT equations using standard numerical techniques \cite{krishnamurthy2022theory, roy2019numerical, eissfeller1992new, eissfeller1994mean, stern2014dynamics, mignacco2020dynamical} (Appendix~\ref{subsec:dmft-for-C}). The DMFT agrees closely with simulations [Fig.~\ref{fig:plasticity-shaped-chaos}(a)].

\subsection{Chaotic states with ${g > 1}$}
\label{subsec:plasticity-shaped-chaos}
\begin{figure}
    \centering
    \includegraphics[width=\columnwidth]{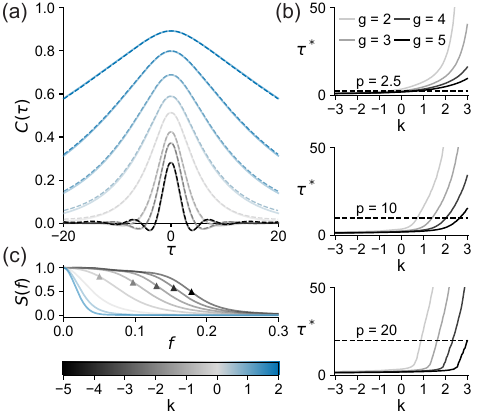}
    \caption{Chaotic states with ${g > 1}$. (a) $C(\tau)$ from the DMFT (solid lines) and in simulations (dashed lines) for $g = 2$, $p = 2.5$, and various values of $k$ (indicated by the lower color bar). (b) Dynamic timescale $\tau^*$ [Eq.~(\ref{eq:dynamic-timescale-def})] as a function of $k$ for various values of $g$. Dotted line indicates $p$. (c) Power spectra (normalized such that $S(0) = 1$) for the autocovariance functions shown in (a). For anti-Hebbian ($k < 0$) power spectra, triangular markers indicate an oscillatory frequency computed from the zero-crossings of $C(\tau)$.}
    \label{fig:plasticity-shaped-chaos}
\end{figure}

We now examine the solutions $C(\tau)$ of the DMFT. In the regime where the nonplastic network is chaotic, ${g > 1}$, Hebbian plasticity slows the activity, broadening $C(\tau)$ (Fig.~\ref{fig:plasticity-shaped-chaos}(a), $k > 0$ solutions). A network with static, symmetric couplings (e.g., the Hopfield network) admits a Lyapunov function that guarantees convergence to fixed points \cite{hopfield1982neural}. The slow activity generated by Hebbian plasticity results from competition between $\bm{J}$, a random, asymmetric matrix that promotes dissipative, chaotic dynamics; and $\bm{A}(t)$, a symmetric matrix that promotes convergence to a drifting fixed point that trails the neuronal state.

This competition has an interesting dependence on the model parameters. For small $k$ and large $p$, neurons fluctuate rapidly relative to the synaptic decay timescale and the effect of plasticity is averaged out. As $k$ is increased, synapses slow neurons by attracting the neuronal state toward its history, permitting a stronger synaptic signal to be encoded. This stronger signal causes further neuronal slowing. The interaction of neurons and synapses in this positive-feedback process causes the timescale of neuronal fluctuations to diverge as $k$ increases. For large $p$ and $k$, finite-size networks can show bistability between a fast state in which plasticity is averaged out and a slow state in which synapses drag neurons (Appendix~\ref{sec:bistability}).

This slowing behavior can also be understood through the DMFT. We quantify the speed of neuronal fluctuations by defining the dynamic timescale
\begin{equation}
    \tau^* = \int_0^{\infty} d\tau \Bigg[\frac{C(\tau)}{C(0)}\Bigg]^2,
    \label{eq:dynamic-timescale-def}
\end{equation}
whose dependence on $g$, $k$, and $p$ is illustrated in Fig.~\ref{fig:plasticity-shaped-chaos}(b). The size of the integral term in the single-site dynamics [Eq.~(\ref{eq:single-site-picture})] is $\ca k \tau^* / p$ for $\tau^* \ll p$ (and $\ca k$ for $\tau^* \gg p$). Increasing this term slows $x(t)$, increasing $\tau^*$. That $\tau^*$ depends on the size of this term, which itself depends on $\tau^*$, produces a positive-feedback loop. Once $k$ is large enough so that the integral term competes with $\eta(t)$, which occurs when $k \tau^*/p \sim g$, $\tau^*$ grows rapidly. The inflection of $\tau^*$ in $k$ is sharpest for large $p$ or $g$, in which case $k \tau^*/p$ and $g$ are well separated at small $k$ (Fig.~\ref{fig:plasticity-shaped-chaos}(b), $p = 20$).

Under anti-Hebbian plasticity, rather than synapses attracting the neuronal state to its history, this effect is repulsive. This quickens the dynamics and adds an oscillatory component to the activity, tightening $C(\tau)$ and creating oscillations during its decay to zero (Fig.~\ref{fig:plasticity-shaped-chaos}(a), $k<0$ solutions). In the single-site picture [Eq.~(\ref{eq:single-site-picture})], plasticity modifies the dynamics of $x(t)$ by introducing time-delayed negative feedback, which generically induces oscillations \cite{beiran2019contrasting, muscinelli2019single}. While finite-size simulations of the model of \cite{sompolinsky1988chaos} show limit cycles, our calculation of $C(\tau)$ for ${N \rightarrow \infty}$ demonstrates that this anti-Hebbian oscillatory component is not merely a finite-size effect. These oscillations are further characterized by the (normalized) power spectrum $S(f) = |\hat{C}(f)|^2 / |\hat{C}(0)|^2$, where $\hat{C}(f)$ denotes the Fourier transform of $C(\tau)$ [Fig.~\ref{fig:plasticity-shaped-chaos}(c)]. Rather than containing a peak at a nonzero frequency, the power spectra corresponding to the anti-Hebbian autocovariance functions in Fig.~\ref{fig:plasticity-shaped-chaos}(a) possess a range of large frequencies (though $S(f)$ develops a nonzero peak for smaller values of $g$; not shown). Point estimates of the dominant oscillatory frequency computed from the first three zero-crossings of $C(\tau)$ roughly capture the frequency scale at which $S(f)$ decays (Fig.~\ref{fig:plasticity-shaped-chaos}(c), triangular markers).

\subsection{Chaotic states for ${g < 1}$}
\label{subsec:chaotic-states-g-less-one}

We next consider the regime ${g < 1}$ in which the nonplastic network is globally quiescent. The plastic network has a trivial fixed point, $(\bm{x}(t), \bm{A}(t)) = (\bm{0}_N, \bm{0}_{N\times N})$, that is stable when all eigenvalues of $\bm{J}$ have real part less than unity. For ${N \rightarrow \infty}$, Girko's circular law implies that this requires ${g < 1}$ \cite{girko1985circular}. In contrast to the nonplastic network, the trivial neuronal-synaptic fixed point coexists with dynamic states for large $k$ as indicated by DMFT solutions (Fig.~\ref{fig:plasticity-induced-chaos}(a), solid lines). \blue{In Sec.~\ref{subsec:lyapunov}, we confirm over a restricted parameter regime that these solutions are chaotic (i.e., have positive maximum Lyapunov exponent).}

% Thus, Hebbian plasticity can induce chaos in an otherwise quiescent network.

\blue{Dynamic states for ${g < 1}$ eventually collapse to the trivial fixed point or, if $k$ is large enough, to a stable nonzero fixed point (Sections~\ref{subsec:lyapunov},~\ref{sec:fixed-points}). In this section, we consider values of $g < 1$ and $k$ leading to chaotic states that are prone to collapsing to the trivial fixed point.} Realizing these states in simulations is nontrivial because random initial conditions typically miss the dynamic attractor and decay to zero. A workaround is to deform a chaotic state with $g > 1$, for which the trivial fixed point is unstable, to a chaotic state for $g < 1$ by reducing $g$ in the spirit of annealing. Using this method, we verified that simulations agree with the DMFT solutions (Fig.~\ref{fig:plasticity-induced-chaos}(a), dashed lines).

After realizing plasticity-induced chaotic states via this procedure, we observe transient activity with a lifetime that is approximately log-normally distributed (Fig.~\ref{fig:plasticity-induced-chaos}(b), right). The median log-lifetime before collapsing is linear in $N$ over several decades, consistent with the typical lifetime diverging exponentially and becoming infinite in the limit ${N \rightarrow \infty}$ in which the DMFT applies (Fig.~\ref{fig:plasticity-induced-chaos}(b), left). This divergence is faster for larger $k$.

% \blue{This phenomenology is reminiscent of \textit{stable chaos,} a state observed in spiking networks in which an exponentially long dynamic transient with temporally stationary properties eventually collapses to a fixed point \cite{zillmer2006desynchronization}.}

\subsection{First-order transition to \blue{nontrivial DMFT solutions} for ${g < 1}$}

\label{subsec:transition-g-greater-one}

\begin{figure}
    \centering
    \includegraphics[width=\columnwidth]{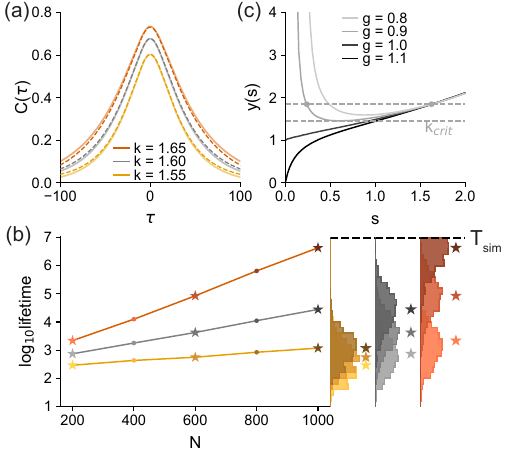}
    \caption{Chaotic states for ${g < 1}$. (a) $C(\tau)$ from the DMFT (lines) and in simulations for $g = 0.9$, $p=2.5$, and various values of $k$. (b) Left: median log-lifetime of transient activity, before collapsing to the trivial fixed point, as a function of $N$ for $g = 0.9$ and values of $k$ from (a). Right: histograms of the log-lifetime of transient chaos, corresponding to stars in the left plot. Simulations were terminated at time $T_\text{sim} = 10^7$. (c) Curves $y(s)$ for solutions of the model of \citet{stern2014dynamics} for various values of $g$. Dashed horizontal lines correspond to different values of $k$, intersecting $y(s)$ at self-consistent solutions of Eq.~(\ref{eq:slow-single-site-picture}).}
    \label{fig:plasticity-induced-chaos}
\end{figure}

For ${g < 1}$, if $k$ is not large enough to induce \blue{dynamic activity}, the plastic network is globally quiescent. We now analyze the boundary between these phases. For order-one values of $p$, we find $\tau^* \gg p$ for dynamic states for ${g < 1}$, which reduces the single-site dynamics to the ``slow'' form
\begin{equation}
    (1 + \partial_t)x(t) = \eta(t) + k C(0) \phi(t).
\label{eq:slow-single-site-picture}
\end{equation}
This single-site picture is related to that of a nonplastic network with order-one self-coupling parameter $s$ for which the single-site dynamics read
\begin{equation}
    (1 + \partial_t) x(t) = \eta(t) + s \phi(t),
    \label{eq:stern-single-site-picture}
\end{equation}
as studied by \citet{stern2014dynamics}. This network has a continuous transition from quiescence to \blue{dynamic activity} at ${g + s = 1}$. We map solutions of the Stern network onto solutions of the plastic network by enforcing ${y(s) = {s}/{C(0)} = k}$. For a given $g < 1$, $C(0)$ becomes nonzero at $s = 1 - g$, grows with $s$, and saturates at unity; thus, $y(s)$ descends from infinity at $s = 1 - g$ and grows as $s$ for large $s$, tracing a U shape [Fig.~\ref{fig:plasticity-induced-chaos}(c)]. Each $k$ draws a horizontal line intersecting $y(s)$ at self-consistent solutions of Eq.~(\ref{eq:slow-single-site-picture}). The critical $k$, defining the boundary between phase-diagram regions iv and v, is the minimum of $y(s)$ (Fig.~\ref{fig:plasticity-induced-chaos}(c), lower dashed line). As $C(0)$ is finite here, this is a discontinuous, first-order transition. Physically, this is because sustaining \blue{dynamic activity} for $g < 1$ requires finite self-coupling. Since the self-coupling depends on $C(0)$, \blue{time-dependent} solutions with arbitrarily small $C(0)$ are not possible. As $g \rightarrow 0$, the dynamic timescale diverges, leaving only fixed points. We show in Sec.~\ref{sec:fixed-points} that fixed points exist for $k > 2.02$ at $g = 0$.

Due to the U shape of $y(s)$ for ${g < 1}$, $y(s) = k$ has two solutions for $k$ larger than its critical value: one with large $s$ and $C(0)$, the other with small $s$ and $C(0)$ (Fig.~\ref{fig:plasticity-induced-chaos}(c), upper dashed line). As ${g \rightarrow 1^-}$, the small-$s$ solution vanishes (Sec.~\ref{subsec:transition-g-less-one}) while the large-$s$ solution remains finite. For ${g > 1}$, $y(s) = k$ has a unique solution whose deformation to $g < 1$ gives the large-$s$ solution (Fig.~\ref{fig:plasticity-induced-chaos}(c), $g = 1.1$ curve). 

\subsection{Second-order transition \blue{at $g = 1$}}
\label{subsec:transition-g-less-one}

Finally, we solve the DMFT for ${g \rightarrow 1^+}$, ${k < 1}$. This limit marks a continuous transition from \blue{dynamic activity} to global quiescence in which $C(0)$ vanishes and $\tau^*$ diverges. Due to the vanishing variance, the self-coupling in Eq.~(\ref{eq:slow-single-site-picture}) can be linearized,
\begin{equation}
    (1  - k C(0) + \partial_t) {x}(t) = \eta(t).
\end{equation}
This is equivalent to the single-site dynamics of the nonplastic network with ${g_{\text{eff}} = g/[1 - kC(0)]}$ and time constant ${\tau_{\text{eff}} = 1/[1 - kC(0)]}$. In the nonplastic network, to leading order in ${\epsilon = g - 1 \ll 1}$, ${C(\tau) = \epsilon \: \text{sech} ( {\epsilon \tau}/{\sqrt{3}} )}$ \cite{sompolinsky1988chaos}.
Enforcing $C(0) = g_{\text{eff}} - 1$ gives, for the plastic network,
\begin{equation}
C(\tau) = \gamma \: \text{sech}\left(\frac{\gamma \tau}{\sqrt{3}} \right), \:\: \gamma = \frac{\epsilon}{1-k},
\label{eq:limiting-autocov}
\end{equation}
to leading order in $\epsilon$. In contrast to the behavior away from this transition, activity becomes faster with increasing $k$. $C(0)$ diverges as $k \rightarrow 1^-$, indicating that solutions with $g = 1$ for $k > 1$ have finite variance. Validity of the solution Eq.~(\ref{eq:limiting-autocov}) requires ${k < 1}$, lest we obtain a nonphysical negative variance. On the other hand, as ${g \rightarrow 1^-}$ for $k > 1$, a positive variance is obtained; this is the small-$s$ solution for ${g < 1}$ described in Sec.~\ref{subsec:transition-g-greater-one}.

\section{High-dimensional analysis}
\label{sec:high-dimensional-analysis}

The DMFT describes the temporal structure of network activity through an effective single-site picture. Importantly, the network dynamics result from a complex interaction of high-dimensional neuronal-synaptic modes. We now probe the high-dimensional origin of the dynamics, first through an analytical study of the spectrum of the Jacobian describing the local, linear dynamics, and then through a numerical study of the Lyapunov spectrum describing the global, nonlinear dynamics. Both the Jacobian and Lyapunov spectra show a topological transition at large $k$ to a form with a slow, synapse-dominated band and a fast, neuron-dominated band, with the former driving network network activity. This suggests a flipped view of the network dynamics as being driven by the synaptic couplings, with neurons serving as the connections. 

% \subsection{Jacobian spectrum}
% \label{subsec:structure-of-the-jacobian}
\subsection{Jacobian spectrum}

\begin{figure}
    \includegraphics[width=\columnwidth]{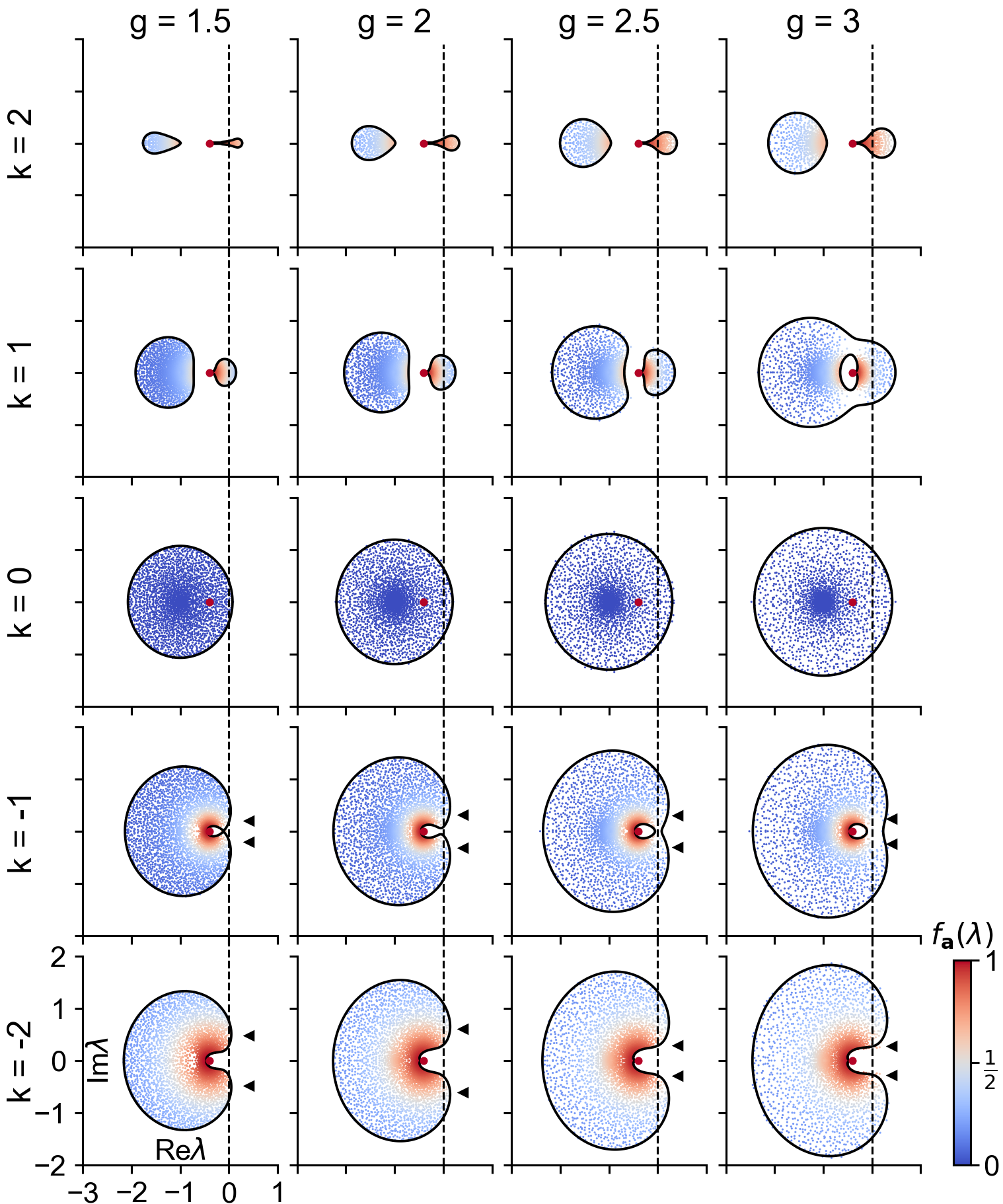}
    \caption{Spectra of the Jacobian for $p = 2.5$ and various values of $g$ (running horizontally) and $k$ (running vertically). Lines: predicted boundary curves from random matrix theory and DMFT. Dots: spectra measured in simulations of chaotic plastic networks. Modes are colored by $f_{\bm{a}}(\lambda)$, the weight on the synaptic part of the corresponding eigenvector of the reduced Jacobian [Eq.~(\ref{eq:weight})]. The red dot at $\lambda = -1/p$ indicates a delta function of $N^2 - N$ synaptic modes. For anti-Hebbian ($k < 0$) spectra, triangular markers indicate an oscillatory frequency computed from the zero-crossings of $C(\tau)$.}
    \label{fig:chaotic-spectra}
\end{figure}

Let us represent Eqs.~(\ref{eq:neuronal-dynamics}--\ref{eq:syn-dynamics}) defining the model in the form
\begin{subequations}
\begin{align}
\partial_t \bm{x}(t) &= \bm{F}(\bm{x}(t), \bm{a}(t)), \\
(1 + p \partial_t) \bm{a}(t) &= k \bm{G}(\bm{x}(t)),
\end{align}
\end{subequations}
where $\bm{a}(t) = \text{vec}\:\bm{A}(t)$ contains all $S = N^2$ elements of $\bm{A}(t)$. We use the notation ${\left[ {\partial \bm{x}} / {\partial \bm{y}}\right]_{ij} = \partial x_i / \partial y_j}$ for vectors $\bm{x}, \bm{y}$. The Jacobian is a ${(N+S)}$-dimensional block matrix,
\begin{equation}
\bm{M} = 
\left(\begin{array}{cc}
\pderiv{\dot{\bm{x}}}{\bm{x}} & \pderiv{\dot{\bm{x}}}{\bm{a}} \\
\pderiv{\dot{\bm{a}}}{\bm{x}} & \pderiv{\dot{\bm{a}}}{\bm{a}}
\end{array}\right)
=
\left(\begin{array}{cc}
\pderiv{\bm{F}}{\bm{x}} & \pderiv{\bm{F}}{\bm{a}} \\
\frac{k}{p}\pderiv{\bm{G}}{\bm{x}} & -\frac{1}{p}\bm{I}_S
\end{array}\right).
\end{equation}
At any instant, the neurons provide $N$-dimensional input to the $S$ synaptic variables, inducing a low-dimensional structure of the Jacobian. Rather than consider $\bm{M}$ directly, it is convenient to study the linear dynamics that it generates,
\begin{subequations}
\begin{align}
\partial_t \delta \bm{x}(t) &= \pderiv{\bm{F}}{\bm{x}} \delta \bm{x}(t) + \pderiv{\bm{F}}{\bm{a}} \delta \bm{a}(t), \\
(1 + p \partial_t) \delta \bm{a}(t) &= k \pderiv{\bm{G}}{\bm{x}} \delta \bm{x}(t) .
\end{align}
\end{subequations}
The input that $\delta \bm{a}$ receives from $\delta \bm{x}$ is confined to a $N$-dimensional subspace spanned by the columns of ${\partial \bm{G}}/{\partial \bm{x}}$. Perturbations to $\delta \bm{a}$ in the $(S-N)$-dimensional orthogonal complement subspace relax with timescale $p$. Due to these relaxational modes, $\bm{M}$ has eigenvalue $ -1/p$ with multiplicity $S - N$. The remaining $2N$ eigenvalues result from the interaction of $\delta \bm{x}$ and the component of $\delta \bm{a}$ in the $N$-dimensional subspace that receives input from $\delta \bm{x}$. Projecting $\delta \bm{a}$ into this subspace via
$
\delta \tilde{\bm{a}} = ( {\partial \bm{G}}/{\partial \bm{x}} )^{+} \delta \bm{a},
$
where $(\cdot)^+$ denotes the pseudoinverse, the $2N$-dimensional dynamics of interest are
\begin{subequations}
\begin{align}
\partial_t \delta \bm{x}(t) &= \pderiv{\bm{F}}{\bm{x}} \delta \bm{x}(t) + \pderiv{\bm{F}}{\bm{a}} \pderiv{\bm{G}}{\bm{x}} \delta \tilde{\bm{a}} (t), \\
(1 + p \partial_t) \delta \tilde{\bm{a}}(t) &= k \delta \bm{x}(t).
\end{align}
\end{subequations}
% Using the pseudoinverse projection, $\delta \tilde{\bm{a}}$ low-pass filters $\delta \bm{x}(t)$. 
The associated $2N$-dimensional dynamics matrix is
\begin{equation}
\tilde{\bm{M}} = 
\left(\begin{array}{cc}
\pderiv{\bm{F}}{\bm{x}} & \pderiv{\bm{F}}{\bm{a}} \pderiv{\bm{G}}{\bm{x}}  \\
\frac{k}{p}\bm{I}_{N} & -\frac{1}{p}\bm{I}_{N}
\end{array}\right).
\end{equation}
In summary, the eigenvalues of $\bm{M}$ are $ -{1}/{p}$ with multiplicity $S - N$ together with the $2N$ eigenvalues of $\tilde{\bm{M}}$. We refer to $\tilde{\bm{M}}$ as the \textit{reduced Jacobian} \footnote{There is a more direct derivation of the spectrum of $\bm{M}$, though it does not provide an interpretation of the reduced $2N$-dimensional dynamics. We use the Schur-complement identity
$
{\left|\begin{array}{cc}
\bm{A} & \bm{B} \\
\bm{C} & \bm{D}
\end{array}\right| = |\bm{D}||\bm{A} - \bm{B}\bm{D}^{-1}\bm{C}  |
}$
to write the characteristic polynomial of $\bm{M}$ as
$
p_{\bm{M}}(\lambda) \propto \left(1 + p\lambda\right)^{S - N}
\left| \lambda^2 p \bm{I}_N + \lambda \left( \bm{I}_N - p\pderiv{\bm{F}}{\bm{x}} \right) - \left(\pderiv{\bm{F}}{\bm{x}} + k \pderiv{\bm{F}}{\bm{a}} \pderiv{\bm{G}}{\bm{x}} \right) \right|
$.
The first term, $(1 + p\lambda)^{S-N}$, encodes the eigenvalue $ -1/p$ with multiplicity $S-N$. The second term encodes an $N$-dimensional quadratic eigenvalue problem (QEP). A QEP can be solved by {linearization}, which entails forming a $2N$-dimensional linear eigenvalue problem whose eigenvalues coincide with those of the QEP \cite{tisseur2001quadratic}. Applying the aforementioned identity to the determinant defining the characteristic polynomial of the reduced Jacobian $\tilde{\bm{M}}$ shows that $\tilde{\bm{M}}$ is one such linearization of the QEP. Thus, we recover the originally derived spectrum. There are infinitely many linearizations given by matrices related to $\tilde{\bm{M}}$ by similarity transformation, so in this sense, the reduced Jacobian is not unique.}.

Each eigenvector of $\tilde{\bm{M}}$ can be written in terms of its $\bm{x}$ and $\bm{a}$ components,
$
    \bm{v} = 
    \left(\bm{v}_{\bm{x}}, \:
    \bm{v}_{\bm{a}}
    \right)
$. From the lower block row of $\tilde{\bm{M}}$, we obtain the relation
$
    k\bm{v}_{\bm{x}} = (1 + p\lambda)\bm{v}_{\bm{a}}
$.
Letting $f_{\bm{a}}(\lambda) = \lVert \bm{v}_{\bm{a}} \rVert^2 / \lVert \bm{v} \rVert^2$ denote the relative weight on the synaptic component, this relation implies
\begin{equation}
    f_{\bm{a}}(\lambda) = \frac{k^2}{k^2 + p^2 |\lambda + 1/p|^2},
\label{eq:weight}
\end{equation}
which falls off radially with distance from $-1/p$. Modes become synapse-dominated as $\lambda \rightarrow -1/p$, giving way to a delta function of $S-N$ purely synaptic modes at this point.

% As $k \rightarrow 0$, the spectrum becomes fully $\bm{x}$-dominated except very close to $-1/p$, namely, $|\lambda + 1/p| \sim k/p$. Note that $f_{\bm{a}}(\lambda)$ and $f_{\bm{x}}$ give the weight on the $\bm{a}$ and $\bm{x}$ variables in a $2N$-dimensional eigenvector of the reduced Jacobian, which is distinct from what one would measure in the corresponding ${(N + S)}$-dimensional eigenvector of the full Jacobian, but provides a useful measure of the neuronal or synaptic content of different modes nevertheless. 

% \subsection{Random matrix theory}
% \label{subsec:random-matrix-theory}
% From this analysis, the eigenvalues of the Jacobian of the plastic network are $-1/p$, with multiplicity $N^2 - N$, and the $2N$ eigenvalues of the reduced Jacobian $\tilde{\bm{M}}$.

The Jacobian analysis thus far holds for any $\bm{F}(\cdot, \cdot)$, $\bm{G}(\cdot)$, and $S > N$. In particular, the simplification to the structure of the Jacobian does not depend on the specific plasticity rule posited in Eq.~(\ref{eq:syn-dynamics}). We now substitute the forms of $\bm{F}(\cdot, \cdot)$ and $\bm{G}(\cdot)$ from Eqs.~(\ref{eq:neuronal-dynamics}--\ref{eq:syn-dynamics}), yielding ${\tilde{\bm{M}} = \tilde{\bm{M}}_{\text{bulk}} + \tilde{\bm{M}}_{\text{low-rank}}}$, where
\begin{subequations}
\begin{align}
&\tilde{\bm{M}}_{\text{bulk}} = 
\tbtmat{
-\bm{I}_N + \bm{J} \diag{\bm{\phi}'}}{C(0) \diag{\bm{\phi}'} }
{\frac{k}{p}\bm{I}_{N}} {-\frac{1}{p}\bm{I}_{N}}, \\
&\tilde{\bm{M}}_{\text{low-rank}} = 
\tbtmat{
\bm{A} \diag{\bm{\phi}'} }{
\frac{1}{N}\bm{\phi}\bm{\phi}^T \diag{\bm{\phi}'}
}{0}{0}.
\end{align}
\end{subequations}
Here, $(\bm{x}, \bm{A})$ is a point in phase space, ${\bm{\phi} = \phi(\bm{x})}$, ${\bm{\phi}' = \phi'(\bm{x})}$, and ${C(0) = \left\lVert \bm{\phi} \right\rVert^2}/N$. $\tilde{\bm{M}}_{\text{bulk}}$ generates the bulk of the spectrum of $\tilde{\bm{M}}$ while $\tilde{\bm{M}}_{\text{low-rank}}$ can, in principle, contribute an intensive number of outlier eigenvalues. However, both in chaotic states and at fixed points, we find that the reduced Jacobian does not have outliers \footnote{In chaotic states, this is a result of the network being locally destabilized by a continuous band of eigenvalues in analogy with the nonplastic network \cite{sompolinsky1988chaos}. At fixed points, this is a result of the marginal stability of typical fixed points, which have a continuous band of Jacobian eigenvalues brushing against the stability line, $\text{Re}(\lambda) = 0$.}. We therefore focus on the spectrum of $\tilde{\bm{M}}_{\text{bulk}}$.

We compute the boundary curve encompassing the compact spectrum of $\tilde{\bm{M}}_{\text{bulk}}$ for ${N \rightarrow \infty}$ using a theorem of \citet{ahmadian2015properties} concerning random matrices expressible as a linear reparameterization of an elementwise independent and identically distributed random matrix (Appendix~\ref{sec:random-matrix-theory}). This shows that the limiting spectral density of $\tilde{\bm{M}}_{\text{bulk}}$ has support at $\lambda \in \mathbb{C}$ if
\begin{equation}
    \aavg{\left| \frac{g(1 + p \lambda)}{(1 + p \lambda)(1 + \lambda) \frac{1}{\phi'(x)} - k C(0) } \right|^2}_x \geq 1,
\label{eq:boundaryformula}
\end{equation}
where $\tavg{\cdot}_x$ is an average over the components of $\bm{x}$.

We use this result to probe the high-dimensional origin of the dynamics in the plastic network, noting that the spectral density of the Jacobian computed at any point on a dynamic trajectory is time-independent and self-averaging as $N \rightarrow \infty$.  We use the DMFT to obtain a Monte-Carlo estimate of $\tavg{\cdot}_x$. The predicted boundary tightly hugs the Jacobian spectra evaluated from simulations in dynamic states (Fig.~\ref{fig:chaotic-spectra}). These simulation spectra do not contain outliers, justifying our focus on $\tilde{\bm{M}}_\text{bulk}$. We place a dot at $-1/p$ to indicate the delta function of $N^2 - N$ synaptic relaxational modes present in the spectrum of $\bm{M}$.
% We color each eigenvalue according to the weight on the neuronal and synaptic components of the corresponding eigenvector of the reduced Jacobian. The dot at $-1/p$ indicates a delta function of $N^2 - N$ synaptic relaxational modes. As per Eq.~(\ref{eq:weight}), the proportion of synaptic weight within a given mode decays radially away from $-1/p$. 
Modes past the stability line, $\text{Re}(\lambda) = 0$, locally destabilize the network and drive the dynamics.

Setting $k = 0$ recovers the circularly symmetric spectrum of the nonplastic network, enclosed by ${|1 + \lambda| \leq g\sqrt{ \tavg{\phi'(x)^2}_x}}$ as predicted by Eq.~(\ref{eq:boundaryformula}) (Fig.~\ref{fig:chaotic-spectra}, $k=0$ row). As $k$ is increased, the delta function at $-1/p$ repels eigenvalues leftward, creating a hole (Fig.~\ref{fig:chaotic-spectra}, $g=3$, $k=1$ spectrum). Further increasing $k$ produces a topological transition to a spectrum with two bands and no holes (Fig.~\ref{fig:chaotic-spectra}, all other spectra with $k> 0$). Unstable modes are increasingly focused along the real axis, leading to slow activity as seen via the DMFT. The slow, destabilizing band is dominantly synaptic, and the fast, relaxational band is dominantly neuronal. The slowest-relaxing of the fast modes have real part close to $-1$, the neuronal decay timescale. This two-band topology therefore reflects a dynamic state in which slow network activity is synapse-driven rather than neuron-driven. 

As $k$ is decreased from zero into the anti-Hebbian regime, the delta function repels eigenvalues rightward, creating a hole to its right (Fig.~\ref{fig:chaotic-spectra}, $k = -1$ row). Further decreasing $k$ produces a topological transition to a form with a single band and no holes (Fig.~\ref{fig:chaotic-spectra}, $k=-2$ row). In contrast to the Hebbian spectra, there are pronounced lobes of dominantly imaginary unstable modes, generating fast, oscillatory activity also seen via the DMFT. The oscillatory frequency computed from zero-crossings of $C(\tau)$, as in Fig.~\ref{fig:plasticity-shaped-chaos}(c), corresponds roughly to the locations of the imaginary lobes of the spectra, particularly for smaller values of $g$ (Fig.~\ref{fig:chaotic-spectra}, triangular markers for $k < 0$ spectra).

% \subsection{Lyapunov spectrum}
% \label{subsec:lyapunov}
\subsection{Lyapunov spectrum}
\label{subsec:lyapunov}
\begin{figure}
    \includegraphics[width=\columnwidth]{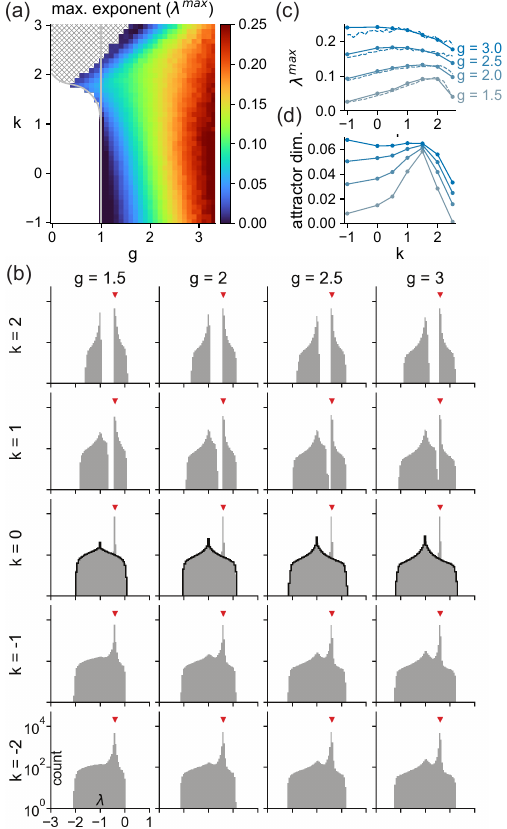}
    \caption{(a) Maximum Lyapunov exponent $\lambda_\text{max}$, computed by a perturbation method, throughout $(g, k)$ parameter space with $p = 2.5$, $N=4000$. \blue{White: quiescence. Hatched: convergence to nonzero fixed points.} (b) Histograms of Lyapunov spectra, computed using tangent-vector propagation with $N = 900$, for $p = 2.5$ and various values of $g$ (running horizontally) and $k$ (running vertically). Black outline for $k = 0$ histograms: spectra of nonplastic network. The red triangle marks $-1/p$, where there are $\mathcal{O}(N^2)$ exponents in the full spectrum. (c) $\lambda_\text{max}$ as a function of $k$ for various values of $g$. Solid lines: estimate from tangent-vector propagation. Dashed lines: estimate from perturbation method. (d) Attractor dimension \blue{divided by $N$} as a function of $k$ for various value of $g$.}
    \label{fig:lyapunov}
\end{figure}While analytically accessible, the Jacobian spectrum describes only the locally linear dynamics. Rigorously characterizing the nonlinear dynamics requires a calculation of the spectrum of Lyapunov exponents, which are defined as follows. Suppose a ball of radius $\epsilon$ is tossed into the flow. As the dynamics unfold, the ball expands and contracts into an ellipsoid. The Lyapunov exponents are the exponential growth or decay rates of the principal axes of this ellipsoid as $t \rightarrow \infty$ (simultaneously, $\epsilon \rightarrow 0$ is taken so that the ellipsoid stays small). A positive Lyapunov exponent implies exponential sensitivity to initial conditions, indicating chaos. The Lyapunov spectrum cannot be derived from the Jacobian spectrum due to both the non-normality of the Jacobian and the time-dependence of its eigenvectors.

We first study the maximum Lyapunov exponent $\lambda^\text{max}$, which dominates trajectory divergence as $t \rightarrow \infty$. We measured $\lambda^\text{max}$ in simulations by injecting a small perturbation to the system and measuring the slope (versus $t$) of the log of the norm of the difference between the perturbed and unperturbed trajectories. For $g < 0$, we realized \blue{dynamic network states} using the deformation method described in Sec.~\ref{subsec:chaotic-states-g-less-one}. \blue{For each setting of $(g, k)$, we simulated 200 random networks.} The output of this analysis is displayed as a heatmap in $(g,k)$ parameter space in Fig.~\ref{fig:lyapunov}(a). \blue{Parameter values that resulted in convergence to nonzero fixed points (Section \ref{sec:fixed-points}) within the simulation time for at least $80\%$ of networks are hatched; values that resulted in quiescence of all networks are white.}

\blue{In regions of parameter space that do not converge to fixed points over the simulation time, $\lambda^\text{max}$ is positive, indicating chaos. This includes part of the region $g < 1$, confirming that plasticity can induce chaos in an otherwise quiescent network.}
% , indicating that the solutions studied using the DMFT are indeed chaotic
This analysis provides a simulation-based confirmation of the boundary marking the first-order transition to nontrivial DMFT solutions for $g < 1$ derived in Sec.~\ref{subsec:transition-g-less-one} [Fig.~\ref{fig:lyapunov}(a), gray lines].

\blue{As $k$ is increased and/or $g$ is decreased, we observe a smaller and smaller Lyapunov exponent that eventually results in simulations reliably collapsing to nonzero fixed points (Fig.~\ref{fig:lyapunov}(a), solid-to-hatched boundary). This crossover occurs in a parameter regime for which phase space is densely filled with stable fixed points (Sec.~\ref{sec:fixed-points}). Using the present finite-$N$ analysis, we are unable to determine whether $\lambda^\text{max}$ in the hatched region in Fig.~\ref{fig:lyapunov}(a) is small and positive, or negative. Additionally, solving the DMFT in the hatched region is numerically difficult due to the diverging dynamic timescale (Sec.~\ref{subsec:plasticity-shaped-chaos}). As $N$ is increased over a decade, the boundary marking this crossover shifts slightly upward (Fig.~\ref{fig:lyap-appdx}).}

We next study the full Lyapunov spectrum for chaotic states. \blue{In general, Lyapunov spectra can be computed numerically by propagating a set of vectors in the tangent space of the flow and periodically orthonormalizing them to prevent their explosion/vanishing and to extract their growth/decay rates, as explained in detail in prior works \cite{geist1990comparison, engelken2023lyapunov, krishnamurthy2022theory}.} The plastic network has $N+N^2$ variables, so propagating a complete basis is prohibitively computationally expensive for large $N$. Fortunately, $\mathcal{O}(N^2)$ exponents concentrate at $-1/p$, so it suffices to compute the $\mathcal{O}(N)$ largest and smallest exponents. We compute the largest exponents using the aforementioned procedure with an undercomplete set of $\mathcal{O}(N)$ tangent-space vectors. We find the smallest exponents by doing the same for the time-reversed dynamics, noting that the smallest exponents are the largest of the time-reversed system. A complication is that the time-reversed dynamics are unstable. We therefore run time-reversed tangent-space dynamics, tamed by orthonormalization, atop a time-reversed trajectory produced by the forward-time dynamics. We verified that this method accurately computes the largest and smallest Lyapunov exponents of the nonplastic network \cite{sompolinsky1988chaos, engelken2023lyapunov}.

Histograms of the Lyapunov spectra are shown in Fig.~\ref{fig:lyapunov}(b). For $k=0$, the spectrum is the same as that of a nonplastic network with a spike at $-1/p$ \footnote{If the whole spectrum were computed, this spike would have $\mathcal{O}(N^2)$ exponents; here, there are fewer since only the ends of the spectrum are computed. Moreover, in principle it should be a delta-function spike, but in the numerics there is some ``leakage'' across modes such that the spike has finite width.}.
We verified that the measurement of $\lambda^\text{max}$ obtained using this method matches the perturbation measurement displayed in the heatmap [Fig.~\ref{fig:lyapunov}(c)].

For large $k$, the Lyapunov spectra recapitulate the topological transition to two bands of the Jacobian spectra, further demonstrating a synapse-driven dynamic state [Fig.~\ref{fig:lyapunov}(b)]. In analogy with the Jacobian spectra, the slow, destabilizing Lyapunov band spans $-1/p$ to $\lambda^\text{max}$, and the fast, relaxational band has an upper limit near $-1$.

% I’m not sure if this kind of topological transition in a Lyapunov spectrum has been observed before.
Our calculation of the Lyapunov spectrum allows for further calculation of diffeomorphic-invariant properties of the strange attractor \cite{engelken2023lyapunov}. We focus on its dimension, shown in Fig.~\ref{fig:lyapunov}(d), defined by the Kaplan–Yorke conjecture as the number of exponents that must be summed, ranked in descending order, to achieve a cumulative sum of zero \cite{kaplan2006chaotic}. \blue{We display an intensive version of this quantity that has been divided by $N$ \cite{clark2023dimension, engelken2023lyapunov}.} Both $\lambda^\text{max}$ and the attractor dimension vary nonmonotonically in $k$ [Fig.~\ref{fig:lyapunov}(c,d)]. This contrasts with the dynamic timescale $\tau^*$, which increases monotonically [Fig.~\ref{fig:plasticity-shaped-chaos}(b)]. \blue{The eventual decrease of the attractor dimension is reminiscent of the behavior of the participation ratio-based dimension of activity in networks with partially symmetric connectivity, which was recently computed analytically \cite{clark2023dimension}.} The decline of $\lambda^\text{max}$ and the attractor dimension at large $k$ likely reflects the proliferation of stable fixed points throughout phase space, the subject of the next section.

\section{Fixed points}
\label{sec:fixed-points}
\begin{figure}
    \centering
    \includegraphics[width=\columnwidth]{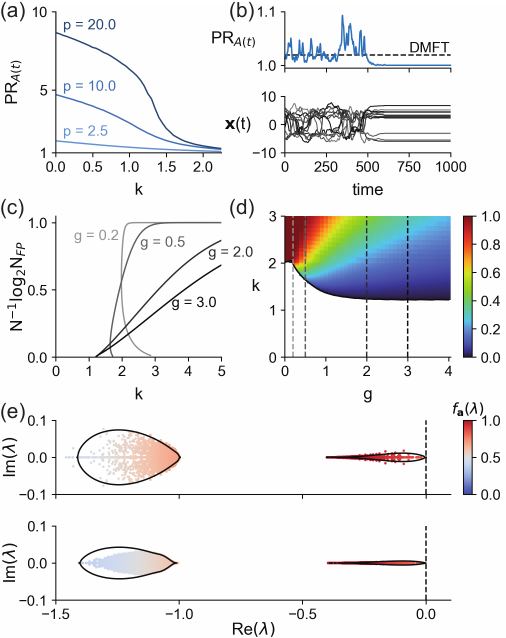}
    \caption{Stable fixed points. (a) DMFT participation ratio [Eq.~(\ref{eq:pr-dmft})] of $\bm{A}(t)$ as a function of $k$ for $g = 3$ and various values of $p$.
    (b) Top: empirical participation ratio [Eq.~(\ref{eq:pr-def})] of $\bm{A}(t)$ with $g = 2$, $k = 2.25$, and $p = 2.5$. The network settles to a stable fixed point. Dashed line: DMFT value. Bottom: neuronal traces during the same settling event.
    (c) Log-number of stable fixed points per neuron as a function of $k$ for various values of $g$. (d) Log-number of stable fixed points per neuron throughout $(g, k)$ parameter space.
    (e) Jacobian spectra at fixed points. Lines: predicted boundary curve from mean-field analysis. Dots: spectra after settling to a fixed point in many simulations. Top: $g = 2$, $k = 2.25$, $p = 2.5$ [as in (b)]. Bottom:  $g = 0.5$, $k = 1.9$, $p = 2.5$.}
    \label{fig:fixed-points}
\end{figure}
For large $k$, the dynamics of the plastic network are shaped by a proliferation of stable fixed points, and finite-size networks settle to these fixed points after transient chaos (Fig.~\ref{fig:phase-diagram}(b), hatched region and vi). We first probe this settling process by analyzing how $\bm{A}(t)$ collapses to a rank-one state, measuring the approximate rank of $\bm{A}(t)$ as the participation ratio of its spectrum, $\{\lambda_i(t)\}$,
\begin{equation}
	\text{PR}_{\bm{A}(t)} = \frac{\left(\sum_i \lambda_i(t)\right)^2}{\sum_i \lambda_i^2(t)} = \frac{\left(\text{tr} \bm{A}(t)\right)^2}{\left\lVert\bm{A}(t)\right\rVert_F^2}.
	\label{eq:pr-def}
\end{equation}
Note that if $\bm{A}(t)$ encodes $P$ decorrelated neuronal states with equal magnitude (i.e., $\lambda_1(t),\ldots,\lambda_P(t) = \text{const.}$ and $\lambda_{P+1}(t), \ldots, \lambda_N(t) = 0$), then $\text{PR}_{\bm{A}(t)} = P$. Evaluating Eq.~(\ref{eq:pr-def}) in the limit ${N \rightarrow \infty}$ gives
\begin{equation}
	\text{PR}_{\bm{A}(t)} = \frac{p}{\mathcal{T}}, \:\: \mathcal{T} = \int_{0}^{\infty} d\tau e^{-\tau/p} \left[ \frac{C(\tau) }{C(0)}\right]^2.
 \label{eq:pr-dmft}
\end{equation}
Thus, $\text{PR}_{\bm{A}(t)}$ is intensive and, as $N \rightarrow \infty$, time-independent. If $\tau^* \ll p$, then $\mathcal{T} \approx \tau^*$, so $\text{PR}_{\bm{A}(t)}$ is the ratio of these timescales. On the other hand, if $\tau^* \gg p$, then $\mathcal{T}$ is slightly less than $p$, so $\text{PR}_{\bm{A}(t)}$ is slightly larger than unity.

Hebbian plasticity tends to slow chaos. As $k$ is increased, $\text{PR}_{\bm{A}(t)}$ therefore drops closer to unity, with temporal fluctuations about the mean-field value in finite-size networks [Fig.~\ref{fig:fixed-points}(b)]. During chaos, neurons continuously escape the synaptic drag. However, in finite-size networks with sufficiently large $k$, synapses can ``win,'' namely, the network fluctuates into fixed point of the form $\left(\bm{x}(t), \bm{A}(t)\right) = \left(\bm{x}, \: \phi(\bm{x})\phi(\bm{x})^T \right)$ with $\text{PR}_{\bm{A}(t)} = 1$ [Fig.~\ref{fig:fixed-points}(b)]. These fixed points are stable with respect to the combined neuronal-synaptic dynamics.
% \subsection{Counting the fixed points}
% \label{sec:counting-the-fixed-points}

We now compute the number of stable fixed points in phase space. As $k \rightarrow \infty$, there is a stable fixed point associated with each of the $2^N$ binary neuronal states. Thus, any initial condition is pulled to a fixed point with high overlap with the initial neuronal state, consistent with the diverging dynamic timescale $\tau^*$ in this regime (Sec.~\ref{subsec:plasticity-shaped-chaos}). For finite $k$, we expect exponentially many stable fixed points. This is in contrast to the nonplastic network \cite{sompolinsky1988chaos}, for which there exist exponentially many fixed points for $g >1$, but they are all unstable \cite{wainrib2013topological, stubenrauch2022phase}. Due to the exponential scaling, we will study the log-number of stable fixed points per neuron, an intensive, self-averaging quantity. 
% Identifying the parameter regime where this order parameter is nonzero allows us to draw the boundary curve separating the parameter regimes with and without stable fixed points (Fig.~\ref{fig:phase-diagram}vi, boundary of hatched region).
% Such fixed points are stable with respect to the combined neuronal-synaptic dynamics, with
Fluctuations around fixed points are governed by the Jacobian spectrum analyzed in Sec.~\ref{sec:high-dimensional-analysis}. Our derivation follows that of \citet{stern2014dynamics}, differing in its initial steps that enforce stability using this neuronal-synaptic spectrum. 

At a fixed point, the DMFT equations reduce to
\begin{equation}
x = \eta + k C(0) \phi(x),
\label{eq:fixed-point-mft}
\end{equation}
where $\eta \sim \mathcal{N}(0, g^2 C(0))$ and $C(0) = \aavg{\phi^2(x)}_{\eta}$. Eq.~(\ref{eq:fixed-point-mft}) can be written ${g(x) = x - k C(0) \phi(x) = \eta}$. The Gaussian distribution over $\eta$ induces a non-Gaussian distribution over $x$ due to the nonlinearity of $g(x)$. If ${|\eta| < \eta_m}$ for some $\eta_m$, there are three solutions to $g(x) = \eta$. The two outer solutions are at points of positive slope of $g(x)$; the central solution has negative slope. The negative-slope solution renders a fixed point unstable for the following reason. The denominator of the averaged quantity in the Jacobian boundary [Eq.~(\ref{eq:boundaryformula})] is
\begin{equation}
\begin{split}
\frac{1}{\phi'(x)} \left( g'(x) + \lambda + p\lambda(1 + \lambda )\right).
\end{split}
\label{eq:denominator}
\end{equation}
Since ${\lambda + p\lambda(1 + \lambda )}$ varies from zero to infinity as $\lambda$ varies from zero to infinity, $g'(x) < 0$ causes the denominator to vanish at for some $\lambda > 0$, precluding stability. Two solutions remain for $|\eta| < \eta_m$. \citet{stern2014dynamics} observed that typical fixed points maximize the combinatorial number of solutions subject to stability and, moreover, are marginally stable, meaning that the spectral boundary sits at $\lambda = 0$. Setting $\lambda = 0$ in Eq.~(\ref{eq:boundaryformula}), marginal stability requires
\begin{equation}
    \tavg{\left( \frac{g}{\cosh^2x - k C(0)} \right)^2}_x = 1.
\label{eq:marginal-stability-cond}
\end{equation}
Since $p$ appears in neither Eq.~(\ref{eq:fixed-point-mft}) nor Eq.~(\ref{eq:marginal-stability-cond}), it drops out of the calculation, indicating that the number of stable fixed points is independent of this timescale [note, however, that the shape of the Jacobian spectrum is $p$-dependent; Fig.~\ref{fig:fixed-points}(e)]. Following Sec.~\ref{sec:dynamical-mean-field-theory}, we map solutions of the nonplastic network of \cite{stern2014dynamics} onto solutions of the plastic network by enforcing $s = kC(0)$.

The structure the solutions is as follows. For each $g$, there is an onset of fixed points at a critical $k$ [Fig.~\ref{fig:fixed-points}(c)]. The log-number grows monotonically with $k$ and saturates at ${\log_2 2^N / N = 1}$. The onset of fixed points is discontinuous for ${g < 0.76}$, in which case there are two solutions of the fixed-point mean-field theory for a given $(g, k)$: one with large $s$ and $C(0)$, the other with small $s$ and $C(0)$. The former are exponentially dominant. In Fig.~\ref{fig:fixed-points}(d), we display the log-number of fixed points per neuron as a heatmap in $(g, k)$ parameter space.

We now study the spectra of fluctuations around fixed points. We used the analytical form of the distribution over $x$ from the mean-field analysis to evaluate $\tavg{\cdot}_x$ in Eq.~\ref{eq:boundaryformula}, yielding a prediction for the Jacobian spectrum at fixed-points. We also allowed many simulations of chaotic networks to settle to fixed points, then measured their Jacobian spectra. We find that the simulation fixed points are indeed marginal, with spectra described accurately by the theoretical prediction [Fig.~\ref{fig:fixed-points}(e)].

For $g = 0$, fixed points satisfy $x_i = k N^{-1}\lVert \phi(\bm{x}) \rVert^2 \phi(x_i)$ and thus $x_i = \pm \chi$ where $\chi = k \phi^3(\chi)$. For fixed points to exist, $k$ must be large enough for this constraint to have a nontrivial solution in $\chi$. The smallest such $k$ satisfies ${1 = 3 k \phi^2(\chi) \phi'(\chi)}$ (upon differentiating both sides in $\chi$). Combining these constraints gives $\phi(\chi)/\phi'(\chi) = 3\chi$ with solution ${\chi = \pm 1.42}$, and thus ${k = 2.02}$. The stability of fixed points with this critical value of $k$ is checked by direct evaluation of the eigenvalues of the reduced Jacobian, which are
${\lambda = -({1 + p \pm \sqrt{(1 + p)^2 - {8p}/{3}}})/{2p} < 0}$ with multiplicity $N-1$ for each sign, comprising the bulk;
${\lambda = -{1}/{p} - {2}/{3} < 0}$ with multiplicity one; and
${\lambda = 0}$ with multiplicity one, indicating marginal stability.
Further increasing $k$ gives proper stability. Here, the marginal eigenvalue is an outlier, while the bulk has finite negative real part. This suggests that the mean-field fixed-point calculation, which assumed marginal stability of the bulk, breaks down for small $g$. Simulations and theory agree down to $g = 0.5$, $k = 1.9$, so such a breakdown would have to occur for smaller values of these parameters.

% Measuring fixed points in this parameter regime is challenging since simulations either settle to a nonzero fixed point during deformation, resulting in an atypical fixed point at the terminal $g$; or collapse to the trivial fixed point, preventing measurement altogether.

\section{Freezable chaos}
\label{sec:freezable-chaos}
\begin{figure*}
\includegraphics[width=\textwidth]{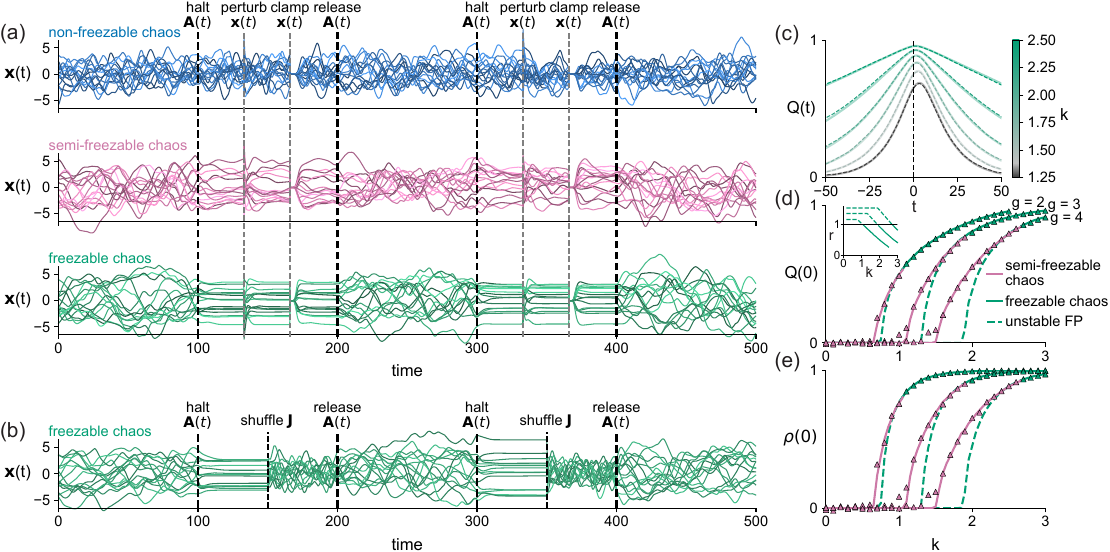}
    \caption{Freezable chaos. (a) Neuronal traces in nonfreezable (top, $k = 0.6$), semifreezable (middle, $k = 1.2$), and freezable (bottom, $k = 1.55$) chaos for $g = 2.25$ and $p = 2.5$. At the first vertical line, $\bm{A}(t)$ is halted. In (semi-)freezable chaos, a memory of the halt-time neuronal state persists. Stability is demonstrated by perturbing the neurons (second line) and clamping them near zero (third line). In (semi-)freezable chaos, neurons relax back to the stored pattern after these manipulations. At the fourth line, $\bm{A}(t)$ is released. This process is repeated at the fifth line. (b) Demonstration of the alignment between $\bm{J}$ and $\bm{A}(t)$. At the first vertical line, $\bm{A}(t)$ is halted. At the second line, $\bm{J}$, but not the halted $\bm{A}(t)$, is shuffled, destroying the neuronal fixed point. At the third line, $\bm{A}(t)$ is released. This process is repeated at the fourth line. $g = 2.5$, $k = 2.55$, and $p = 2.5$ [as in the freezable-chaos plot in (a)]. (c) Time-dependent overlap $Q(t)$ from the two-replica DMFT (solid lines) and in simulations (dashed lines) for $g = 2$, $p=2.5$, and various values of $k$. (d) Halt-time overlap $Q(0)$ from the two-replica DMFT (lines) and in simulations (triangular markers) as a function of $k$ for $p = 2.5$ and various values of $g$. Markers are colored according to whether simulations show semifreezable or freezable activity. (e) Same as (d), but using the zero-time correlation coefficient $\rho(0)$.
    In (c---e), We show only the positive solutions for $Q(t)$ and $\rho(t)$, which are realized upon halting synapses; there is also a symmetric negative solution. 
    }
    \label{fig:freezable}
\end{figure*}
The previous section characterized stable fixed points of the dynamics, as is typical dynamical-systems studies. Another question is whether a subsystem can have a stable fixed point that is unstable in the context of full system. We now study \textit{freezable chaos}, a state where a stable fixed point of neuronal dynamics is continuously destabilized through synaptic dynamics, generating chaos.

As described in Sec.~\ref{sec:phase-diagram}, for chaotic states with Hebbian plasticity, we define nonfreezable, semifreezable, and freezable chaos depending on the neuronal dynamics that ensue after halting synaptic dynamics. In \text{(semi-)freezable} chaos, neurons retain a stable memory of the halt-time state as we illustrate in Fig.~\ref{fig:freezable}(a).
% Picking the halt time to be $t= 0$, the couplings are left in the state $\bm{W}(0)$ and $\bm{\phi}(0)$ as the halt-time couplings and neuronal state, respectively. In nonfreezable chaos, upon halting synaptic dynamics, neuronal dynamics remain chaotic with no trace of $\bm{\phi}(0)$ in the activity [Fig.~\ref{fig:freezable}(a)]. In semifreezable chaos, neuronal dynamics remain chaotic with neurons fluctuating about a baseline that has nonvanishing overlap with $\bm{\phi}(0)$ [Fig.~\ref{fig:freezable}(b)]. In freezable chaos, neurons settle to a stable fixed point with nonvanishing overlap with $\bm{\phi}(0)$ [Fig.~\ref{fig:freezable}(c)].
% \begin{align}
% W_{ij}(0) &= J_{ij} + A_{ij}(0)\nonumber \\
% &= J_{ij} + \frac{k}{p} \int_0^\infty dt e^{-{t}/{p}} \phi_i(-t)\phi_j(-t).
% \label{eq:frozen}
% \end{align}
% We refer to 
% \footnote{We note that \citet{mastrogiuseppe2018linking} studied phases analogous to nonfreezable, semifreezable, and freezable chaos arising from static (nonplastic) structure in the couplings and termed them homogeneous chaotic, structured chaotic, and structured stationary, respectively.}.

Picking the halt time to be $t= 0$, the couplings remain at $\bm{W}(0)=\bm{J}+\bm{A}(0)$. Networks with such ``random-plus-low-rank'' couplings have been studied in the context of nonplastic networks by \citet{mastrogiuseppe2018linking}, who found chaotic, structured-chaotic, and structured-homogeneous phases that are qualitatively similar to nonfreezable, semifreezable, and freezable chaos in plastic networks. Crucially, our analysis in plastic networks is complicated by $\bm{A}(0)$ arising due to synaptic plasticity dependent upon neuronal activity driven by $\bm{J}$. More specifically, the neuronal states comprising $\bm{A}(0)$ are largely confined to a subspace spanned by dominant eigenvectors of $\bm{J}$ \cite{clark2023dimension}, inducing alignment between $\bm{J}$ and $\bm{A}(0)$. We demonstrate the importance of this alignment by running a simulation in which we halt $\bm{A}(t)$, storing the halt-time neuronal state as a stable fixed point [Fig.~\ref{fig:freezable}(b)]. We then shuffle $\bm{J}$ while keeping the halted $\bm{A}(t)$ fixed. This preserves the statistics of $\bm{J}$, but destroys correlations between $\bm{J}$ and the halted $\bm{A}(t)$. If these correlations were negligible, the neuronal fixed point would reorganize to a different fixed point. Instead, shuffling destroys the fixed point, causing the network to switch to neuronal chaos. Upon releasing $\bm{A}(t)$, the network returns to neuronal-synaptic chaos, and $\bm{A}(t)$ adapts to the permuted version of $\bm{J}$. At a later time, we halt synaptic dynamics again. As $\bm{A}(t)$ has adapted to the shuffled $\bm{J}$, a stable neuronal fixed point is created.

% \subsection{Replica dynamical mean-field theory}
% \label{subsec:replica-dynamical-mean-field-theory}
We handle this alignment through a replica mean-field analysis involving two networks, A and B, with neuronal states $\bm{\phi}^A(t) = \phi(\bm{x}^A(t))$ and $\bm{\phi}^B(t) = \phi(\bm{x}^B(t))$. Network A has neuronal-synaptic dynamics for all $t$ with quenched random couplings $\bm{J}$. Network B has neuronal dynamics for all $t$ with halt-time couplings $\bm{W}(0)$ constructed from the same $\bm{J}$ and neuronal states $\bm{\phi}^A(t)$ as network A,
\begin{equation}
W_{ij}(0) = J_{ij} + \frac{k}{p} \int_0^\infty dt e^{-{t}/{p}} \phi^A_i(-t)\phi^A_j(-t).
\label{eq:frozen}
\end{equation}
We define order parameters to characterize the phases of interest, starting with the overlap
\begin{equation}
    Q(t) = \lim_{t' \rightarrow \infty} \tavg{\phi^A_i(t) \phi_i^B(t')}_{\bm{J}}.
    \label{eq:order-param-def-Q}
\end{equation}
Evaluated at $t = 0$, this parameter indicates how accurately the halt-time neuronal state is retained as a memory. We also define the autocovariance function of network B,
\begin{equation}
    D(\tau) = \lim_{t \rightarrow \infty} \tavg{ \phi_i^B(t) \phi_i^B(t+\tau)  }_{\bm{J}}, \label{eq:order-param-def-D}
\end{equation}
where we assume statistical stationary in time. The autocovariance function of network A is $C(\tau)$, derived previously. Finally, assessing stability of the neuronal fixed point, when one exists, requires the stability matrix ${-\bm{I}_N + \bm{W}(0) \text{diag}\left(\phi'\left(\bm{x}^B\right)\right)}$, where $\bm{x}^B$ is a fixed point in network B. Stability requires that the spectrum has negative real part. Instability is dominated by the circularly symmetric bulk, so it suffices to compute its radius $r$,
\begin{equation}
    r^2 = g^2\tavg{ \phi'(x^B_i)^2 }_{\bm{J}}.
\end{equation}
% Calculating $r$ requires the non-Gaussian distribution over $x_i^B$, samples from which can be generated once the other order parameters have been determined. 
which follows from random matrix theory \cite{ahmadian2015properties}.

Having established these order parameters, we can define the phases of interest quantitatively. Chaos is nonfreezable when $Q(t) = 0$ is the only solution. Chaos is semifreezable when there is a nontrivial solution for $Q(t)$ associated with a solution for $D(\tau)$ that decays in $\tau$ (to a nonzero value). In this case, there may or may not be a fixed-point solution in which there is a distinct nontrivial solution for $Q(t)$ associated with $D(\tau) = \text{const}$. If there is a fixed point, it is unstable, $r > 1$. Finally, chaos is freezable when only a fixed-point solution for $Q(t)$ and $D(\tau)$ exists. In this case, it is stable, $r < 1$. Because the fixed point of the halted-synapse system is stable, the lifetime of the memory is infinite. Due to the $\bm{x} \rightarrow -\bm{x}$ symmetry of the network, if $Q(t)$ is a solution, so is $-Q(t)$. Upon halting synapses, the positive solution is realized, barring ``flips'' that occur in finite-size networks near the onset of semifreezable chaos.

We derive a two-replica DMFT that permits calculation of these order parameters. The high-dimensional equations describing the two replicas are
\begin{subequations}
\begin{align}
&(1 + \partial_t)x^A_i(t) = \sum_j J_{ij}\phi^A_j(t) \nonumber \\
&+ \frac{k}{p}\int_{-\infty}^t dt' e^{-({t-t'})/{p}} \left[\frac{1}{N}\sum_j \phi^A_j(t)\phi^A_j(t') \right] \phi^A_i(t'),
	\label{eq:highd2replicaA} \\
&(1 + \partial_t) x^B_i(t) =  \sum_j J_{ij}\phi_j^B(t) \nonumber\\
&+ \frac{k}{p} \int_{0}^\infty dt' e^{-t'/p} \left[\frac{1}{N}\sum_j \phi^A_j(-t')\phi^B_j(t) \right] \phi^A_i(-t').
	\label{eq:highd2replicaB}
\end{align}
\end{subequations}
Sending ${N \rightarrow \infty}$ and taking the limit where the time coordinate of network B is much greater than zero, we obtain the single-site picture
\begin{subequations}
\begin{align}
\left(1 + \partial_t\right)x^A(t) &=  \eta^A(t)\nonumber\\
&\hspace{-2em} + \frac{k}{p}\int_{-\infty}^t {dt'}e^{-({t-t'})/{p}} C(t-t') \phi^A(t'), \label{eq:two-replica-single-site-picture-A} \\
(1 + \partial_t)x^B(t) &= \eta^B(t) \nonumber \\
&\hspace{-2em}+ \frac{k}{p}\int_{0}^\infty dt e^{-t/p} Q(-t) \phi^A(-t),
\label{eq:two-replica-single-site-picture-B}
\end{align}
\end{subequations}
where $\eta^A(t)$ and $\eta^B(t)$ are Gaussian fields with zero mean and second-order statistics
\begin{multline}
    \tavg{\tbomat{\eta^A(t)}{\eta^B(t)}
    \obtmat{\eta^A(t')}{\eta^B(t')}}_{\eta^A, \eta^B} \\
    = g^2\tbtmat{C(t-t')}{Q(t)}{Q(t')}{D(t-t')}.
\end{multline}
The off-diagonal covariances effectively encode the alignment between $\bm{J}$ and $\bm{A}(0)$. These off-diagonals do not depend on the time coordinates of network B in accordance with the limits taken in Eqs.~(\ref{eq:order-param-def-Q}) and (\ref{eq:order-param-def-D}). The system is closed by the self-consistency conditions
\begin{subequations}
\begin{align}
Q(t) &= \tavg{\phi^A(t) \phi^B(t')}_{\eta^A, \eta^B}, \\
D(\tau) &= \tavg{\phi^B(t) \phi^B(t + \tau))}_{\eta^A, \eta^B}.
\end{align}
\end{subequations}
We solve the DMFT numerically; imposing $\partial_t x^B(t) = 0$ and $D(\tau) = \text{const.}$ gives the fixed-point solution (Appendix~\ref{subsec:replica-dmft-for-Q-and-D}). We find excellent agreement between theory and simulations [Fig.~\ref{fig:freezable}(c---e)].

We now examine the solutions of the two-replica DMFT. In the freezable-chaos regime, $Q(t)$ peaks at $t > 0$, indicating that the fixed point is more aligned with neuronal states that would have unfolded after the halt time than with the halt-time state itself [Fig.~\ref{fig:freezable}(c)]. This reflects a tendency of neurons to continue with ``momentum'' before becoming trapped in a fixed point [Fig.~\ref{fig:freezable}(a)].

Increasing $k$ from zero yields a sequence of continuous phase transitions that we analyze by plotting $Q(0)$ and $r$ against $k$ [Fig.~\ref{fig:freezable}(d)]. For small $k$, ${Q(0) = 0}$, indicating nonfreezable chaos. As $k$ is increased, $Q(0)$ develops a nonzero solution associated with a decaying $D(\tau)$, marking the onset of semifreezable chaos.
% Since ${Q(0) = 0}$, $r^\text{FP}$ is not meaningful, but is given by ${r^\text{FP}_{0} = g\sqrt{ {\int} \mathcal{D}z \phi'(g \sqrt{D_0} z)^2 } > 1}$, where $D_0$ is the nontrivial solution of $D_0 = {\int} \mathcal{D}z \phi^2(g \sqrt{D_0} z)$ and $\mathcal{D}z$ denotes Gaussian measure.
As $k$ is increased further, $Q(0)$ develops an additional nonzero solution associated with $D(\tau) = \text{const.}$, marking the onset of an unstable fixed point. Instability is signaled by $r > 0$, with $r$ computed under the fixed-point solution. Continuing to increase $k$ causes $r$ to decrease and drop below unity, marking the onset freezable chaos, at which point the dynamic and fixed-point solutions converge. The convergence of the dynamic and fixed-point solutions at $r = 1$ is expected on physical grounds and can also be derived through the DMFT: when $r$ is computed under the dynamic solution, the decay timescale of $D(\tau)$ diverges as $\ca 1/\sqrt{r-1}$ as $r \rightarrow 1^+$, giving $D(\tau) = \text{const.}$ at $r = 1$ (Appendix~\ref{sec:relationshio-between-r-and-D}). The quality of memory retention can be measured by the correlation coefficient
\begin{equation}
    \rho(t) = \frac{Q(t)}{\sqrt{D(0)  C(0)}},
\end{equation}
which varies between zero and unity [Fig.~\ref{fig:freezable}(e)]. 

% \subsection{Limiting behavior}
% \label{subsec:limiting-behavior}
The two-replica DMFT can be solved analytically as ${g \rightarrow 1^+}$. In this limit, $C(t),D(t),Q(t)\sim g-1$ and the decay timescales of $C(t)$ and $Q(t)$ diverge. We consider fixed-point solutions, $D(\tau) = D$. At leading order in $\epsilon = g-1$, the single-site equations reduce to
\begin{subequations}
\begin{align}
    (1 + \partial_t)x^A(t) &= \eta^A(t) + k C(0) x^A(t), \label{eq:slow-replica-single-site-picture-A} \\
    x^B &= \eta^B + k Q(0) x^A(0).
    \label{eq:slow-replica-single-site-picture-B}
\end{align}
\end{subequations}
To determine $Q(0)$, we square Eq.~(\ref{eq:slow-replica-single-site-picture-B}) and average over $\eta^A(t)$ and $\eta^B$. This gives, to order $\epsilon^2$, ${D^2 - \epsilon D - k Q^2(0) = 0}$,
with the solution
\begin{equation}
    Q(0) = \pm \sqrt{\frac{D(D - \epsilon)}{k} },
\end{equation}
implying $D \geq \epsilon$, with strict equality when ${Q(0) = 0}$. Next, we multiply Eqs.~(\ref{eq:slow-replica-single-site-picture-A}) and (\ref{eq:slow-replica-single-site-picture-B}) and average over $\eta^A(t)$ and $\eta^B$. This gives, gives to order $\epsilon^2$, ${(D - \epsilon + \partial_t)Q(t) = k Q(0) C(t)}$, with the solution
\begin{equation}
    Q(t) = k Q(0) \int_{-\infty}^t dt' e^{-(D - \epsilon)(t - t')} C(t').
\end{equation}
This causal filtering makes $Q(t)$ peaked at $t > 0$. Setting $t = 0$ and doing the integral using Eq.~(\ref{eq:limiting-autocov}) gives a condition for determining $D$, namely, ${h( {D}/{\epsilon}, k ) = k}$, where
\begin{multline}
    h(u, k) = \frac{2}{\sqrt{3}}\Bigg[ \psi\left(\frac{\sqrt{3}(1-k)(u-1)+3}{4} \right)  \\
    -\psi\left(\frac{\sqrt{3}(1-k)(u-1)+1}{4}\right) \Bigg]^{-1},
\end{multline}
and $\psi(\cdot)$ is the digamma function. Because $h(u, k)$ is monotonically increasing in $u$ and, in our case, ${u = {D}/{\epsilon} \geq 1}$, a nontrivial solution exists only for ${k > h(1, k) = 2/\sqrt{3}\pi \approx 0.37}$ ($h(1, k)$ does not depend on $k$). Stability is checked by computing the fixed-point spectral radius to first order in $\epsilon$, $r = 1 - D + \epsilon \leq 1$,
implying that the fixed-point and dynamic solutions emerge together at $k = 0.37$ as ${g \rightarrow 1^+}$ [Fig.~\ref{fig:phase-diagram}(b)]. 

% We start with estimates of $Q(t)$ and $D$, and the known solution for $C(\tau)$. At each iteration, we sample Gaussian fields $\eta^{\alpha}(t)$ according to Eq.~(\ref{eq:eta-distr}). These fields are passed through the single-site dynamics of Eq.~(\ref{eq:tworeplicassa}) (equivalently, Eq.~(\ref{eq:single-site-picture})) to obtain non-Gaussian fields $x^{\alpha}(t)$. For each $\eta^{\alpha}(t)$, the conditional Gaussian distribution of $\eta^{\text{FP}}$ follows from Eq.~(\ref{eq:blockcov}) according to the Schur-complement rule for conditional Gaussians. The conditional mean and variance are computed using the Fourier transforms of $\eta^{\alpha}(t)$ and $Q(t)$. We compute updated estimates of $Q(t)$ and $D$ through Eq.~(\ref{eq:tworeplicassb}), where expectations over $\eta(t)$ are done by empirical averaging over $\eta^{\alpha}(t)$, and expectations over $\eta^{\text{FP}}$ are done by numerical evaluation of Gaussian integrals using the conditional mean and variance. 

\section{Discussion}
\label{sec:discussion}

We characterized the dynamics of $N$ neurons coupled to $N^2$ dynamic synapses. Strong Hebbian plasticity causes the timescales of the system, measured through the Jacobian or Lyapunov spectra, to segregate into a slow, synapse-dominated band and and a fast, neuron-dominated band. The synapse-dominated band drives the dynamics. It is possible that this two-band structure could be detected through \textit{in-vivo} recordings of neuronal activity. Takens' embedding theorem implies that it is possible, in principle, to extract the spectrum of neuronal-synaptic timescales from neuronal activity alone \cite{brunton2017chaos}. This segregation of timescales could also be examined during task execution. If the dynamics are synapse-driven, neurons may revert to their trial-average trajectories upon optogenetic or electrophysiological perturbation. Prior studies have attributed such robustness to neuronal mechanisms such as excitatory-inhibitory balance \cite{o2022direct}, but our study invites reevaluation of such data with an emphasis on synaptic dynamics. Indeed, Hebbian plasticity can enhance the robustness of an attractor manifold against distractors \cite{sandberg2003working}.

Increasing the strength of Hebbian plasticity initially enriches network dynamics, indicated by an increased maximum Lyapunov exponent and attractor dimension. Beyond a certain plasticity strength, these metrics decrease, likely due to the increased presence of stable fixed points throughout phase space. This implies that there may be an optimal level of plasticity for task performance---one that is robust enough to enrich the dynamics compared to a nonplastic network but not so overpowering that it simplifies the dynamics through the overabundance of fixed points. This could be investigated by training plastic networks to solve tasks, e.g., using FORCE or backpropagation \cite{sussillo2009generating}.

\blue{Our analyses point to a region of parameter space with large $k$ and/or small $g$ where it is possible that the behaviors of the fixed-point density and maximum Lyapunov exponent are more interesting than what we have explored. In particular, as $(g,k) \rightarrow (0, 2.02)$, fixed points are marginally stable by virtue of outliers, not the bulk (Sec.~\ref{sec:fixed-points}). Additionally, our Lyapunov analysis leaves open the possibility that the maximum exponent is negative for sufficiently large $k$ even as $N\rightarrow \infty$ (Sections~\ref{subsec:lyapunov},~\ref{sec:lyap-appdx}; Fig.~\ref{fig:lyap-appdx}). Given that phase space is densely filled with stable fixed points in this parameter regime, it is possible that these features signal a novel glassy phase of the model. Such phases in dynamic networks remain poorly understood and are an important direction for future research \cite{marti2018correlations, berlemont2022glassy}.}

Our study addresses chaotic networks that either generate activity autonomously or produce complex responses to inputs \cite{rajan2010stimulus}. A potentially desirable alternative property is stability, defined by the ability of a network to generate input-driven trajectories that are robust against perturbations; however, such networks cannot generate rich activity autonomously. \citet{kozachkov2020achieving} analyzed a plastic network with the same governing equations as our model, but without quenched disorder, establishing conditions for stable dynamics \cite{kozachkov2022note}. A key finding was that anti-Hebbian plasticity can promote stable dynamics, pointing to a possible function for ongoing plasticity in input-driven computations.

We considered plastic synapses with strengths weaker than random synapses by a factor of $1/\sqrt{N}$. In experiments measuring synaptic strength changes, such plasticity may easily be overlooked despite its order-one network-level impact. Our model assumes all-to-all connectivity; if neurons receive $K < N$ inputs, the structure-to-randomness scaling is $1/\sqrt{K}$, making detection of plasticity more feasible if $K$ is not too large.

Humans and animals can remember a stimulus over a delay period, implying a form of rapid information storage in neural circuits, i.e., working memory (WM). Freezable chaos provides a new WM mechanism that we now compare to prior models. Most WM models rely on either cell-intrinsic or network-level mechanisms that support self-sustained activity. These ``persistent activity'' models are supported by some experimental studies, but undermined by others showing ``activity-silent'' WM \cite{stokes2015activity}. These latter studies suggest that information can be rapidly stored in synapses, requiring fast synaptic plasticity. Synaptic WM models typically use short-term facilitation (STF) due to the convention that fast plasticity is presynaptic \cite{zucker2002short, wang2006heterogeneity}. Because such plasticity cannot create attractor states of neuronal dynamics, STF models require existing symmetric structure in the synapses, potentially formed through prior Hebbian plasticity. A prototypical example is the model of \citet{mongillo2008synaptic} in which clusters of excitatory neurons with broad inhibition prime a network to function in a metastable regime. Due to STF, an activity pattern can be selectively sustained by providing transient external input to one of the clusters. A key requirement of this class of proposals is that the possible neuronal states to be stored are known in advance.

The inability of STF models to store novel patterns suggests the existence of fast Hebbian plasticity. This is at odds with conventional wisdom, but supported experimentally \cite{gustafsson1989onset, malenka1991postsynaptic, malenka1993nmda, volianskis2003transient, erickson2010single, volianskis2013different, driesen2013impact, park2014nmda, lisman2017glutamatergic}. For examples of Hebbian WM models, see \cite{von1986neural, sandberg2003working, polyn2009context, fiebig2017spiking, manohar2019neural, fiebig2020indexing, huang2021computational, bocincova2022neural, kozachkov2022robust}. Due to its Hebbian nature and ability to store novel patterns, freezable chaos aligns more with these proposals than with STF models. A crucial feature distinguishing freezable chaos from both STF and Hebbian WM models is that plasticity is deactivated, rather than activated, to store a pattern. Whereas other models require an external input carrying the pattern to be stored, this feature allows our model to store the neuronal state while it is engaged in strongly recurrent dynamics (in our random-network model, chaos).

Hinton and collaborators considered the possibility of a network performing a computation, saving its state in synapses, using neurons to perform a subroutine, and resuming computation from the saved state. This was termed ``true recursion'' \cite{hinton1987using, ba2016using, bengio2021deep}. Freezable chaos provides a minimal example of this: the neuronal state can be saved by halting plasticity, allowing neurons to engage in arbitrary dynamics before returning to the saved state. In our model, halting plasticity leaves a globally stable fixed point, so neuronal dynamics during the subroutine must be driven by external inputs. An interesting question is whether halting plasticity can leave the network with a fixed point that coexists with a dynamic regime that can be used for recurrent computation. This could be implemented in an ad-hoc manner by turning on feedback loops upon halting plasticity.

This feature of freezable chaos suggests a method of detecting it \textit{in-vivo}, namely, by ``interupting'' a task requiring strong recurrent dynamics, such as evidence integration, for variable periods of time \cite{mante2013context}. Finding that neurons involved in the computation show continuity in their activities at the beginning and end of the interruption period would be suggestive of freezable dynamics. This conclusion would be further supported if task performance degrades when synaptic plasticity is disabled by genetic or pharmacological manipulations \cite{driesen2013impact}. The activity expressed by the neurons during the interruption period would depend on whether and how they are recruited in this interval.

In other Hebbian WM models, an external neuronal or neuromodulatory signal is typically required to erase information stored in the synapses and return the network to a dynamic state. Freezable chaos avoids this requirement by leveraging fixed points that are stable with respect to neuronal, but not neuronal-synaptic dynamics. The presence or absence of a resetting signal could enable experimental disambiguation of our proposal.

Our model offers a mechanism for WM, but lacks a mechanism for long-term memory. This could be addressed by introducing slow Hebbian dynamics into $\bm{J}$. For example, prior DMFT studies have taken $\bm{J}$ to be a static result of associative plasticity, resulting in various combinations of chaos and long-term memory retrieval \cite{tirozzi1991chaos, mastrogiuseppe2018linking, pereira2018attractor, pereira2023forgetting}. Models like these could be extended by incorporating fast Hebbian synaptic dynamics atop this static structure. It would be particularly interesting if the short- and long-term dynamics could be made to interact, e.g., to implement memory consolidation. For example, during frozen chaos, if long-term plasticity was activated while the shorter-term plasticity of $\bm{A}(t)$ was disabled, the frozen state could be consolidated into $\bm{J}$.

Humans have a WM capacity of $\sim\! 7$ items, but freezable chaos can store just one item because synaptic plasticity is halted once a pattern is stored. One way of overcoming this limitation would be to use multiple $\bm{A}(t)$ matrices that can be independently modulated, corresponding either to different biophysical plasticity mechanisms or disjoint sets of synapses. Organisms might benefit from these different sets of synaptic variables possessing a hierarchy of timescales.

Large language models display a remarkable capacity for \textit{in-context learning}: producing output that incorporates information or algorithms contained in the input \cite{akyurek2022learning}. This is surprising because, in both neuroscience and machine learning, such learning is generally thought to require weight updates. One possibility is that these models have such capabilities because they emulate weight dynamics through {attention layers} \cite{schlag2021linear}. Nevertheless, explicit weight dynamics could benefit machine-learning models, for example by enabling in-context learning with fewer model parameters \cite{ba2016using} or on longer sequences \cite{katharopoulos2020transformers}. An impediment to work in this direction is that weight dynamics are computationally costly. An important direction of work therefore pertains to ameliorating this burden, e.g., by formulating new forms of weight dynamics that are both expressive and have low computational complexity, or by exploring low-power neuromorphic architectures with dynamic weights.

\section*{Acknowledgments}
We thank Tala Fakhoury and Albert Wakhloo for insightful discussions and valuable feedback during the development of this work. We thank Ashok Litwin-Kumar, Manuel Beiran, and Patricia Cooney for comments on a manuscript. The authors were supported by the Gatsby Charitable Foundation and NSF NeuroNex award DBI-1707398.

\appendix

\section{Extensions and limits}
\label{sec:extensions-and-limits}

Here, we derive how the single-site picture of Eq.~(\ref{eq:single-site-picture}) changes under various modifications of the synaptic dynamics of Eq.~(\ref{eq:syn-dynamics}). We begin with two modifications that change the single-site dynamics simply by modifying the plasticity kernel,
\begin{equation}
    K(\tau) = \frac{k}{p}e^{-\tau/p}C(\tau).
    \label{eq:plasticity-kernel}
\end{equation}
\begin{enumerate}
\item 
First, we take the limit $k , p \rightarrow \infty$ in Eq.~(\ref{eq:syn-dynamics}) while keeping $k/p = \tilde{k}$ constant. This is equivalent to eliminating the decay term in the synaptic dynamics,
\begin{equation}
    \partial_t {A}_{ij}(t) = \tilde{k} {\phi}_i(t) {\phi}_j(t). 
\end{equation}
In this case, the kernel Eq.~(\ref{eq:plasticity-kernel}) becomes
\begin{equation}
    K(\tau) = \tilde{k} C(\tau).
\end{equation}
Thus, the decay timescale of the kernel is determined entirely though self-consistency.

\item
We next analyze the case where each synapse has a different time constant $p$ that follows a heavy-tailed distribution with a power-law decay. We assume an inverse-gamma distribution, $f(p; \alpha, \beta) = N(\alpha, \beta) \exp(-\beta / p) p^{-(\alpha + 1)} $ where $N(\alpha, \beta) = \beta^{\alpha} / \Gamma(\alpha)$ is a normalization constant with $\Gamma(\cdot)$ the gamma function. This distribution decays as $\ca 1/p^{\alpha + 1}$ for large $p$ and is exponentially suppressed as $p \rightarrow 0$ according to $\ca \exp(-\beta / p)$. In this case, the kernel of Eq.~(\ref{eq:plasticity-kernel}) becomes
\begin{equation}
    K(\tau) = \frac{k \alpha}{\beta}\left(\frac{\beta}{\beta + \tau}\right)^{\alpha+1} C(\tau). 
\end{equation}
Thus, the kernel inherits the power-law decay of the distribution of time constants.

\item
We next consider a form of plasticity with a presynaptic dependence,
\begin{equation}
    (1 + p\partial_t){A}_{ij}(t) = \frac{k}{N} {\phi}_j(t).
\end{equation}
For this plasticity rule, the single-site dynamics are
\begin{equation}
    (1 + \partial_t)x(t) = \eta(t) + I,
\end{equation}
where $I$ is a time-independent input given self-consistently by
\begin{equation}
    I = \frac{k}{p}\int_{0}^{\infty} d\tau e^{-\tau/p} C(\tau).
\end{equation}

\item 
Next, by introducing a time delay $d$, we consider a form of temporally asymmetric Hebbian plasticity,
\begin{multline}
    (1 + p\partial_t) {A}_{ij}(t) \\
    = \frac{k}{N} \left[ {\phi}_i(t) {\phi}_j(t - d)
     - {\phi}_i(t - d){\phi}_j(t) \right].
\end{multline}
This single-site problem for this plasticity rule is
\begin{multline}
    (1 + \partial_t)x(t) = \eta(t) \\
    + \frac{k}{p}\int_{-\infty}^t dt'
    e^{-(t-t')/p} \left[C(t - t' + d)\phi(t') \right.\\
    \left. - C(t - t')\phi(t' - d) \right].
\end{multline}

\item
Finally, we consider the case in which plasticity depends on arbitrary functions of pre- and postsynaptic activity, $f_{\text{pre}}(\cdot)$ and $f_{\text{post}}(\cdot)$,
\begin{equation}
    (1 + p\partial_t) {A}_{ij}(t) = \frac{k}{N} f_{\text{post}}({\phi}_i(t))
    {f}_{\text{pre}}({\phi}_j(t)).
\end{equation}
In this case, the single-site dynamics are
\begin{subequations}
\begin{align}
    &(1 + \partial_t)x(t) = \eta(t)\nonumber  \\
    &+ \frac{k}{p}\int^{t}_{-\infty} e^{-(t-t')/p}
    C_{\text{pre},\phi}(t-t') f_{\text{post}}(\phi(t')), \\
    &\text{where}\:\: C_{\text{pre},\phi}(\tau) = \tavg{ f_{\text{pre}}(\phi(t)) \phi(t+\tau) }_{\eta}.
\end{align}
\end{subequations}
\end{enumerate}

\section{Bistable attractors for finite $N$}
\label{sec:bistability}
\begin{figure}
    \includegraphics[width=\columnwidth]{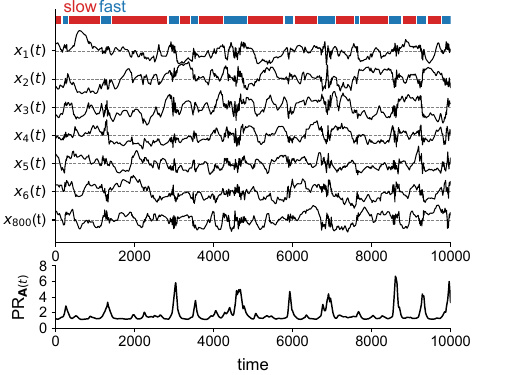}
    \caption{
        Bistable chaotic attractors in a simulation with ${g = 2}$, ${k = 1.1}$, $p = 45$, and $N = 800$. Top: example neuronal traces $x_i(t)$. Bottom: participation ratio of $\bm{A}(t)$ (as discussed in Sec. \ref{sec:fixed-points}). ``Fast'' and ``slow'' states are defined by thresholding a low-pass filtered version of $\lVert \bm{\dot{x}} \rVert$.
    }
    \label{fig:bistability}
\end{figure}
For large $p$ and $k$, the plastic network has two approximate dynamic solutions: a fast solution in which plasticity is averaged out, with dynamic timescale $\tau^* = \mathcal{O}(1)$; and a slow solution in which synapses drag neurons, with $\tau^* \gg p$. In finite-size networks, both behaviors can be realized depending on initial conditions. Moreover, for appropriately tuned values of the parameters, the system can switch between these behaviors in a bistable manner (Fig.~\ref{fig:bistability}). 

\section{Dynamical mean-field theory numerics}
We solve the DMFT equations in this paper using iterative Monte-Carlo methods \cite{krishnamurthy2022theory, roy2019numerical, eissfeller1992new, eissfeller1994mean, stern2014dynamics, mignacco2020dynamical}.

\subsection{Solution for $C(\tau)$}
\label{subsec:dmft-for-C}
We first describe solving for $ C(\tau) $. At each iteration, we sample a field $ \eta(t) $ from a Gaussian process with zero mean and autocovariance $ C(\tau) $. This is achieved by independently drawing Fourier coefficients from Gaussian distributions with the appropriate variances, while ensuring Hermitian symmetry. The sampled field $ \eta(t) $ is then processed using the single-site dynamics given by Eq.~\eqref{eq:single-site-picture} to produce $ x(t) $ and consequently $ \phi(t) $. An updated estimate of $ C(\tau) $ is obtained from the empirical autocovariance of $ \phi(t) $. This procedure is repeated until $ C(\tau) $ converges. Multiple fields $ \eta(t) $ are drawn and processed in parallel at each iteration.

\subsection{Solution for $Q(t)$ and $D(\tau)$}
\label{subsec:replica-dmft-for-Q-and-D}
The replica DFMT for freezable chaos follows a similar procedure, but is more complicated due to the requirement to sample correlated fields $ \eta^A(t) $ and $ \eta^B(t) $. Given that $ C(\tau) $ is known, our task is to determine $ Q(t) $ and $ D(\tau) $. First, we sample $ \eta^A(t) $ with the correct marginal statistics, namely, zero mean and autocovariance $ C(\tau) $. Given $ \eta^A(t) $, the conditional distribution for $ \eta^B(t) $ is Gaussian with mean $ \mu^B[\eta^A] $ and autocovariance $ \Sigma^B(\tau) $, given by
\begin{subequations}
\begin{align}
    \mu^B[\eta^A] &= \int dt \int dt' C^{-1}(t - t')Q(t)\eta^A(t'), \label{eq:cond-mean} \\
    \Sigma^B(\tau) &= g^2(D(\tau) - \delta), \label{eq:cond-variance} \\
    \text{where}\:\: \delta &= \int dt \int dt' C^{-1}(t - t')Q(t)Q(t') .
\end{align}
\end{subequations}
These integrals are straightforwardly evaluated in Fourier space. To find the dynamic solution where $ D(\tau) $ decays as a function of $ \tau $, we sample $ \eta^B(t) $ from this conditional distribution and process it through the single-site dynamics given by Eq.~\eqref{eq:two-replica-single-site-picture-B} to obtain $x^B(t)$ and thus $\phi^B(t)$. We then update the estimates of $ Q(t) $ and $ D(\tau) $ based on the empirical statistics of $\phi^A(t)$ and $\phi^B(t)$. For fixed-point solutions with $ D(\tau) = D $, low-variance estimates of $ Q(t) $ and $ D $ can be efficiently obtained via numerical evaluation of Gaussian integrals using the conditional mean and variance. In both the fixed-point and dynamic cases, an effective increase in the number of samples is achieved by shifting $ \eta^A(t) $, which multiplies the conditional mean by a phase factor and leaves the conditional variance unchanged. Finally, similar to the solution for $ C(\tau) $, multiple fields $ \eta^A(t) $ are sampled during each iteration of the solver.

\section{Relating $r$ and the decay of $D(\tau)$}
\label{sec:relationshio-between-r-and-D}
Given $\eta^A(t)$, the conditional distribution of $\eta^B(t)$ is Gaussian with mean Eq.~(\ref{eq:cond-mean}) and autocovariance Eq.~(\ref{eq:cond-variance}). Denote the integral term in Eq.~(\ref{eq:two-replica-single-site-picture-B}) by $I^B[\eta^A]$. Then, Eq.~(\ref{eq:two-replica-single-site-picture-B}) can be expressed as $x^B(t) = z(t) + \mu^B[\eta^A] + I^B[\eta^A]$, where $z(t)$ is a zero-mean Gaussian field with autocovariance $\tilde{\Sigma}^B(\tau) = g^2(\tilde{D}(\tau) - \delta)$. Here, $\tilde{f}(\tau)$ is related to $f(\tau)$ via
\begin{equation}
    (1 - \partial_\tau^2)\tilde{f}(\tau) = f(\tau).
    \label{eq:def-of-f}
\end{equation}
The self-consistent condition for $D(\tau)$ becomes
\begin{multline}
    D(\tau) = \left\langle \int \mathcal{D}u \left[ \int \mathcal{D}x \phi\left( \sqrt{\tilde{\Sigma}^B(0) - \tilde{\Sigma}^B(\tau)} x \right.  \right. \right. \\
    \left. \left. \left. + \sqrt{\tilde{\Sigma}^B(\tau)}u + \mu^B[\eta^A] + I^B[\eta^A] \right) \right]^2 \right\rangle_{\eta^A}.
    \label{eq:self-consistent-condition}
\end{multline}
The right-hand side depends on $\tau$ only through $\tilde{\Sigma}^B(\tau)$, which in turn depends on $\tau$ only via $\tilde{D}(\tau)$. Differentiating both sides of Eq.~\ref{eq:self-consistent-condition} twice in $\tau$ and applying Price's theorem yields
\begin{multline}
    \partial_\tau^2 D(\tau) = g^4 D^{\phi''}(\tau) \left[\partial_\tau \tilde{D}(\tau)\right]^2 \\
    + g^2 D^{\phi'}(\tau) \partial_\tau^2 \tilde{D}(\tau),
\end{multline}
where $D^{\phi^{(n)}}(\tau)$ is given by Eq.~\ref{eq:self-consistent-condition} with $\phi(\cdot) \rightarrow \phi^{(n)}(\cdot)$. Setting $\tau = 0$ and using Eq.~(\ref{eq:def-of-f}) gives
\begin{equation}
    \left. \partial_\tau^2 D(\tau)\right|_{\tau = 0} = r^2\left[\tilde{D}(0) - D(0)\right],
\end{equation}
noting that $\left. \partial_{\tau} \tilde{D}(\tau)\right|_{\tau=0} = 0$ and $r^2 = g^2 D^{\phi'}(0)$. Taylor expanding $D(\tau)$ about $\tau = 0$, this can be expressed as
\begin{equation}
    \left. \left[(r^2 - 1)\partial_\tau^2 + r^2\left(\partial_\tau^4 + \partial_\tau^6 + \cdots\right)\right]D(\tau) \right|_{\tau=0} = 0.
    \label{eq:derivs-and-radius}
\end{equation}
Assuming that $D(\tau)$ decays on a timescale $T \gg 1$, dimensional analysis implies that ${\left. \partial_\tau^n D(\tau)\right|_{\tau=0} = c_n / T^n}$, where $c_n$ are order-one coefficients. Keeping terms up to $1/T^4$ in Eq.~(\ref{eq:derivs-and-radius}) results in
\begin{equation}
    T =  \sqrt{-\frac{c_4}{c_2}} \frac{r}{\sqrt{r^2 - 1}},
\end{equation}
where $c_2 < 0$ and $c_4 > 0$ for a generic decaying autocovariance function with smoothness at $\tau = 0$. Thus, as $r \rightarrow 1^+$, $T$ diverges as $T \sim {1}/{\sqrt{r - 1}}$.

\section{Random matrix theory}
\label{sec:random-matrix-theory}
We study the spectrum of the Jacobian by applying a result of \citet{ahmadian2015properties} concerning a $D$-dimensional random matrix of the form $\bm{\mu} + \bm{L} \bm{N}\bm{R}$ where $\bm{\mu}$, $\bm{L}$, and $\bm{R}$ are deterministic square matrices with $\bm{L}$ and $\bm{R}$ invertible. $\bm{N}$ is a random square matrix with entries drawn from $\mathcal{N}(0, 1/D)$. It was shown that the limiting spectral density of this matrix has support at a point $\lambda\in \mathbb{C}$ when $\lim_{D \rightarrow \infty} D^{-1} \lVert \bm{R} \left( \bm{\mu} - \lambda \bm{I}_N \right)^{-1} \bm{L} \rVert_F^2 \geq 1
$.
We can write $\tilde{\bm{M}}_{\text{bulk}} = \bm{\mu} + \bm{L} \bm{N}\bm{R}$ by setting $D = 2N$ and choosing
\begin{subequations}
\begin{align}
& \bm{\mu} = \tbtmat{-\bm{I}_N}{ C(0) \diag{ \phi'(\bm{x})} }{\frac{k}{p}\bm{I}_N }{-\frac{1}{p}\bm{I}_N}, \nonumber \\
& \bm{L} = \tbtmat{\bm{I}_N}{\bm{0}_N}{\bm{0}_N}{\bm{0}_N}, \:\: \bm{R} = \sqrt{2}\tbtmat{\diag{ \phi'(\bm{x})}}{\bm{0}_N}{\bm{0}_N}{\bm{0}_N}. \nonumber 
\end{align}
\end{subequations}
In this parameterization, 3 of the 4 $N \times N$ blocks of the random matrix $\bm{N}$ have zero contribution to the result, reflecting the fact that the randomness of $\tilde{\bm{M}}_{\text{bulk}}$, a $2N \times  2N$ matrix, is generated through $\bm{J}$, an $N \times N$ matrix. Additionally, while $\bm{L}$ and $\bm{R}$ are singular, a violation of the assumptions of the theorem, one can add $\epsilon \bm{I}_N$ to each of $\bm{L}$ and $\bm{R}$ and safely compute the resulting spectral boundary curve, then take $\epsilon \rightarrow 0$ at the end.
% . Because this gives an accurate curve in the limit $D \rightarrow \infty $ for all $\epsilon > 0$, we can
% (this is not viable for arbitrary noninvertible $\bm{L}$ and $\bm{R}$, for example those that cause the Frobenius norm given previously to be subextensive).
We proceed to compute the normalized Frobenius norm, yielding Eq.~(\ref{eq:boundaryformula}).

\section{\blue{Lyapunov numerics}}
\label{sec:lyap-appdx}

\begin{figure}
    \centering
    \includegraphics[width=\columnwidth]{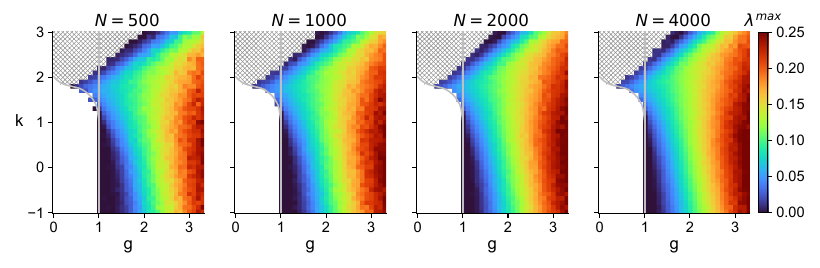}
    \caption{\blue{Same as Fig.~\ref{fig:lyapunov}(a) with different network sizes $N$.}}
    \label{fig:lyap-appdx}
\end{figure}

\blue{For the maximum Lyapunov exponent heatmaps in Fig.~\ref{fig:lyapunov}(a) and Fig.~\ref{fig:lyap-appdx}, the dynamics were run for time $T_\text{sim} = 850$ with a random perturbation of magnitude $10^{-2}$ applied at time $250$.

Fig.~\ref{fig:lyap-appdx} shows how the heatmap in Fig.~\ref{fig:lyapunov}(a) changes for different values of $N$. We find empirically that the boundary boundary between regions of parameter space producing small, positive $\lambda^\text{max}$ (dark blue) and convergence to stable nonzero fixed points (in at least 80\% of simulations; hatched) is well fit by an $N$-dependent isocontour of the log-number of fixed points [Fig.~\ref{fig:fixed-points}(d)].

To compute the Lyapunov spectra in Fig.~\ref{fig:lyapunov}(b), we use the forward- and backward-pass method described in Sec.~\ref{subsec:lyapunov}. In both the forward and backward directions, we computed 1600 exponents for networks of size $N = 900$. Dynamics were run for time $T_\text{sim} = 800$ with a burn-in period of duration 100. We used orthonormalization intervals of 20 and 5 for the forward and backward directions, respectively. For each setting of the parameters, we computed the spectra for five different network realizations and combined the results.}

% \section{Freezable chaos clamping procedure}
% \label{sec:clamping}
% In Fig. \ref{fig:freezable}, when neurons are clamped near zero, the $i$-th neuron is clamped at $x_i = 10^{-4} s_i$ where $s_i = \pm 1$ in sign agreement with the pre-clamp state of $x_i$ for 75\% of neurons and in sign disagreement for the remaining 25\%. This guarantees that the neurons spring back to the halt-time neuronal state and not its negative, which occurs half the time when neurons are clamped at zero plus a small amount of random noise. If neurons are clamped exactly at zero, they will remain there indefinitely in simulations since this is a fixed point, albeit an unstable one. 

%apsrev4-2.bst 2019-01-14 (MD) hand-edited version of apsrev4-1.bst
%Control: key (0)
%Control: author (8) initials jnrlst
%Control: editor formatted (1) identically to author
%Control: production of article title (0) allowed
%Control: page (0) single
%Control: year (1) truncated
%Control: production of eprint (0) enabled
%


\begin{thebibliography}{92}%
\makeatletter
\providecommand \@ifxundefined [1]{%
 \@ifx{#1\undefined}
}%
\providecommand \@ifnum [1]{%
 \ifnum #1\expandafter \@firstoftwo
 \else \expandafter \@secondoftwo
 \fi
}%
\providecommand \@ifx [1]{%
 \ifx #1\expandafter \@firstoftwo
 \else \expandafter \@secondoftwo
 \fi
}%
\providecommand \natexlab [1]{#1}%
\providecommand \enquote  [1]{``#1''}%
\providecommand \bibnamefont  [1]{#1}%
\providecommand \bibfnamefont [1]{#1}%
\providecommand \citenamefont [1]{#1}%
\providecommand \href@noop [0]{\@secondoftwo}%
\providecommand \href [0]{\begingroup \@sanitize@url \@href}%
\providecommand \@href[1]{\@@startlink{#1}\@@href}%
\providecommand \@@href[1]{\endgroup#1\@@endlink}%
\providecommand \@sanitize@url [0]{\catcode `\\12\catcode `\$12\catcode
  `\&12\catcode `\#12\catcode `\^12\catcode `\_12\catcode `\%12\relax}%
\providecommand \@@startlink[1]{}%
\providecommand \@@endlink[0]{}%
\providecommand \url  [0]{\begingroup\@sanitize@url \@url }%
\providecommand \@url [1]{\endgroup\@href {#1}{\urlprefix }}%
\providecommand \urlprefix  [0]{URL }%
\providecommand \Eprint [0]{\href }%
\providecommand \doibase [0]{https://doi.org/}%
\providecommand \selectlanguage [0]{\@gobble}%
\providecommand \bibinfo  [0]{\@secondoftwo}%
\providecommand \bibfield  [0]{\@secondoftwo}%
\providecommand \translation [1]{[#1]}%
\providecommand \BibitemOpen [0]{}%
\providecommand \bibitemStop [0]{}%
\providecommand \bibitemNoStop [0]{.\EOS\space}%
\providecommand \EOS [0]{\spacefactor3000\relax}%
\providecommand \BibitemShut  [1]{\csname bibitem#1\endcsname}%
\let\auto@bib@innerbib\@empty
%</preamble>
\bibitem [{\citenamefont {Gao}\ and\ \citenamefont
  {Ganguli}(2015)}]{gao2015simplicity}%
  \BibitemOpen
  \bibfield  {author} {\bibinfo {author} {\bibfnamefont {P.}~\bibnamefont
  {Gao}}\ and\ \bibinfo {author} {\bibfnamefont {S.}~\bibnamefont {Ganguli}},\
  }\bibfield  {title} {\bibinfo {title} {On simplicity and complexity in the
  brave new world of large-scale neuroscience},\ }\href@noop {} {\bibfield
  {journal} {\bibinfo  {journal} {Current Opinion in Neurobiology}\ }\textbf
  {\bibinfo {volume} {32}},\ \bibinfo {pages} {148} (\bibinfo {year}
  {2015})}\BibitemShut {NoStop}%
\bibitem [{\citenamefont {Pandarinath}\ \emph {et~al.}(2018)\citenamefont
  {Pandarinath}, \citenamefont {O’Shea}, \citenamefont {Collins},
  \citenamefont {Jozefowicz}, \citenamefont {Stavisky}, \citenamefont {Kao},
  \citenamefont {Trautmann}, \citenamefont {Kaufman}, \citenamefont {Ryu},
  \citenamefont {Hochberg} \emph {et~al.}}]{pandarinath2018inferring}%
  \BibitemOpen
  \bibfield  {author} {\bibinfo {author} {\bibfnamefont {C.}~\bibnamefont
  {Pandarinath}}, \bibinfo {author} {\bibfnamefont {D.~J.}\ \bibnamefont
  {O’Shea}}, \bibinfo {author} {\bibfnamefont {J.}~\bibnamefont {Collins}},
  \bibinfo {author} {\bibfnamefont {R.}~\bibnamefont {Jozefowicz}}, \bibinfo
  {author} {\bibfnamefont {S.~D.}\ \bibnamefont {Stavisky}}, \bibinfo {author}
  {\bibfnamefont {J.~C.}\ \bibnamefont {Kao}}, \bibinfo {author} {\bibfnamefont
  {E.~M.}\ \bibnamefont {Trautmann}}, \bibinfo {author} {\bibfnamefont {M.~T.}\
  \bibnamefont {Kaufman}}, \bibinfo {author} {\bibfnamefont {S.~I.}\
  \bibnamefont {Ryu}}, \bibinfo {author} {\bibfnamefont {L.~R.}\ \bibnamefont
  {Hochberg}}, \emph {et~al.},\ }\bibfield  {title} {\bibinfo {title}
  {Inferring single-trial neural population dynamics using sequential
  auto-encoders},\ }\href@noop {} {\bibfield  {journal} {\bibinfo  {journal}
  {Nature Methods}\ }\textbf {\bibinfo {volume} {15}},\ \bibinfo {pages} {805}
  (\bibinfo {year} {2018})}\BibitemShut {NoStop}%
\bibitem [{\citenamefont {Vyas}\ \emph {et~al.}(2020)\citenamefont {Vyas},
  \citenamefont {Golub}, \citenamefont {Sussillo},\ and\ \citenamefont
  {Shenoy}}]{vyas2020computation}%
  \BibitemOpen
  \bibfield  {author} {\bibinfo {author} {\bibfnamefont {S.}~\bibnamefont
  {Vyas}}, \bibinfo {author} {\bibfnamefont {M.~D.}\ \bibnamefont {Golub}},
  \bibinfo {author} {\bibfnamefont {D.}~\bibnamefont {Sussillo}},\ and\
  \bibinfo {author} {\bibfnamefont {K.~V.}\ \bibnamefont {Shenoy}},\ }\bibfield
   {title} {\bibinfo {title} {Computation through neural population dynamics},\
  }\href@noop {} {\bibfield  {journal} {\bibinfo  {journal} {Annual Review of
  Neuroscience}\ }\textbf {\bibinfo {volume} {43}},\ \bibinfo {pages} {249}
  (\bibinfo {year} {2020})}\BibitemShut {NoStop}%
\bibitem [{\citenamefont {Benna}\ and\ \citenamefont
  {Fusi}(2016)}]{benna2016computational}%
  \BibitemOpen
  \bibfield  {author} {\bibinfo {author} {\bibfnamefont {M.~K.}\ \bibnamefont
  {Benna}}\ and\ \bibinfo {author} {\bibfnamefont {S.}~\bibnamefont {Fusi}},\
  }\bibfield  {title} {\bibinfo {title} {Computational principles of synaptic
  memory consolidation},\ }\href@noop {} {\bibfield  {journal} {\bibinfo
  {journal} {Nature Neuroscience}\ }\textbf {\bibinfo {volume} {19}},\ \bibinfo
  {pages} {1697} (\bibinfo {year} {2016})}\BibitemShut {NoStop}%
\bibitem [{\citenamefont {Abbott}\ and\ \citenamefont
  {Regehr}(2004)}]{abbott2004synaptic}%
  \BibitemOpen
  \bibfield  {author} {\bibinfo {author} {\bibfnamefont {L.~F.}\ \bibnamefont
  {Abbott}}\ and\ \bibinfo {author} {\bibfnamefont {W.~G.}\ \bibnamefont
  {Regehr}},\ }\bibfield  {title} {\bibinfo {title} {Synaptic computation},\
  }\href@noop {} {\bibfield  {journal} {\bibinfo  {journal} {Nature}\ }\textbf
  {\bibinfo {volume} {431}},\ \bibinfo {pages} {796} (\bibinfo {year}
  {2004})}\BibitemShut {NoStop}%
\bibitem [{\citenamefont {Abbott}\ \emph {et~al.}(1997)\citenamefont {Abbott},
  \citenamefont {Varela}, \citenamefont {Sen},\ and\ \citenamefont
  {Nelson}}]{abbott1997synaptic}%
  \BibitemOpen
  \bibfield  {author} {\bibinfo {author} {\bibfnamefont {L.~F.}\ \bibnamefont
  {Abbott}}, \bibinfo {author} {\bibfnamefont {J.}~\bibnamefont {Varela}},
  \bibinfo {author} {\bibfnamefont {K.}~\bibnamefont {Sen}},\ and\ \bibinfo
  {author} {\bibfnamefont {S.}~\bibnamefont {Nelson}},\ }\bibfield  {title}
  {\bibinfo {title} {Synaptic depression and cortical gain control},\
  }\href@noop {} {\bibfield  {journal} {\bibinfo  {journal} {Science}\ }\textbf
  {\bibinfo {volume} {275}},\ \bibinfo {pages} {221} (\bibinfo {year}
  {1997})}\BibitemShut {NoStop}%
\bibitem [{\citenamefont {Maass}\ and\ \citenamefont
  {Zador}(1997)}]{maass1997dynamic}%
  \BibitemOpen
  \bibfield  {author} {\bibinfo {author} {\bibfnamefont {W.}~\bibnamefont
  {Maass}}\ and\ \bibinfo {author} {\bibfnamefont {A.}~\bibnamefont {Zador}},\
  }\bibfield  {title} {\bibinfo {title} {Dynamic stochastic synapses as
  computational units},\ }\href@noop {} {\bibfield  {journal} {\bibinfo
  {journal} {Advances in Neural Information Processing Systems}\ }\textbf
  {\bibinfo {volume} {10}} (\bibinfo {year} {1997})}\BibitemShut {NoStop}%
\bibitem [{\citenamefont {Barak}\ and\ \citenamefont
  {Tsodyks}(2007)}]{barak2007persistent}%
  \BibitemOpen
  \bibfield  {author} {\bibinfo {author} {\bibfnamefont {O.}~\bibnamefont
  {Barak}}\ and\ \bibinfo {author} {\bibfnamefont {M.}~\bibnamefont
  {Tsodyks}},\ }\bibfield  {title} {\bibinfo {title} {Persistent activity in
  neural networks with dynamic synapses},\ }\href@noop {} {\bibfield  {journal}
  {\bibinfo  {journal} {PLoS Computational Biology}\ }\textbf {\bibinfo
  {volume} {3}},\ \bibinfo {pages} {e35} (\bibinfo {year} {2007})}\BibitemShut
  {NoStop}%
\bibitem [{\citenamefont {Mongillo}\ \emph {et~al.}(2008)\citenamefont
  {Mongillo}, \citenamefont {Barak},\ and\ \citenamefont
  {Tsodyks}}]{mongillo2008synaptic}%
  \BibitemOpen
  \bibfield  {author} {\bibinfo {author} {\bibfnamefont {G.}~\bibnamefont
  {Mongillo}}, \bibinfo {author} {\bibfnamefont {O.}~\bibnamefont {Barak}},\
  and\ \bibinfo {author} {\bibfnamefont {M.}~\bibnamefont {Tsodyks}},\
  }\bibfield  {title} {\bibinfo {title} {Synaptic theory of working memory},\
  }\href@noop {} {\bibfield  {journal} {\bibinfo  {journal} {Science}\ }\textbf
  {\bibinfo {volume} {319}},\ \bibinfo {pages} {1543} (\bibinfo {year}
  {2008})}\BibitemShut {NoStop}%
\bibitem [{\citenamefont {Buonomano}\ and\ \citenamefont
  {Maass}(2009)}]{buonomano2009state}%
  \BibitemOpen
  \bibfield  {author} {\bibinfo {author} {\bibfnamefont {D.~V.}\ \bibnamefont
  {Buonomano}}\ and\ \bibinfo {author} {\bibfnamefont {W.}~\bibnamefont
  {Maass}},\ }\bibfield  {title} {\bibinfo {title} {State-dependent
  computations: spatiotemporal processing in cortical networks},\ }\href@noop
  {} {\bibfield  {journal} {\bibinfo  {journal} {Nature Reviews Neuroscience}\
  }\textbf {\bibinfo {volume} {10}},\ \bibinfo {pages} {113} (\bibinfo {year}
  {2009})}\BibitemShut {NoStop}%
\bibitem [{\citenamefont {Mi}\ \emph {et~al.}(2017)\citenamefont {Mi},
  \citenamefont {Katkov},\ and\ \citenamefont {Tsodyks}}]{mi2017synaptic}%
  \BibitemOpen
  \bibfield  {author} {\bibinfo {author} {\bibfnamefont {Y.}~\bibnamefont
  {Mi}}, \bibinfo {author} {\bibfnamefont {M.}~\bibnamefont {Katkov}},\ and\
  \bibinfo {author} {\bibfnamefont {M.}~\bibnamefont {Tsodyks}},\ }\bibfield
  {title} {\bibinfo {title} {Synaptic correlates of working memory capacity},\
  }\href@noop {} {\bibfield  {journal} {\bibinfo  {journal} {Neuron}\ }\textbf
  {\bibinfo {volume} {93}},\ \bibinfo {pages} {323} (\bibinfo {year}
  {2017})}\BibitemShut {NoStop}%
\bibitem [{\citenamefont {Penney}\ \emph {et~al.}(1993)\citenamefont {Penney},
  \citenamefont {Coolen},\ and\ \citenamefont
  {Sherrington}}]{penney1993coupled}%
  \BibitemOpen
  \bibfield  {author} {\bibinfo {author} {\bibfnamefont {R.}~\bibnamefont
  {Penney}}, \bibinfo {author} {\bibfnamefont {A.}~\bibnamefont {Coolen}},\
  and\ \bibinfo {author} {\bibfnamefont {D.}~\bibnamefont {Sherrington}},\
  }\bibfield  {title} {\bibinfo {title} {Coupled dynamics of fast spins and
  slow interactions in neural networks and spin systems},\ }\href@noop {}
  {\bibfield  {journal} {\bibinfo  {journal} {Journal of Physics A:
  Mathematical and General}\ }\textbf {\bibinfo {volume} {26}},\ \bibinfo
  {pages} {3681} (\bibinfo {year} {1993})}\BibitemShut {NoStop}%
\bibitem [{\citenamefont {Penney}\ and\ \citenamefont
  {Sherrington}(1994)}]{penney1994slow}%
  \BibitemOpen
  \bibfield  {author} {\bibinfo {author} {\bibfnamefont {R.}~\bibnamefont
  {Penney}}\ and\ \bibinfo {author} {\bibfnamefont {D.}~\bibnamefont
  {Sherrington}},\ }\bibfield  {title} {\bibinfo {title} {Slow interaction
  dynamics in spin-glass models},\ }\href@noop {} {\bibfield  {journal}
  {\bibinfo  {journal} {Journal of Physics A: Mathematical and General}\
  }\textbf {\bibinfo {volume} {27}},\ \bibinfo {pages} {4027} (\bibinfo {year}
  {1994})}\BibitemShut {NoStop}%
\bibitem [{\citenamefont {Roberts}(2000)}]{roberts2000dynamics}%
  \BibitemOpen
  \bibfield  {author} {\bibinfo {author} {\bibfnamefont {P.~D.}\ \bibnamefont
  {Roberts}},\ }\bibfield  {title} {\bibinfo {title} {Dynamics of temporal
  learning rules},\ }\href@noop {} {\bibfield  {journal} {\bibinfo  {journal}
  {Physical Review E}\ }\textbf {\bibinfo {volume} {62}},\ \bibinfo {pages}
  {4077} (\bibinfo {year} {2000})}\BibitemShut {NoStop}%
\bibitem [{\citenamefont {Ocker}\ \emph {et~al.}(2015)\citenamefont {Ocker},
  \citenamefont {Litwin-Kumar},\ and\ \citenamefont {Doiron}}]{ocker2015self}%
  \BibitemOpen
  \bibfield  {author} {\bibinfo {author} {\bibfnamefont {G.~K.}\ \bibnamefont
  {Ocker}}, \bibinfo {author} {\bibfnamefont {A.}~\bibnamefont
  {Litwin-Kumar}},\ and\ \bibinfo {author} {\bibfnamefont {B.}~\bibnamefont
  {Doiron}},\ }\bibfield  {title} {\bibinfo {title} {Self-organization of
  microcircuits in networks of spiking neurons with plastic synapses},\
  }\href@noop {} {\bibfield  {journal} {\bibinfo  {journal} {PLoS Computational
  Biology}\ }\textbf {\bibinfo {volume} {11}},\ \bibinfo {pages} {e1004458}
  (\bibinfo {year} {2015})}\BibitemShut {NoStop}%
\bibitem [{\citenamefont {Gustafsson}\ \emph {et~al.}(1989)\citenamefont
  {Gustafsson}, \citenamefont {Asztely}, \citenamefont {Hanse},\ and\
  \citenamefont {Wigstr{\"o}m}}]{gustafsson1989onset}%
  \BibitemOpen
  \bibfield  {author} {\bibinfo {author} {\bibfnamefont {B.}~\bibnamefont
  {Gustafsson}}, \bibinfo {author} {\bibfnamefont {F.}~\bibnamefont {Asztely}},
  \bibinfo {author} {\bibfnamefont {E.}~\bibnamefont {Hanse}},\ and\ \bibinfo
  {author} {\bibfnamefont {H.}~\bibnamefont {Wigstr{\"o}m}},\ }\bibfield
  {title} {\bibinfo {title} {Onset characteristics of long-term potentiation in
  the guinea-pig hippocampal ca1 region in vitro},\ }\href@noop {} {\bibfield
  {journal} {\bibinfo  {journal} {European Journal of Neuroscience}\ }\textbf
  {\bibinfo {volume} {1}},\ \bibinfo {pages} {382} (\bibinfo {year}
  {1989})}\BibitemShut {NoStop}%
\bibitem [{\citenamefont {Malenka}(1991)}]{malenka1991postsynaptic}%
  \BibitemOpen
  \bibfield  {author} {\bibinfo {author} {\bibfnamefont {R.~C.}\ \bibnamefont
  {Malenka}},\ }\bibfield  {title} {\bibinfo {title} {Postsynaptic factors
  control the duration of synaptic enhancement in area ca1 of the
  hippocampus},\ }\href@noop {} {\bibfield  {journal} {\bibinfo  {journal}
  {Neuron}\ }\textbf {\bibinfo {volume} {6}},\ \bibinfo {pages} {53} (\bibinfo
  {year} {1991})}\BibitemShut {NoStop}%
\bibitem [{\citenamefont {Malenka}\ and\ \citenamefont
  {Nicoll}(1993)}]{malenka1993nmda}%
  \BibitemOpen
  \bibfield  {author} {\bibinfo {author} {\bibfnamefont {R.~C.}\ \bibnamefont
  {Malenka}}\ and\ \bibinfo {author} {\bibfnamefont {R.~A.}\ \bibnamefont
  {Nicoll}},\ }\bibfield  {title} {\bibinfo {title} {Nmda-receptor-dependent
  synaptic plasticity: multiple forms and mechanisms},\ }\href@noop {}
  {\bibfield  {journal} {\bibinfo  {journal} {Trends in Neurosciences}\
  }\textbf {\bibinfo {volume} {16}},\ \bibinfo {pages} {521} (\bibinfo {year}
  {1993})}\BibitemShut {NoStop}%
\bibitem [{\citenamefont {Volianskis}\ and\ \citenamefont
  {Jensen}(2003)}]{volianskis2003transient}%
  \BibitemOpen
  \bibfield  {author} {\bibinfo {author} {\bibfnamefont {A.}~\bibnamefont
  {Volianskis}}\ and\ \bibinfo {author} {\bibfnamefont {M.~S.}\ \bibnamefont
  {Jensen}},\ }\bibfield  {title} {\bibinfo {title} {Transient and sustained
  types of long-term potentiation in the ca1 area of the rat hippocampus},\
  }\href@noop {} {\bibfield  {journal} {\bibinfo  {journal} {The Journal of
  Physiology}\ }\textbf {\bibinfo {volume} {550}},\ \bibinfo {pages} {459}
  (\bibinfo {year} {2003})}\BibitemShut {NoStop}%
\bibitem [{\citenamefont {Erickson}\ \emph {et~al.}(2010)\citenamefont
  {Erickson}, \citenamefont {Maramara},\ and\ \citenamefont
  {Lisman}}]{erickson2010single}%
  \BibitemOpen
  \bibfield  {author} {\bibinfo {author} {\bibfnamefont {M.~A.}\ \bibnamefont
  {Erickson}}, \bibinfo {author} {\bibfnamefont {L.~A.}\ \bibnamefont
  {Maramara}},\ and\ \bibinfo {author} {\bibfnamefont {J.}~\bibnamefont
  {Lisman}},\ }\bibfield  {title} {\bibinfo {title} {A single brief burst
  induces glur1-dependent associative short-term potentiation: a potential
  mechanism for short-term memory},\ }\href@noop {} {\bibfield  {journal}
  {\bibinfo  {journal} {Journal of Cognitive Neuroscience}\ }\textbf {\bibinfo
  {volume} {22}},\ \bibinfo {pages} {2530} (\bibinfo {year}
  {2010})}\BibitemShut {NoStop}%
\bibitem [{\citenamefont {Volianskis}\ \emph {et~al.}(2013)\citenamefont
  {Volianskis}, \citenamefont {Bannister}, \citenamefont {Collett},
  \citenamefont {Irvine}, \citenamefont {Monaghan}, \citenamefont {Fitzjohn},
  \citenamefont {Jensen}, \citenamefont {Jane},\ and\ \citenamefont
  {Collingridge}}]{volianskis2013different}%
  \BibitemOpen
  \bibfield  {author} {\bibinfo {author} {\bibfnamefont {A.}~\bibnamefont
  {Volianskis}}, \bibinfo {author} {\bibfnamefont {N.}~\bibnamefont
  {Bannister}}, \bibinfo {author} {\bibfnamefont {V.~J.}\ \bibnamefont
  {Collett}}, \bibinfo {author} {\bibfnamefont {M.~W.}\ \bibnamefont {Irvine}},
  \bibinfo {author} {\bibfnamefont {D.~T.}\ \bibnamefont {Monaghan}}, \bibinfo
  {author} {\bibfnamefont {S.~M.}\ \bibnamefont {Fitzjohn}}, \bibinfo {author}
  {\bibfnamefont {M.~S.}\ \bibnamefont {Jensen}}, \bibinfo {author}
  {\bibfnamefont {D.~E.}\ \bibnamefont {Jane}},\ and\ \bibinfo {author}
  {\bibfnamefont {G.~L.}\ \bibnamefont {Collingridge}},\ }\bibfield  {title}
  {\bibinfo {title} {Different nmda receptor subtypes mediate induction of
  long-term potentiation and two forms of short-term potentiation at ca1
  synapses in rat hippocampus in vitro},\ }\href@noop {} {\bibfield  {journal}
  {\bibinfo  {journal} {The Journal of Physiology}\ }\textbf {\bibinfo {volume}
  {591}},\ \bibinfo {pages} {955} (\bibinfo {year} {2013})}\BibitemShut
  {NoStop}%
\bibitem [{\citenamefont {Driesen}\ \emph {et~al.}(2013)\citenamefont
  {Driesen}, \citenamefont {McCarthy}, \citenamefont {Bhagwagar}, \citenamefont
  {Bloch}, \citenamefont {Calhoun}, \citenamefont {D'souza}, \citenamefont
  {Gueorguieva}, \citenamefont {He}, \citenamefont {Leung}, \citenamefont
  {Ramani} \emph {et~al.}}]{driesen2013impact}%
  \BibitemOpen
  \bibfield  {author} {\bibinfo {author} {\bibfnamefont {N.~R.}\ \bibnamefont
  {Driesen}}, \bibinfo {author} {\bibfnamefont {G.}~\bibnamefont {McCarthy}},
  \bibinfo {author} {\bibfnamefont {Z.}~\bibnamefont {Bhagwagar}}, \bibinfo
  {author} {\bibfnamefont {M.~H.}\ \bibnamefont {Bloch}}, \bibinfo {author}
  {\bibfnamefont {V.~D.}\ \bibnamefont {Calhoun}}, \bibinfo {author}
  {\bibfnamefont {D.~C.}\ \bibnamefont {D'souza}}, \bibinfo {author}
  {\bibfnamefont {R.}~\bibnamefont {Gueorguieva}}, \bibinfo {author}
  {\bibfnamefont {G.}~\bibnamefont {He}}, \bibinfo {author} {\bibfnamefont
  {H.-C.}\ \bibnamefont {Leung}}, \bibinfo {author} {\bibfnamefont
  {R.}~\bibnamefont {Ramani}}, \emph {et~al.},\ }\bibfield  {title} {\bibinfo
  {title} {The impact of nmda receptor blockade on human working memory-related
  prefrontal function and connectivity},\ }\href@noop {} {\bibfield  {journal}
  {\bibinfo  {journal} {Neuropsychopharmacology}\ }\textbf {\bibinfo {volume}
  {38}},\ \bibinfo {pages} {2613} (\bibinfo {year} {2013})}\BibitemShut
  {NoStop}%
\bibitem [{\citenamefont {Park}\ \emph {et~al.}(2014)\citenamefont {Park},
  \citenamefont {Volianskis}, \citenamefont {Sanderson}, \citenamefont
  {Bortolotto}, \citenamefont {Jane}, \citenamefont {Zhuo}, \citenamefont
  {Kaang},\ and\ \citenamefont {Collingridge}}]{park2014nmda}%
  \BibitemOpen
  \bibfield  {author} {\bibinfo {author} {\bibfnamefont {P.}~\bibnamefont
  {Park}}, \bibinfo {author} {\bibfnamefont {A.}~\bibnamefont {Volianskis}},
  \bibinfo {author} {\bibfnamefont {T.~M.}\ \bibnamefont {Sanderson}}, \bibinfo
  {author} {\bibfnamefont {Z.~A.}\ \bibnamefont {Bortolotto}}, \bibinfo
  {author} {\bibfnamefont {D.~E.}\ \bibnamefont {Jane}}, \bibinfo {author}
  {\bibfnamefont {M.}~\bibnamefont {Zhuo}}, \bibinfo {author} {\bibfnamefont
  {B.-K.}\ \bibnamefont {Kaang}},\ and\ \bibinfo {author} {\bibfnamefont
  {G.~L.}\ \bibnamefont {Collingridge}},\ }\bibfield  {title} {\bibinfo {title}
  {Nmda receptor-dependent long-term potentiation comprises a family of
  temporally overlapping forms of synaptic plasticity that are induced by
  different patterns of stimulation},\ }\href@noop {} {\bibfield  {journal}
  {\bibinfo  {journal} {Philosophical Transactions of the Royal Society B:
  Biological Sciences}\ }\textbf {\bibinfo {volume} {369}},\ \bibinfo {pages}
  {20130131} (\bibinfo {year} {2014})}\BibitemShut {NoStop}%
\bibitem [{\citenamefont {Lisman}(2017)}]{lisman2017glutamatergic}%
  \BibitemOpen
  \bibfield  {author} {\bibinfo {author} {\bibfnamefont {J.}~\bibnamefont
  {Lisman}},\ }\bibfield  {title} {\bibinfo {title} {Glutamatergic synapses are
  structurally and biochemically complex because of multiple plasticity
  processes: long-term potentiation, long-term depression, short-term
  potentiation and scaling},\ }\href@noop {} {\bibfield  {journal} {\bibinfo
  {journal} {Philosophical Transactions of the Royal Society B: Biological
  Sciences}\ }\textbf {\bibinfo {volume} {372}},\ \bibinfo {pages} {20160260}
  (\bibinfo {year} {2017})}\BibitemShut {NoStop}%
\bibitem [{\citenamefont {Lansner}\ \emph {et~al.}(2023)\citenamefont
  {Lansner}, \citenamefont {Fiebig},\ and\ \citenamefont
  {Herman}}]{lansner2023hebbian}%
  \BibitemOpen
  \bibfield  {author} {\bibinfo {author} {\bibfnamefont {A.}~\bibnamefont
  {Lansner}}, \bibinfo {author} {\bibfnamefont {F.}~\bibnamefont {Fiebig}},\
  and\ \bibinfo {author} {\bibfnamefont {P.}~\bibnamefont {Herman}},\
  }\bibfield  {title} {\bibinfo {title} {Hebbian fast plasticity and working
  memory},\ }\href@noop {} {\bibfield  {journal} {\bibinfo  {journal} {arXiv
  preprint arXiv:2304.06626}\ } (\bibinfo {year} {2023})}\BibitemShut {NoStop}%
\bibitem [{\citenamefont {Hopfield}(1982)}]{hopfield1982neural}%
  \BibitemOpen
  \bibfield  {author} {\bibinfo {author} {\bibfnamefont {J.~J.}\ \bibnamefont
  {Hopfield}},\ }\bibfield  {title} {\bibinfo {title} {Neural networks and
  physical systems with emergent collective computational abilities},\
  }\href@noop {} {\bibfield  {journal} {\bibinfo  {journal} {Proceedings of the
  National Academy of Sciences}\ }\textbf {\bibinfo {volume} {79}},\ \bibinfo
  {pages} {2554} (\bibinfo {year} {1982})}\BibitemShut {NoStop}%
\bibitem [{\citenamefont {Miconi}\ \emph
  {et~al.}(2018{\natexlab{a}})\citenamefont {Miconi}, \citenamefont {Rawal},
  \citenamefont {Clune},\ and\ \citenamefont
  {Stanley}}]{miconi2018backpropamine}%
  \BibitemOpen
  \bibfield  {author} {\bibinfo {author} {\bibfnamefont {T.}~\bibnamefont
  {Miconi}}, \bibinfo {author} {\bibfnamefont {A.}~\bibnamefont {Rawal}},
  \bibinfo {author} {\bibfnamefont {J.}~\bibnamefont {Clune}},\ and\ \bibinfo
  {author} {\bibfnamefont {K.~O.}\ \bibnamefont {Stanley}},\ }\bibfield
  {title} {\bibinfo {title} {Backpropamine: training self-modifying neural
  networks with differentiable neuromodulated plasticity},\ }in\ \href@noop {}
  {\emph {\bibinfo {booktitle} {International Conference on Learning
  Representations}}}\ (\bibinfo {year} {2018})\BibitemShut {NoStop}%
\bibitem [{\citenamefont {Miconi}\ \emph
  {et~al.}(2018{\natexlab{b}})\citenamefont {Miconi}, \citenamefont {Stanley},\
  and\ \citenamefont {Clune}}]{miconi2018differentiable}%
  \BibitemOpen
  \bibfield  {author} {\bibinfo {author} {\bibfnamefont {T.}~\bibnamefont
  {Miconi}}, \bibinfo {author} {\bibfnamefont {K.}~\bibnamefont {Stanley}},\
  and\ \bibinfo {author} {\bibfnamefont {J.}~\bibnamefont {Clune}},\ }\bibfield
   {title} {\bibinfo {title} {Differentiable plasticity: training plastic
  neural networks with backpropagation},\ }in\ \href@noop {} {\emph {\bibinfo
  {booktitle} {International Conference on Machine Learning}}}\ (\bibinfo
  {organization} {PMLR},\ \bibinfo {year} {2018})\ pp.\ \bibinfo {pages}
  {3559--3568}\BibitemShut {NoStop}%
\bibitem [{\citenamefont {Hinton}\ and\ \citenamefont
  {Plaut}(1987)}]{hinton1987using}%
  \BibitemOpen
  \bibfield  {author} {\bibinfo {author} {\bibfnamefont {G.~E.}\ \bibnamefont
  {Hinton}}\ and\ \bibinfo {author} {\bibfnamefont {D.~C.}\ \bibnamefont
  {Plaut}},\ }\bibfield  {title} {\bibinfo {title} {Using fast weights to
  deblur old memories},\ }in\ \href@noop {} {\emph {\bibinfo {booktitle}
  {Proceedings of the 9th annual conference of the cognitive science
  society}}}\ (\bibinfo {year} {1987})\ pp.\ \bibinfo {pages}
  {177--186}\BibitemShut {NoStop}%
\bibitem [{\citenamefont {Miconi}(2022)}]{miconi2022learning}%
  \BibitemOpen
  \bibfield  {author} {\bibinfo {author} {\bibfnamefont {T.}~\bibnamefont
  {Miconi}},\ }\bibfield  {title} {\bibinfo {title} {Learning to acquire novel
  cognitive tasks with evolution, plasticity and meta-meta-learning},\ }in\
  \href@noop {} {\emph {\bibinfo {booktitle} {Proceedings of the Genetic and
  Evolutionary Computation Conference Companion}}}\ (\bibinfo {year} {2022})\
  pp.\ \bibinfo {pages} {1971--1978}\BibitemShut {NoStop}%
\bibitem [{\citenamefont {Tyulmankov}\ \emph {et~al.}(2022)\citenamefont
  {Tyulmankov}, \citenamefont {Yang},\ and\ \citenamefont
  {Abbott}}]{tyulmankov2022meta}%
  \BibitemOpen
  \bibfield  {author} {\bibinfo {author} {\bibfnamefont {D.}~\bibnamefont
  {Tyulmankov}}, \bibinfo {author} {\bibfnamefont {G.~R.}\ \bibnamefont
  {Yang}},\ and\ \bibinfo {author} {\bibfnamefont {L.~F.}\ \bibnamefont
  {Abbott}},\ }\bibfield  {title} {\bibinfo {title} {Meta-learning synaptic
  plasticity and memory addressing for continual familiarity detection},\
  }\href@noop {} {\bibfield  {journal} {\bibinfo  {journal} {Neuron}\ }\textbf
  {\bibinfo {volume} {110}},\ \bibinfo {pages} {544} (\bibinfo {year}
  {2022})}\BibitemShut {NoStop}%
\bibitem [{\citenamefont {Masse}\ \emph {et~al.}(2019)\citenamefont {Masse},
  \citenamefont {Yang}, \citenamefont {Song}, \citenamefont {Wang},\ and\
  \citenamefont {Freedman}}]{masse2019circuit}%
  \BibitemOpen
  \bibfield  {author} {\bibinfo {author} {\bibfnamefont {N.~Y.}\ \bibnamefont
  {Masse}}, \bibinfo {author} {\bibfnamefont {G.~R.}\ \bibnamefont {Yang}},
  \bibinfo {author} {\bibfnamefont {H.~F.}\ \bibnamefont {Song}}, \bibinfo
  {author} {\bibfnamefont {X.-J.}\ \bibnamefont {Wang}},\ and\ \bibinfo
  {author} {\bibfnamefont {D.~J.}\ \bibnamefont {Freedman}},\ }\bibfield
  {title} {\bibinfo {title} {Circuit mechanisms for the maintenance and
  manipulation of information in working memory},\ }\href@noop {} {\bibfield
  {journal} {\bibinfo  {journal} {Nature Neuroscience}\ }\textbf {\bibinfo
  {volume} {22}},\ \bibinfo {pages} {1159} (\bibinfo {year}
  {2019})}\BibitemShut {NoStop}%
\bibitem [{\citenamefont {Ba}\ \emph {et~al.}(2016)\citenamefont {Ba},
  \citenamefont {Hinton}, \citenamefont {Mnih}, \citenamefont {Leibo},\ and\
  \citenamefont {Ionescu}}]{ba2016using}%
  \BibitemOpen
  \bibfield  {author} {\bibinfo {author} {\bibfnamefont {J.}~\bibnamefont
  {Ba}}, \bibinfo {author} {\bibfnamefont {G.~E.}\ \bibnamefont {Hinton}},
  \bibinfo {author} {\bibfnamefont {V.}~\bibnamefont {Mnih}}, \bibinfo {author}
  {\bibfnamefont {J.~Z.}\ \bibnamefont {Leibo}},\ and\ \bibinfo {author}
  {\bibfnamefont {C.}~\bibnamefont {Ionescu}},\ }\bibfield  {title} {\bibinfo
  {title} {Using fast weights to attend to the recent past},\ }\href@noop {}
  {\bibfield  {journal} {\bibinfo  {journal} {Advances in Neural Information
  Processing Systems}\ }\textbf {\bibinfo {volume} {29}} (\bibinfo {year}
  {2016})}\BibitemShut {NoStop}%
\bibitem [{\citenamefont {Vaswani}\ \emph {et~al.}(2017)\citenamefont
  {Vaswani}, \citenamefont {Shazeer}, \citenamefont {Parmar}, \citenamefont
  {Uszkoreit}, \citenamefont {Jones}, \citenamefont {Gomez}, \citenamefont
  {Kaiser},\ and\ \citenamefont {Polosukhin}}]{vaswani2017attention}%
  \BibitemOpen
  \bibfield  {author} {\bibinfo {author} {\bibfnamefont {A.}~\bibnamefont
  {Vaswani}}, \bibinfo {author} {\bibfnamefont {N.}~\bibnamefont {Shazeer}},
  \bibinfo {author} {\bibfnamefont {N.}~\bibnamefont {Parmar}}, \bibinfo
  {author} {\bibfnamefont {J.}~\bibnamefont {Uszkoreit}}, \bibinfo {author}
  {\bibfnamefont {L.}~\bibnamefont {Jones}}, \bibinfo {author} {\bibfnamefont
  {A.~N.}\ \bibnamefont {Gomez}}, \bibinfo {author} {\bibfnamefont
  {{\L}.}~\bibnamefont {Kaiser}},\ and\ \bibinfo {author} {\bibfnamefont
  {I.}~\bibnamefont {Polosukhin}},\ }\bibfield  {title} {\bibinfo {title}
  {Attention is all you need},\ }\href@noop {} {\bibfield  {journal} {\bibinfo
  {journal} {Advances in Neural Information Processing Systems}\ }\textbf
  {\bibinfo {volume} {30}} (\bibinfo {year} {2017})}\BibitemShut {NoStop}%
\bibitem [{\citenamefont {Schlag}\ \emph {et~al.}(2021)\citenamefont {Schlag},
  \citenamefont {Irie},\ and\ \citenamefont {Schmidhuber}}]{schlag2021linear}%
  \BibitemOpen
  \bibfield  {author} {\bibinfo {author} {\bibfnamefont {I.}~\bibnamefont
  {Schlag}}, \bibinfo {author} {\bibfnamefont {K.}~\bibnamefont {Irie}},\ and\
  \bibinfo {author} {\bibfnamefont {J.}~\bibnamefont {Schmidhuber}},\
  }\bibfield  {title} {\bibinfo {title} {Linear transformers are secretly fast
  weight programmers},\ }in\ \href@noop {} {\emph {\bibinfo {booktitle}
  {International Conference on Machine Learning}}}\ (\bibinfo {organization}
  {PMLR},\ \bibinfo {year} {2021})\ pp.\ \bibinfo {pages}
  {9355--9366}\BibitemShut {NoStop}%
\bibitem [{\citenamefont {Katharopoulos}\ \emph {et~al.}(2020)\citenamefont
  {Katharopoulos}, \citenamefont {Vyas}, \citenamefont {Pappas},\ and\
  \citenamefont {Fleuret}}]{katharopoulos2020transformers}%
  \BibitemOpen
  \bibfield  {author} {\bibinfo {author} {\bibfnamefont {A.}~\bibnamefont
  {Katharopoulos}}, \bibinfo {author} {\bibfnamefont {A.}~\bibnamefont {Vyas}},
  \bibinfo {author} {\bibfnamefont {N.}~\bibnamefont {Pappas}},\ and\ \bibinfo
  {author} {\bibfnamefont {F.}~\bibnamefont {Fleuret}},\ }\bibfield  {title}
  {\bibinfo {title} {Transformers are rnns: Fast autoregressive transformers
  with linear attention},\ }in\ \href@noop {} {\emph {\bibinfo {booktitle}
  {International Conference on Machine Learning}}}\ (\bibinfo {organization}
  {PMLR},\ \bibinfo {year} {2020})\ pp.\ \bibinfo {pages}
  {5156--5165}\BibitemShut {NoStop}%
\bibitem [{\citenamefont {Schmidhuber}(1992)}]{schmidhuber1992learning}%
  \BibitemOpen
  \bibfield  {author} {\bibinfo {author} {\bibfnamefont {J.}~\bibnamefont
  {Schmidhuber}},\ }\bibfield  {title} {\bibinfo {title} {Learning to control
  fast-weight memories: An alternative to dynamic recurrent networks},\
  }\href@noop {} {\bibfield  {journal} {\bibinfo  {journal} {Neural
  Computation}\ }\textbf {\bibinfo {volume} {4}},\ \bibinfo {pages} {131}
  (\bibinfo {year} {1992})}\BibitemShut {NoStop}%
\bibitem [{Note1()}]{Note1}%
  \BibitemOpen
  \bibinfo {note} {This equivalence arises due to the fact that \protect
  \textit {attention}, the core component of transformers, is based on forming
  linear combinations of neuronal states weighted by inner products between
  pairs of neuronal states; this can be mimicked by multiplying a neuronal
  state by a coupling matrix that has been shaped through activity-dependent
  plasticity to encode outer products of neuronal states.}\BibitemShut {Stop}%
\bibitem [{\citenamefont {Magnasco}\ \emph {et~al.}(2009)\citenamefont
  {Magnasco}, \citenamefont {Piro},\ and\ \citenamefont
  {Cecchi}}]{magnasco2009self}%
  \BibitemOpen
  \bibfield  {author} {\bibinfo {author} {\bibfnamefont {M.~O.}\ \bibnamefont
  {Magnasco}}, \bibinfo {author} {\bibfnamefont {O.}~\bibnamefont {Piro}},\
  and\ \bibinfo {author} {\bibfnamefont {G.~A.}\ \bibnamefont {Cecchi}},\
  }\bibfield  {title} {\bibinfo {title} {Self-tuned critical anti-hebbian
  networks},\ }\href@noop {} {\bibfield  {journal} {\bibinfo  {journal}
  {Physical Review Letters}\ }\textbf {\bibinfo {volume} {102}},\ \bibinfo
  {pages} {258102} (\bibinfo {year} {2009})}\BibitemShut {NoStop}%
\bibitem [{\citenamefont {Sussillo}\ and\ \citenamefont
  {Abbott}(2009)}]{sussillo2009generating}%
  \BibitemOpen
  \bibfield  {author} {\bibinfo {author} {\bibfnamefont {D.}~\bibnamefont
  {Sussillo}}\ and\ \bibinfo {author} {\bibfnamefont {L.~F.}\ \bibnamefont
  {Abbott}},\ }\bibfield  {title} {\bibinfo {title} {Generating coherent
  patterns of activity from chaotic neural networks},\ }\href@noop {}
  {\bibfield  {journal} {\bibinfo  {journal} {Neuron}\ }\textbf {\bibinfo
  {volume} {63}},\ \bibinfo {pages} {544} (\bibinfo {year} {2009})}\BibitemShut
  {NoStop}%
\bibitem [{\citenamefont {Sompolinsky}\ \emph {et~al.}(1988)\citenamefont
  {Sompolinsky}, \citenamefont {Crisanti},\ and\ \citenamefont
  {Sommers}}]{sompolinsky1988chaos}%
  \BibitemOpen
  \bibfield  {author} {\bibinfo {author} {\bibfnamefont {H.}~\bibnamefont
  {Sompolinsky}}, \bibinfo {author} {\bibfnamefont {A.}~\bibnamefont
  {Crisanti}},\ and\ \bibinfo {author} {\bibfnamefont {H.-J.}\ \bibnamefont
  {Sommers}},\ }\bibfield  {title} {\bibinfo {title} {Chaos in random neural
  networks},\ }\href@noop {} {\bibfield  {journal} {\bibinfo  {journal}
  {Physical Review Letters}\ }\textbf {\bibinfo {volume} {61}},\ \bibinfo
  {pages} {259} (\bibinfo {year} {1988})}\BibitemShut {NoStop}%
\bibitem [{\citenamefont {Clark}\ \emph {et~al.}(2023)\citenamefont {Clark},
  \citenamefont {Abbott},\ and\ \citenamefont
  {Litwin-Kumar}}]{clark2023dimension}%
  \BibitemOpen
  \bibfield  {author} {\bibinfo {author} {\bibfnamefont {D.~G.}\ \bibnamefont
  {Clark}}, \bibinfo {author} {\bibfnamefont {L.~F.}\ \bibnamefont {Abbott}},\
  and\ \bibinfo {author} {\bibfnamefont {A.}~\bibnamefont {Litwin-Kumar}},\
  }\bibfield  {title} {\bibinfo {title} {Dimension of activity in random neural
  networks},\ }\href@noop {} {\bibfield  {journal} {\bibinfo  {journal}
  {Physical Review Letters}\ }\textbf {\bibinfo {volume} {131}},\ \bibinfo
  {pages} {118401} (\bibinfo {year} {2023})}\BibitemShut {NoStop}%
\bibitem [{\citenamefont {Kadmon}\ and\ \citenamefont
  {Sompolinsky}(2015)}]{kadmon2015transition}%
  \BibitemOpen
  \bibfield  {author} {\bibinfo {author} {\bibfnamefont {J.}~\bibnamefont
  {Kadmon}}\ and\ \bibinfo {author} {\bibfnamefont {H.}~\bibnamefont
  {Sompolinsky}},\ }\bibfield  {title} {\bibinfo {title} {Transition to chaos
  in random neuronal networks},\ }\href@noop {} {\bibfield  {journal} {\bibinfo
   {journal} {Physical Review X}\ }\textbf {\bibinfo {volume} {5}},\ \bibinfo
  {pages} {041030} (\bibinfo {year} {2015})}\BibitemShut {NoStop}%
\bibitem [{\citenamefont {Mastrogiuseppe}\ and\ \citenamefont
  {Ostojic}(2018)}]{mastrogiuseppe2018linking}%
  \BibitemOpen
  \bibfield  {author} {\bibinfo {author} {\bibfnamefont {F.}~\bibnamefont
  {Mastrogiuseppe}}\ and\ \bibinfo {author} {\bibfnamefont {S.}~\bibnamefont
  {Ostojic}},\ }\bibfield  {title} {\bibinfo {title} {Linking connectivity,
  dynamics, and computations in low-rank recurrent neural networks},\
  }\href@noop {} {\bibfield  {journal} {\bibinfo  {journal} {Neuron}\ }\textbf
  {\bibinfo {volume} {99}},\ \bibinfo {pages} {609} (\bibinfo {year}
  {2018})}\BibitemShut {NoStop}%
\bibitem [{\citenamefont {Schuessler}\ \emph
  {et~al.}(2020{\natexlab{a}})\citenamefont {Schuessler}, \citenamefont
  {Dubreuil}, \citenamefont {Mastrogiuseppe}, \citenamefont {Ostojic},\ and\
  \citenamefont {Barak}}]{schuessler2020dynamics}%
  \BibitemOpen
  \bibfield  {author} {\bibinfo {author} {\bibfnamefont {F.}~\bibnamefont
  {Schuessler}}, \bibinfo {author} {\bibfnamefont {A.}~\bibnamefont
  {Dubreuil}}, \bibinfo {author} {\bibfnamefont {F.}~\bibnamefont
  {Mastrogiuseppe}}, \bibinfo {author} {\bibfnamefont {S.}~\bibnamefont
  {Ostojic}},\ and\ \bibinfo {author} {\bibfnamefont {O.}~\bibnamefont
  {Barak}},\ }\bibfield  {title} {\bibinfo {title} {Dynamics of random
  recurrent networks with correlated low-rank structure},\ }\href@noop {}
  {\bibfield  {journal} {\bibinfo  {journal} {Physical Review Research}\
  }\textbf {\bibinfo {volume} {2}},\ \bibinfo {pages} {013111} (\bibinfo {year}
  {2020}{\natexlab{a}})}\BibitemShut {NoStop}%
\bibitem [{\citenamefont {Schuessler}\ \emph
  {et~al.}(2020{\natexlab{b}})\citenamefont {Schuessler}, \citenamefont
  {Mastrogiuseppe}, \citenamefont {Dubreuil}, \citenamefont {Ostojic},\ and\
  \citenamefont {Barak}}]{schuessler2020interplay}%
  \BibitemOpen
  \bibfield  {author} {\bibinfo {author} {\bibfnamefont {F.}~\bibnamefont
  {Schuessler}}, \bibinfo {author} {\bibfnamefont {F.}~\bibnamefont
  {Mastrogiuseppe}}, \bibinfo {author} {\bibfnamefont {A.}~\bibnamefont
  {Dubreuil}}, \bibinfo {author} {\bibfnamefont {S.}~\bibnamefont {Ostojic}},\
  and\ \bibinfo {author} {\bibfnamefont {O.}~\bibnamefont {Barak}},\ }\bibfield
   {title} {\bibinfo {title} {The interplay between randomness and structure
  during learning in rnns},\ }\href@noop {} {\bibfield  {journal} {\bibinfo
  {journal} {Advances in Neural Information Processing Systems}\ }\textbf
  {\bibinfo {volume} {33}},\ \bibinfo {pages} {13352} (\bibinfo {year}
  {2020}{\natexlab{b}})}\BibitemShut {NoStop}%
\bibitem [{\citenamefont {Landau}\ and\ \citenamefont
  {Sompolinsky}(2018)}]{landau2018coherent}%
  \BibitemOpen
  \bibfield  {author} {\bibinfo {author} {\bibfnamefont {I.~D.}\ \bibnamefont
  {Landau}}\ and\ \bibinfo {author} {\bibfnamefont {H.}~\bibnamefont
  {Sompolinsky}},\ }\bibfield  {title} {\bibinfo {title} {Coherent chaos in a
  recurrent neural network with structured connectivity},\ }\href@noop {}
  {\bibfield  {journal} {\bibinfo  {journal} {PLoS Computational Biology}\
  }\textbf {\bibinfo {volume} {14}},\ \bibinfo {pages} {e1006309} (\bibinfo
  {year} {2018})}\BibitemShut {NoStop}%
\bibitem [{\citenamefont {Baik}\ \emph {et~al.}(2005)\citenamefont {Baik},
  \citenamefont {{Ben Arous}},\ and\ \citenamefont
  {P{\'e}ch{\'e}}}]{baik2005phase}%
  \BibitemOpen
  \bibfield  {author} {\bibinfo {author} {\bibfnamefont {J.}~\bibnamefont
  {Baik}}, \bibinfo {author} {\bibfnamefont {G.}~\bibnamefont {{Ben Arous}}},\
  and\ \bibinfo {author} {\bibfnamefont {S.}~\bibnamefont {P{\'e}ch{\'e}}},\
  }\bibfield  {title} {\bibinfo {title} {{Phase transition of the largest
  eigenvalue for nonnull complex sample covariance matrices}},\ }\href
  {https://doi.org/10.1214/009117905000000233} {\bibfield  {journal} {\bibinfo
  {journal} {The Annals of Probability}\ }\textbf {\bibinfo {volume} {33}},\
  \bibinfo {pages} {1643 } (\bibinfo {year} {2005})}\BibitemShut {NoStop}%
\bibitem [{\citenamefont {Krishnamurthy}\ \emph {et~al.}(2022)\citenamefont
  {Krishnamurthy}, \citenamefont {Can},\ and\ \citenamefont
  {Schwab}}]{krishnamurthy2022theory}%
  \BibitemOpen
  \bibfield  {author} {\bibinfo {author} {\bibfnamefont {K.}~\bibnamefont
  {Krishnamurthy}}, \bibinfo {author} {\bibfnamefont {T.}~\bibnamefont {Can}},\
  and\ \bibinfo {author} {\bibfnamefont {D.~J.}\ \bibnamefont {Schwab}},\
  }\bibfield  {title} {\bibinfo {title} {Theory of gating in recurrent neural
  networks},\ }\href@noop {} {\bibfield  {journal} {\bibinfo  {journal}
  {Physical Review X}\ }\textbf {\bibinfo {volume} {12}},\ \bibinfo {pages}
  {011011} (\bibinfo {year} {2022})}\BibitemShut {NoStop}%
\bibitem [{\citenamefont {Roy}\ \emph {et~al.}(2019)\citenamefont {Roy},
  \citenamefont {Biroli}, \citenamefont {Bunin},\ and\ \citenamefont
  {Cammarota}}]{roy2019numerical}%
  \BibitemOpen
  \bibfield  {author} {\bibinfo {author} {\bibfnamefont {F.}~\bibnamefont
  {Roy}}, \bibinfo {author} {\bibfnamefont {G.}~\bibnamefont {Biroli}},
  \bibinfo {author} {\bibfnamefont {G.}~\bibnamefont {Bunin}},\ and\ \bibinfo
  {author} {\bibfnamefont {C.}~\bibnamefont {Cammarota}},\ }\bibfield  {title}
  {\bibinfo {title} {Numerical implementation of dynamical mean field theory
  for disordered systems: Application to the lotka--volterra model of
  ecosystems},\ }\href@noop {} {\bibfield  {journal} {\bibinfo  {journal}
  {Journal of Physics A: Mathematical and Theoretical}\ }\textbf {\bibinfo
  {volume} {52}},\ \bibinfo {pages} {484001} (\bibinfo {year}
  {2019})}\BibitemShut {NoStop}%
\bibitem [{\citenamefont {Eissfeller}\ and\ \citenamefont
  {Opper}(1992)}]{eissfeller1992new}%
  \BibitemOpen
  \bibfield  {author} {\bibinfo {author} {\bibfnamefont {H.}~\bibnamefont
  {Eissfeller}}\ and\ \bibinfo {author} {\bibfnamefont {M.}~\bibnamefont
  {Opper}},\ }\bibfield  {title} {\bibinfo {title} {New method for studying the
  dynamics of disordered spin systems without finite-size effects},\
  }\href@noop {} {\bibfield  {journal} {\bibinfo  {journal} {Physical Review
  Letters}\ }\textbf {\bibinfo {volume} {68}},\ \bibinfo {pages} {2094}
  (\bibinfo {year} {1992})}\BibitemShut {NoStop}%
\bibitem [{\citenamefont {Eissfeller}\ and\ \citenamefont
  {Opper}(1994)}]{eissfeller1994mean}%
  \BibitemOpen
  \bibfield  {author} {\bibinfo {author} {\bibfnamefont {H.}~\bibnamefont
  {Eissfeller}}\ and\ \bibinfo {author} {\bibfnamefont {M.}~\bibnamefont
  {Opper}},\ }\bibfield  {title} {\bibinfo {title} {Mean-field monte carlo
  approach to the sherrington-kirkpatrick model with asymmetric couplings},\
  }\href@noop {} {\bibfield  {journal} {\bibinfo  {journal} {Physical Review
  E}\ }\textbf {\bibinfo {volume} {50}},\ \bibinfo {pages} {709} (\bibinfo
  {year} {1994})}\BibitemShut {NoStop}%
\bibitem [{\citenamefont {Stern}\ \emph {et~al.}(2014)\citenamefont {Stern},
  \citenamefont {Sompolinsky},\ and\ \citenamefont
  {Abbott}}]{stern2014dynamics}%
  \BibitemOpen
  \bibfield  {author} {\bibinfo {author} {\bibfnamefont {M.}~\bibnamefont
  {Stern}}, \bibinfo {author} {\bibfnamefont {H.}~\bibnamefont {Sompolinsky}},\
  and\ \bibinfo {author} {\bibfnamefont {L.~F.}\ \bibnamefont {Abbott}},\
  }\bibfield  {title} {\bibinfo {title} {Dynamics of random neural networks
  with bistable units},\ }\href@noop {} {\bibfield  {journal} {\bibinfo
  {journal} {Physical Review E}\ }\textbf {\bibinfo {volume} {90}},\ \bibinfo
  {pages} {062710} (\bibinfo {year} {2014})}\BibitemShut {NoStop}%
\bibitem [{\citenamefont {Mignacco}\ \emph {et~al.}(2020)\citenamefont
  {Mignacco}, \citenamefont {Krzakala}, \citenamefont {Urbani},\ and\
  \citenamefont {Zdeborov{\'a}}}]{mignacco2020dynamical}%
  \BibitemOpen
  \bibfield  {author} {\bibinfo {author} {\bibfnamefont {F.}~\bibnamefont
  {Mignacco}}, \bibinfo {author} {\bibfnamefont {F.}~\bibnamefont {Krzakala}},
  \bibinfo {author} {\bibfnamefont {P.}~\bibnamefont {Urbani}},\ and\ \bibinfo
  {author} {\bibfnamefont {L.}~\bibnamefont {Zdeborov{\'a}}},\ }\bibfield
  {title} {\bibinfo {title} {Dynamical mean-field theory for stochastic
  gradient descent in gaussian mixture classification},\ }\href@noop {}
  {\bibfield  {journal} {\bibinfo  {journal} {Advances in Neural Information
  Processing Systems}\ }\textbf {\bibinfo {volume} {33}},\ \bibinfo {pages}
  {9540} (\bibinfo {year} {2020})}\BibitemShut {NoStop}%
\bibitem [{\citenamefont {Beiran}\ and\ \citenamefont
  {Ostojic}(2019)}]{beiran2019contrasting}%
  \BibitemOpen
  \bibfield  {author} {\bibinfo {author} {\bibfnamefont {M.}~\bibnamefont
  {Beiran}}\ and\ \bibinfo {author} {\bibfnamefont {S.}~\bibnamefont
  {Ostojic}},\ }\bibfield  {title} {\bibinfo {title} {Contrasting the effects
  of adaptation and synaptic filtering on the timescales of dynamics in
  recurrent networks},\ }\href@noop {} {\bibfield  {journal} {\bibinfo
  {journal} {PLoS Computational Biology}\ }\textbf {\bibinfo {volume} {15}},\
  \bibinfo {pages} {e1006893} (\bibinfo {year} {2019})}\BibitemShut {NoStop}%
\bibitem [{\citenamefont {Muscinelli}\ \emph {et~al.}(2019)\citenamefont
  {Muscinelli}, \citenamefont {Gerstner},\ and\ \citenamefont
  {Schwalger}}]{muscinelli2019single}%
  \BibitemOpen
  \bibfield  {author} {\bibinfo {author} {\bibfnamefont {S.~P.}\ \bibnamefont
  {Muscinelli}}, \bibinfo {author} {\bibfnamefont {W.}~\bibnamefont
  {Gerstner}},\ and\ \bibinfo {author} {\bibfnamefont {T.}~\bibnamefont
  {Schwalger}},\ }\bibfield  {title} {\bibinfo {title} {How single neuron
  properties shape chaotic dynamics and signal transmission in random neural
  networks},\ }\href@noop {} {\bibfield  {journal} {\bibinfo  {journal} {PLoS
  Computational Biology}\ }\textbf {\bibinfo {volume} {15}},\ \bibinfo {pages}
  {e1007122} (\bibinfo {year} {2019})}\BibitemShut {NoStop}%
\bibitem [{\citenamefont {Girko}(1985)}]{girko1985circular}%
  \BibitemOpen
  \bibfield  {author} {\bibinfo {author} {\bibfnamefont {V.~L.}\ \bibnamefont
  {Girko}},\ }\bibfield  {title} {\bibinfo {title} {Circular law},\ }\href@noop
  {} {\bibfield  {journal} {\bibinfo  {journal} {Theory of Probability \& Its
  Applications}\ }\textbf {\bibinfo {volume} {29}},\ \bibinfo {pages} {694}
  (\bibinfo {year} {1985})}\BibitemShut {NoStop}%
\bibitem [{Note2()}]{Note2}%
  \BibitemOpen
  \bibinfo {note} {There is a more direct derivation of the spectrum of
  $\protect \bm {M}$, though it does not provide an interpretation of the
  reduced $2N$-dimensional dynamics. We use the Schur-complement identity $
  {\left |\begin {array}{cc} \protect \bm {A} & \protect \bm {B} \\ \protect
  \bm {C} & \protect \bm {D} \end {array}\right | = |\protect \bm {D}||\protect
  \bm {A} - \protect \bm {B}\protect \bm {D}^{-1}\protect \bm {C} | }$ to write
  the characteristic polynomial of $\protect \bm {M}$ as $ p_{\protect \bm
  {M}}(\lambda ) \propto \left (1 + p\lambda \right )^{S - N} \left | \lambda
  ^2 p \protect \bm {I}_N + \lambda \left ( \protect \bm {I}_N - p\protect
  \frac {\partial \protect \bm {F}}{\partial \protect \bm {x}} \right ) - \left
  (\protect \frac {\partial \protect \bm {F}}{\partial \protect \bm {x}} + k
  \protect \frac {\partial \protect \bm {F}}{\partial \protect \bm {a}}
  \protect \frac {\partial \protect \bm {G}}{\partial \protect \bm {x}} \right
  ) \right | $. The first term, $(1 + p\lambda )^{S-N}$, encodes the eigenvalue
  $ -1/p$ with multiplicity $S-N$. The second term encodes an $N$-dimensional
  quadratic eigenvalue problem (QEP). A QEP can be solved by {linearization},
  which entails forming a $2N$-dimensional linear eigenvalue problem whose
  eigenvalues coincide with those of the QEP \cite {tisseur2001quadratic}.
  Applying the aforementioned identity to the determinant defining the
  characteristic polynomial of the reduced Jacobian $\protect \tilde {\protect
  \bm {M}}$ shows that $\protect \tilde {\protect \bm {M}}$ is one such
  linearization of the QEP. Thus, we recover the originally derived spectrum.
  There are infinitely many linearizations given by matrices related to
  $\protect \tilde {\protect \bm {M}}$ by similarity transformation, so in this
  sense, the reduced Jacobian is not unique.}\BibitemShut {Stop}%
\bibitem [{Note3()}]{Note3}%
  \BibitemOpen
  \bibinfo {note} {In chaotic states, this is a result of the network being
  locally destabilized by a continuous band of eigenvalues in analogy with the
  nonplastic network \cite {sompolinsky1988chaos}. At fixed points, this is a
  result of the marginal stability of typical fixed points, which have a
  continuous band of Jacobian eigenvalues brushing against the stability line,
  $\protect \text {Re}(\lambda ) = 0$.}\BibitemShut {Stop}%
\bibitem [{\citenamefont {Ahmadian}\ \emph {et~al.}(2015)\citenamefont
  {Ahmadian}, \citenamefont {Fumarola},\ and\ \citenamefont
  {Miller}}]{ahmadian2015properties}%
  \BibitemOpen
  \bibfield  {author} {\bibinfo {author} {\bibfnamefont {Y.}~\bibnamefont
  {Ahmadian}}, \bibinfo {author} {\bibfnamefont {F.}~\bibnamefont {Fumarola}},\
  and\ \bibinfo {author} {\bibfnamefont {K.~D.}\ \bibnamefont {Miller}},\
  }\bibfield  {title} {\bibinfo {title} {Properties of networks with partially
  structured and partially random connectivity},\ }\href@noop {} {\bibfield
  {journal} {\bibinfo  {journal} {Physical Review E}\ }\textbf {\bibinfo
  {volume} {91}},\ \bibinfo {pages} {012820} (\bibinfo {year}
  {2015})}\BibitemShut {NoStop}%
\bibitem [{\citenamefont {Geist}\ \emph {et~al.}(1990)\citenamefont {Geist},
  \citenamefont {Parlitz},\ and\ \citenamefont
  {Lauterborn}}]{geist1990comparison}%
  \BibitemOpen
  \bibfield  {author} {\bibinfo {author} {\bibfnamefont {K.}~\bibnamefont
  {Geist}}, \bibinfo {author} {\bibfnamefont {U.}~\bibnamefont {Parlitz}},\
  and\ \bibinfo {author} {\bibfnamefont {W.}~\bibnamefont {Lauterborn}},\
  }\bibfield  {title} {\bibinfo {title} {Comparison of different methods for
  computing lyapunov exponents},\ }\href@noop {} {\bibfield  {journal}
  {\bibinfo  {journal} {Progress of theoretical physics}\ }\textbf {\bibinfo
  {volume} {83}},\ \bibinfo {pages} {875} (\bibinfo {year} {1990})}\BibitemShut
  {NoStop}%
\bibitem [{\citenamefont {Engelken}\ \emph {et~al.}(2023)\citenamefont
  {Engelken}, \citenamefont {Wolf},\ and\ \citenamefont
  {Abbott}}]{engelken2023lyapunov}%
  \BibitemOpen
  \bibfield  {author} {\bibinfo {author} {\bibfnamefont {R.}~\bibnamefont
  {Engelken}}, \bibinfo {author} {\bibfnamefont {F.}~\bibnamefont {Wolf}},\
  and\ \bibinfo {author} {\bibfnamefont {L.~F.}\ \bibnamefont {Abbott}},\
  }\bibfield  {title} {\bibinfo {title} {Lyapunov spectra of chaotic recurrent
  neural networks},\ }\href@noop {} {\bibfield  {journal} {\bibinfo  {journal}
  {Physical Review Research}\ }\textbf {\bibinfo {volume} {5}},\ \bibinfo
  {pages} {043044} (\bibinfo {year} {2023})}\BibitemShut {NoStop}%
\bibitem [{Note4()}]{Note4}%
  \BibitemOpen
  \bibinfo {note} {If the whole spectrum were computed, this spike would have
  $\protect \mathcal {O}(N^2)$ exponents; here, there are fewer since only the
  ends of the spectrum are computed. Moreover, in principle it should be a
  delta-function spike, but in the numerics there is some ``leakage'' across
  modes such that the spike has finite width.}\BibitemShut {Stop}%
\bibitem [{\citenamefont {Kaplan}\ and\ \citenamefont
  {Yorke}(2006)}]{kaplan2006chaotic}%
  \BibitemOpen
  \bibfield  {author} {\bibinfo {author} {\bibfnamefont {J.~L.}\ \bibnamefont
  {Kaplan}}\ and\ \bibinfo {author} {\bibfnamefont {J.~A.}\ \bibnamefont
  {Yorke}},\ }\bibfield  {title} {\bibinfo {title} {Chaotic behavior of
  multidimensional difference equations},\ }in\ \href@noop {} {\emph {\bibinfo
  {booktitle} {Functional Differential Equations and Approximation of Fixed
  Points: Proceedings, Bonn, July 1978}}}\ (\bibinfo  {publisher} {Springer},\
  \bibinfo {year} {2006})\ pp.\ \bibinfo {pages} {204--227}\BibitemShut
  {NoStop}%
\bibitem [{\citenamefont {Wainrib}\ and\ \citenamefont
  {Touboul}(2013)}]{wainrib2013topological}%
  \BibitemOpen
  \bibfield  {author} {\bibinfo {author} {\bibfnamefont {G.}~\bibnamefont
  {Wainrib}}\ and\ \bibinfo {author} {\bibfnamefont {J.}~\bibnamefont
  {Touboul}},\ }\bibfield  {title} {\bibinfo {title} {Topological and dynamical
  complexity of random neural networks},\ }\href@noop {} {\bibfield  {journal}
  {\bibinfo  {journal} {Physical Review Letters}\ }\textbf {\bibinfo {volume}
  {110}},\ \bibinfo {pages} {118101} (\bibinfo {year} {2013})}\BibitemShut
  {NoStop}%
\bibitem [{\citenamefont {Stubenrauch}\ \emph {et~al.}(2022)\citenamefont
  {Stubenrauch}, \citenamefont {Keup}, \citenamefont {Kurth}, \citenamefont
  {Helias},\ and\ \citenamefont {van Meegen}}]{stubenrauch2022phase}%
  \BibitemOpen
  \bibfield  {author} {\bibinfo {author} {\bibfnamefont {J.}~\bibnamefont
  {Stubenrauch}}, \bibinfo {author} {\bibfnamefont {C.}~\bibnamefont {Keup}},
  \bibinfo {author} {\bibfnamefont {A.~C.}\ \bibnamefont {Kurth}}, \bibinfo
  {author} {\bibfnamefont {M.}~\bibnamefont {Helias}},\ and\ \bibinfo {author}
  {\bibfnamefont {A.}~\bibnamefont {van Meegen}},\ }\bibfield  {title}
  {\bibinfo {title} {Phase space analysis of chaotic neural networks},\
  }\href@noop {} {\bibfield  {journal} {\bibinfo  {journal} {arXiv preprint
  arXiv:2210.07877}\ } (\bibinfo {year} {2022})}\BibitemShut {NoStop}%
\bibitem [{\citenamefont {Brunton}\ \emph {et~al.}(2017)\citenamefont
  {Brunton}, \citenamefont {Brunton}, \citenamefont {Proctor}, \citenamefont
  {Kaiser},\ and\ \citenamefont {Kutz}}]{brunton2017chaos}%
  \BibitemOpen
  \bibfield  {author} {\bibinfo {author} {\bibfnamefont {S.~L.}\ \bibnamefont
  {Brunton}}, \bibinfo {author} {\bibfnamefont {B.~W.}\ \bibnamefont
  {Brunton}}, \bibinfo {author} {\bibfnamefont {J.~L.}\ \bibnamefont
  {Proctor}}, \bibinfo {author} {\bibfnamefont {E.}~\bibnamefont {Kaiser}},\
  and\ \bibinfo {author} {\bibfnamefont {J.~N.}\ \bibnamefont {Kutz}},\
  }\bibfield  {title} {\bibinfo {title} {Chaos as an intermittently forced
  linear system},\ }\href@noop {} {\bibfield  {journal} {\bibinfo  {journal}
  {Nature Communications}\ }\textbf {\bibinfo {volume} {8}},\ \bibinfo {pages}
  {19} (\bibinfo {year} {2017})}\BibitemShut {NoStop}%
\bibitem [{\citenamefont {O{\textquoteright}Shea}\ \emph
  {et~al.}(2022)\citenamefont {O{\textquoteright}Shea}, \citenamefont
  {Duncker}, \citenamefont {Goo}, \citenamefont {Sun}, \citenamefont {Vyas},
  \citenamefont {Trautmann}, \citenamefont {Diester}, \citenamefont
  {Ramakrishnan}, \citenamefont {Deisseroth}, \citenamefont {Sahani},\ and\
  \citenamefont {Shenoy}}]{o2022direct}%
  \BibitemOpen
  \bibfield  {author} {\bibinfo {author} {\bibfnamefont {D.~J.}\ \bibnamefont
  {O{\textquoteright}Shea}}, \bibinfo {author} {\bibfnamefont {L.}~\bibnamefont
  {Duncker}}, \bibinfo {author} {\bibfnamefont {W.}~\bibnamefont {Goo}},
  \bibinfo {author} {\bibfnamefont {X.}~\bibnamefont {Sun}}, \bibinfo {author}
  {\bibfnamefont {S.}~\bibnamefont {Vyas}}, \bibinfo {author} {\bibfnamefont
  {E.~M.}\ \bibnamefont {Trautmann}}, \bibinfo {author} {\bibfnamefont
  {I.}~\bibnamefont {Diester}}, \bibinfo {author} {\bibfnamefont
  {C.}~\bibnamefont {Ramakrishnan}}, \bibinfo {author} {\bibfnamefont
  {K.}~\bibnamefont {Deisseroth}}, \bibinfo {author} {\bibfnamefont
  {M.}~\bibnamefont {Sahani}},\ and\ \bibinfo {author} {\bibfnamefont {K.~V.}\
  \bibnamefont {Shenoy}},\ }\bibfield  {title} {\bibinfo {title} {Direct neural
  perturbations reveal a dynamical mechanism for robust computation},\
  }\bibfield  {journal} {\bibinfo  {journal} {bioRxiv}\ }\href
  {https://doi.org/10.1101/2022.12.16.520768} {10.1101/2022.12.16.520768}
  (\bibinfo {year} {2022})\BibitemShut {NoStop}%
\bibitem [{\citenamefont {Sandberg}\ \emph {et~al.}(2003)\citenamefont
  {Sandberg}, \citenamefont {Tegn{\'e}r},\ and\ \citenamefont
  {Lansner}}]{sandberg2003working}%
  \BibitemOpen
  \bibfield  {author} {\bibinfo {author} {\bibfnamefont {A.}~\bibnamefont
  {Sandberg}}, \bibinfo {author} {\bibfnamefont {J.}~\bibnamefont
  {Tegn{\'e}r}},\ and\ \bibinfo {author} {\bibfnamefont {A.}~\bibnamefont
  {Lansner}},\ }\bibfield  {title} {\bibinfo {title} {A working memory model
  based on fast hebbian learning},\ }\href@noop {} {\bibfield  {journal}
  {\bibinfo  {journal} {Network: Computation in Neural Systems}\ }\textbf
  {\bibinfo {volume} {14}},\ \bibinfo {pages} {789} (\bibinfo {year}
  {2003})}\BibitemShut {NoStop}%
\bibitem [{\citenamefont {Mart{\'\i}}\ \emph {et~al.}(2018)\citenamefont
  {Mart{\'\i}}, \citenamefont {Brunel},\ and\ \citenamefont
  {Ostojic}}]{marti2018correlations}%
  \BibitemOpen
  \bibfield  {author} {\bibinfo {author} {\bibfnamefont {D.}~\bibnamefont
  {Mart{\'\i}}}, \bibinfo {author} {\bibfnamefont {N.}~\bibnamefont {Brunel}},\
  and\ \bibinfo {author} {\bibfnamefont {S.}~\bibnamefont {Ostojic}},\
  }\bibfield  {title} {\bibinfo {title} {Correlations between synapses in pairs
  of neurons slow down dynamics in randomly connected neural networks},\
  }\href@noop {} {\bibfield  {journal} {\bibinfo  {journal} {Physical Review
  E}\ }\textbf {\bibinfo {volume} {97}},\ \bibinfo {pages} {062314} (\bibinfo
  {year} {2018})}\BibitemShut {NoStop}%
\bibitem [{\citenamefont {Berlemont}\ and\ \citenamefont
  {Mongillo}(2022)}]{berlemont2022glassy}%
  \BibitemOpen
  \bibfield  {author} {\bibinfo {author} {\bibfnamefont {K.}~\bibnamefont
  {Berlemont}}\ and\ \bibinfo {author} {\bibfnamefont {G.}~\bibnamefont
  {Mongillo}},\ }\bibfield  {title} {\bibinfo {title} {Glassy phase in
  dynamically-balanced neuronal networks},\ }\bibfield  {journal} {\bibinfo
  {journal} {bioRxiv}\ }\href {https://doi.org/10.1101/2022.03.14.484348}
  {10.1101/2022.03.14.484348} (\bibinfo {year} {2022})\BibitemShut {NoStop}%
\bibitem [{\citenamefont {Rajan}\ \emph {et~al.}(2010)\citenamefont {Rajan},
  \citenamefont {Abbott},\ and\ \citenamefont
  {Sompolinsky}}]{rajan2010stimulus}%
  \BibitemOpen
  \bibfield  {author} {\bibinfo {author} {\bibfnamefont {K.}~\bibnamefont
  {Rajan}}, \bibinfo {author} {\bibfnamefont {L.~F.}\ \bibnamefont {Abbott}},\
  and\ \bibinfo {author} {\bibfnamefont {H.}~\bibnamefont {Sompolinsky}},\
  }\bibfield  {title} {\bibinfo {title} {Stimulus-dependent suppression of
  chaos in recurrent neural networks},\ }\href@noop {} {\bibfield  {journal}
  {\bibinfo  {journal} {Physical Review E}\ }\textbf {\bibinfo {volume} {82}},\
  \bibinfo {pages} {011903} (\bibinfo {year} {2010})}\BibitemShut {NoStop}%
\bibitem [{\citenamefont {Kozachkov}\ \emph {et~al.}(2020)\citenamefont
  {Kozachkov}, \citenamefont {Lundqvist}, \citenamefont {Slotine},\ and\
  \citenamefont {Miller}}]{kozachkov2020achieving}%
  \BibitemOpen
  \bibfield  {author} {\bibinfo {author} {\bibfnamefont {L.}~\bibnamefont
  {Kozachkov}}, \bibinfo {author} {\bibfnamefont {M.}~\bibnamefont
  {Lundqvist}}, \bibinfo {author} {\bibfnamefont {J.-J.}\ \bibnamefont
  {Slotine}},\ and\ \bibinfo {author} {\bibfnamefont {E.~K.}\ \bibnamefont
  {Miller}},\ }\bibfield  {title} {\bibinfo {title} {Achieving stable dynamics
  in neural circuits},\ }\href@noop {} {\bibfield  {journal} {\bibinfo
  {journal} {PLoS Computational Biology}\ }\textbf {\bibinfo {volume} {16}},\
  \bibinfo {pages} {e1007659} (\bibinfo {year} {2020})}\BibitemShut {NoStop}%
\bibitem [{\citenamefont {Kozachkov}\ and\ \citenamefont
  {Slotine}(2022)}]{kozachkov2022note}%
  \BibitemOpen
  \bibfield  {author} {\bibinfo {author} {\bibfnamefont {L.}~\bibnamefont
  {Kozachkov}}\ and\ \bibinfo {author} {\bibfnamefont {J.-J.}\ \bibnamefont
  {Slotine}},\ }\bibfield  {title} {\bibinfo {title} {A note on matrix measure
  flows, with applications to the contraction analysis of plastic neural
  networks},\ }\href@noop {} {\bibfield  {journal} {\bibinfo  {journal} {arXiv
  preprint arXiv:2212.12639}\ } (\bibinfo {year} {2022})}\BibitemShut {NoStop}%
\bibitem [{\citenamefont {Stokes}(2015)}]{stokes2015activity}%
  \BibitemOpen
  \bibfield  {author} {\bibinfo {author} {\bibfnamefont {M.~G.}\ \bibnamefont
  {Stokes}},\ }\bibfield  {title} {\bibinfo {title}
  {‘activity-silent’working memory in prefrontal cortex: a dynamic coding
  framework},\ }\href@noop {} {\bibfield  {journal} {\bibinfo  {journal}
  {Trends in Cognitive Sciences}\ }\textbf {\bibinfo {volume} {19}},\ \bibinfo
  {pages} {394} (\bibinfo {year} {2015})}\BibitemShut {NoStop}%
\bibitem [{\citenamefont {Zucker}\ and\ \citenamefont
  {Regehr}(2002)}]{zucker2002short}%
  \BibitemOpen
  \bibfield  {author} {\bibinfo {author} {\bibfnamefont {R.~S.}\ \bibnamefont
  {Zucker}}\ and\ \bibinfo {author} {\bibfnamefont {W.~G.}\ \bibnamefont
  {Regehr}},\ }\bibfield  {title} {\bibinfo {title} {Short-term synaptic
  plasticity},\ }\href@noop {} {\bibfield  {journal} {\bibinfo  {journal}
  {Annual Review of Physiology}\ }\textbf {\bibinfo {volume} {64}},\ \bibinfo
  {pages} {355} (\bibinfo {year} {2002})}\BibitemShut {NoStop}%
\bibitem [{\citenamefont {Wang}\ \emph {et~al.}(2006)\citenamefont {Wang},
  \citenamefont {Markram}, \citenamefont {Goodman}, \citenamefont {Berger},
  \citenamefont {Ma},\ and\ \citenamefont
  {Goldman-Rakic}}]{wang2006heterogeneity}%
  \BibitemOpen
  \bibfield  {author} {\bibinfo {author} {\bibfnamefont {Y.}~\bibnamefont
  {Wang}}, \bibinfo {author} {\bibfnamefont {H.}~\bibnamefont {Markram}},
  \bibinfo {author} {\bibfnamefont {P.~H.}\ \bibnamefont {Goodman}}, \bibinfo
  {author} {\bibfnamefont {T.~K.}\ \bibnamefont {Berger}}, \bibinfo {author}
  {\bibfnamefont {J.}~\bibnamefont {Ma}},\ and\ \bibinfo {author}
  {\bibfnamefont {P.~S.}\ \bibnamefont {Goldman-Rakic}},\ }\bibfield  {title}
  {\bibinfo {title} {Heterogeneity in the pyramidal network of the medial
  prefrontal cortex},\ }\href@noop {} {\bibfield  {journal} {\bibinfo
  {journal} {Nature Neuroscience}\ }\textbf {\bibinfo {volume} {9}},\ \bibinfo
  {pages} {534} (\bibinfo {year} {2006})}\BibitemShut {NoStop}%
\bibitem [{\citenamefont {Von Der~Malsburg}\ and\ \citenamefont
  {Schneider}(1986)}]{von1986neural}%
  \BibitemOpen
  \bibfield  {author} {\bibinfo {author} {\bibfnamefont {C.}~\bibnamefont {Von
  Der~Malsburg}}\ and\ \bibinfo {author} {\bibfnamefont {W.}~\bibnamefont
  {Schneider}},\ }\bibfield  {title} {\bibinfo {title} {A neural cocktail-party
  processor},\ }\href@noop {} {\bibfield  {journal} {\bibinfo  {journal}
  {Biological Cybernetics}\ }\textbf {\bibinfo {volume} {54}},\ \bibinfo
  {pages} {29} (\bibinfo {year} {1986})}\BibitemShut {NoStop}%
\bibitem [{\citenamefont {Polyn}\ \emph {et~al.}(2009)\citenamefont {Polyn},
  \citenamefont {Norman},\ and\ \citenamefont {Kahana}}]{polyn2009context}%
  \BibitemOpen
  \bibfield  {author} {\bibinfo {author} {\bibfnamefont {S.~M.}\ \bibnamefont
  {Polyn}}, \bibinfo {author} {\bibfnamefont {K.~A.}\ \bibnamefont {Norman}},\
  and\ \bibinfo {author} {\bibfnamefont {M.~J.}\ \bibnamefont {Kahana}},\
  }\bibfield  {title} {\bibinfo {title} {A context maintenance and retrieval
  model of organizational processes in free recall.},\ }\href@noop {}
  {\bibfield  {journal} {\bibinfo  {journal} {Psychological Review}\ }\textbf
  {\bibinfo {volume} {116}},\ \bibinfo {pages} {129} (\bibinfo {year}
  {2009})}\BibitemShut {NoStop}%
\bibitem [{\citenamefont {Fiebig}\ and\ \citenamefont
  {Lansner}(2017)}]{fiebig2017spiking}%
  \BibitemOpen
  \bibfield  {author} {\bibinfo {author} {\bibfnamefont {F.}~\bibnamefont
  {Fiebig}}\ and\ \bibinfo {author} {\bibfnamefont {A.}~\bibnamefont
  {Lansner}},\ }\bibfield  {title} {\bibinfo {title} {A spiking working memory
  model based on hebbian short-term potentiation},\ }\href@noop {} {\bibfield
  {journal} {\bibinfo  {journal} {Journal of Neuroscience}\ }\textbf {\bibinfo
  {volume} {37}},\ \bibinfo {pages} {83} (\bibinfo {year} {2017})}\BibitemShut
  {NoStop}%
\bibitem [{\citenamefont {Manohar}\ \emph {et~al.}(2019)\citenamefont
  {Manohar}, \citenamefont {Zokaei}, \citenamefont {Fallon}, \citenamefont
  {Vogels},\ and\ \citenamefont {Husain}}]{manohar2019neural}%
  \BibitemOpen
  \bibfield  {author} {\bibinfo {author} {\bibfnamefont {S.~G.}\ \bibnamefont
  {Manohar}}, \bibinfo {author} {\bibfnamefont {N.}~\bibnamefont {Zokaei}},
  \bibinfo {author} {\bibfnamefont {S.~J.}\ \bibnamefont {Fallon}}, \bibinfo
  {author} {\bibfnamefont {T.~P.}\ \bibnamefont {Vogels}},\ and\ \bibinfo
  {author} {\bibfnamefont {M.}~\bibnamefont {Husain}},\ }\bibfield  {title}
  {\bibinfo {title} {Neural mechanisms of attending to items in working
  memory},\ }\href@noop {} {\bibfield  {journal} {\bibinfo  {journal}
  {Neuroscience \& Biobehavioral Reviews}\ }\textbf {\bibinfo {volume} {101}},\
  \bibinfo {pages} {1} (\bibinfo {year} {2019})}\BibitemShut {NoStop}%
\bibitem [{\citenamefont {Fiebig}\ \emph {et~al.}(2020)\citenamefont {Fiebig},
  \citenamefont {Herman},\ and\ \citenamefont {Lansner}}]{fiebig2020indexing}%
  \BibitemOpen
  \bibfield  {author} {\bibinfo {author} {\bibfnamefont {F.}~\bibnamefont
  {Fiebig}}, \bibinfo {author} {\bibfnamefont {P.}~\bibnamefont {Herman}},\
  and\ \bibinfo {author} {\bibfnamefont {A.}~\bibnamefont {Lansner}},\
  }\bibfield  {title} {\bibinfo {title} {An indexing theory for working memory
  based on fast hebbian plasticity},\ }\href@noop {} {\bibfield  {journal}
  {\bibinfo  {journal} {eNeuro}\ }\textbf {\bibinfo {volume} {7}} (\bibinfo
  {year} {2020})}\BibitemShut {NoStop}%
\bibitem [{\citenamefont {Huang}\ and\ \citenamefont
  {Wei}(2021)}]{huang2021computational}%
  \BibitemOpen
  \bibfield  {author} {\bibinfo {author} {\bibfnamefont {Q.-S.}\ \bibnamefont
  {Huang}}\ and\ \bibinfo {author} {\bibfnamefont {H.}~\bibnamefont {Wei}},\
  }\bibfield  {title} {\bibinfo {title} {A computational model of working
  memory based on spike-timing-dependent plasticity},\ }\href@noop {}
  {\bibfield  {journal} {\bibinfo  {journal} {Frontiers in Computational
  Neuroscience}\ }\textbf {\bibinfo {volume} {15}},\ \bibinfo {pages} {630999}
  (\bibinfo {year} {2021})}\BibitemShut {NoStop}%
\bibitem [{\citenamefont {Bocincova}\ \emph {et~al.}(2022)\citenamefont
  {Bocincova}, \citenamefont {Buschman}, \citenamefont {Stokes},\ and\
  \citenamefont {Manohar}}]{bocincova2022neural}%
  \BibitemOpen
  \bibfield  {author} {\bibinfo {author} {\bibfnamefont {A.}~\bibnamefont
  {Bocincova}}, \bibinfo {author} {\bibfnamefont {T.~J.}\ \bibnamefont
  {Buschman}}, \bibinfo {author} {\bibfnamefont {M.~G.}\ \bibnamefont
  {Stokes}},\ and\ \bibinfo {author} {\bibfnamefont {S.~G.}\ \bibnamefont
  {Manohar}},\ }\bibfield  {title} {\bibinfo {title} {Neural signature of
  flexible coding in prefrontal cortex},\ }\href@noop {} {\bibfield  {journal}
  {\bibinfo  {journal} {Proceedings of the National Academy of Sciences}\
  }\textbf {\bibinfo {volume} {119}},\ \bibinfo {pages} {e2200400119} (\bibinfo
  {year} {2022})}\BibitemShut {NoStop}%
\bibitem [{\citenamefont {Kozachkov}\ \emph {et~al.}(2022)\citenamefont
  {Kozachkov}, \citenamefont {Tauber}, \citenamefont {Lundqvist}, \citenamefont
  {Brincat}, \citenamefont {Slotine},\ and\ \citenamefont
  {Miller}}]{kozachkov2022robust}%
  \BibitemOpen
  \bibfield  {author} {\bibinfo {author} {\bibfnamefont {L.}~\bibnamefont
  {Kozachkov}}, \bibinfo {author} {\bibfnamefont {J.}~\bibnamefont {Tauber}},
  \bibinfo {author} {\bibfnamefont {M.}~\bibnamefont {Lundqvist}}, \bibinfo
  {author} {\bibfnamefont {S.~L.}\ \bibnamefont {Brincat}}, \bibinfo {author}
  {\bibfnamefont {J.-J.}\ \bibnamefont {Slotine}},\ and\ \bibinfo {author}
  {\bibfnamefont {E.~K.}\ \bibnamefont {Miller}},\ }\bibfield  {title}
  {\bibinfo {title} {Robust and brain-like working memory through short-term
  synaptic plasticity},\ }\href@noop {} {\bibfield  {journal} {\bibinfo
  {journal} {PLoS Computational Biology}\ }\textbf {\bibinfo {volume} {18}},\
  \bibinfo {pages} {e1010776} (\bibinfo {year} {2022})}\BibitemShut {NoStop}%
\bibitem [{\citenamefont {Bengio}\ \emph {et~al.}(2021)\citenamefont {Bengio},
  \citenamefont {Lecun},\ and\ \citenamefont {Hinton}}]{bengio2021deep}%
  \BibitemOpen
  \bibfield  {author} {\bibinfo {author} {\bibfnamefont {Y.}~\bibnamefont
  {Bengio}}, \bibinfo {author} {\bibfnamefont {Y.}~\bibnamefont {Lecun}},\ and\
  \bibinfo {author} {\bibfnamefont {G.}~\bibnamefont {Hinton}},\ }\bibfield
  {title} {\bibinfo {title} {Deep learning for ai},\ }\href@noop {} {\bibfield
  {journal} {\bibinfo  {journal} {Communications of the ACM}\ }\textbf
  {\bibinfo {volume} {64}},\ \bibinfo {pages} {58} (\bibinfo {year}
  {2021})}\BibitemShut {NoStop}%
\bibitem [{\citenamefont {Mante}\ \emph {et~al.}(2013)\citenamefont {Mante},
  \citenamefont {Sussillo}, \citenamefont {Shenoy},\ and\ \citenamefont
  {Newsome}}]{mante2013context}%
  \BibitemOpen
  \bibfield  {author} {\bibinfo {author} {\bibfnamefont {V.}~\bibnamefont
  {Mante}}, \bibinfo {author} {\bibfnamefont {D.}~\bibnamefont {Sussillo}},
  \bibinfo {author} {\bibfnamefont {K.~V.}\ \bibnamefont {Shenoy}},\ and\
  \bibinfo {author} {\bibfnamefont {W.~T.}\ \bibnamefont {Newsome}},\
  }\bibfield  {title} {\bibinfo {title} {Context-dependent computation by
  recurrent dynamics in prefrontal cortex},\ }\href@noop {} {\bibfield
  {journal} {\bibinfo  {journal} {Nature}\ }\textbf {\bibinfo {volume} {503}},\
  \bibinfo {pages} {78} (\bibinfo {year} {2013})}\BibitemShut {NoStop}%
\bibitem [{\citenamefont {Tirozzi}\ and\ \citenamefont
  {Tsodyks}(1991)}]{tirozzi1991chaos}%
  \BibitemOpen
  \bibfield  {author} {\bibinfo {author} {\bibfnamefont {B.}~\bibnamefont
  {Tirozzi}}\ and\ \bibinfo {author} {\bibfnamefont {M.}~\bibnamefont
  {Tsodyks}},\ }\bibfield  {title} {\bibinfo {title} {Chaos in highly diluted
  neural networks},\ }\href@noop {} {\bibfield  {journal} {\bibinfo  {journal}
  {Europhysics Letters}\ }\textbf {\bibinfo {volume} {14}},\ \bibinfo {pages}
  {727} (\bibinfo {year} {1991})}\BibitemShut {NoStop}%
\bibitem [{\citenamefont {Pereira}\ and\ \citenamefont
  {Brunel}(2018)}]{pereira2018attractor}%
  \BibitemOpen
  \bibfield  {author} {\bibinfo {author} {\bibfnamefont {U.}~\bibnamefont
  {Pereira}}\ and\ \bibinfo {author} {\bibfnamefont {N.}~\bibnamefont
  {Brunel}},\ }\bibfield  {title} {\bibinfo {title} {Attractor dynamics in
  networks with learning rules inferred from in vivo data},\ }\href@noop {}
  {\bibfield  {journal} {\bibinfo  {journal} {Neuron}\ }\textbf {\bibinfo
  {volume} {99}},\ \bibinfo {pages} {227} (\bibinfo {year} {2018})}\BibitemShut
  {NoStop}%
\bibitem [{\citenamefont {Pereira-Obilinovic}\ \emph
  {et~al.}(2023)\citenamefont {Pereira-Obilinovic}, \citenamefont {Aljadeff},\
  and\ \citenamefont {Brunel}}]{pereira2023forgetting}%
  \BibitemOpen
  \bibfield  {author} {\bibinfo {author} {\bibfnamefont {U.}~\bibnamefont
  {Pereira-Obilinovic}}, \bibinfo {author} {\bibfnamefont {J.}~\bibnamefont
  {Aljadeff}},\ and\ \bibinfo {author} {\bibfnamefont {N.}~\bibnamefont
  {Brunel}},\ }\bibfield  {title} {\bibinfo {title} {Forgetting leads to chaos
  in attractor networks},\ }\href@noop {} {\bibfield  {journal} {\bibinfo
  {journal} {Physical Review X}\ }\textbf {\bibinfo {volume} {13}},\ \bibinfo
  {pages} {011009} (\bibinfo {year} {2023})}\BibitemShut {NoStop}%
\bibitem [{\citenamefont {Aky{\"u}rek}\ \emph {et~al.}(2022)\citenamefont
  {Aky{\"u}rek}, \citenamefont {Schuurmans}, \citenamefont {Andreas},
  \citenamefont {Ma},\ and\ \citenamefont {Zhou}}]{akyurek2022learning}%
  \BibitemOpen
  \bibfield  {author} {\bibinfo {author} {\bibfnamefont {E.}~\bibnamefont
  {Aky{\"u}rek}}, \bibinfo {author} {\bibfnamefont {D.}~\bibnamefont
  {Schuurmans}}, \bibinfo {author} {\bibfnamefont {J.}~\bibnamefont {Andreas}},
  \bibinfo {author} {\bibfnamefont {T.}~\bibnamefont {Ma}},\ and\ \bibinfo
  {author} {\bibfnamefont {D.}~\bibnamefont {Zhou}},\ }\bibfield  {title}
  {\bibinfo {title} {What learning algorithm is in-context learning?
  investigations with linear models},\ }\href@noop {} {\bibfield  {journal}
  {\bibinfo  {journal} {arXiv preprint arXiv:2211.15661}\ } (\bibinfo {year}
  {2022})}\BibitemShut {NoStop}%
\bibitem [{\citenamefont {Tisseur}\ and\ \citenamefont
  {Meerbergen}(2001)}]{tisseur2001quadratic}%
  \BibitemOpen
  \bibfield  {author} {\bibinfo {author} {\bibfnamefont {F.}~\bibnamefont
  {Tisseur}}\ and\ \bibinfo {author} {\bibfnamefont {K.}~\bibnamefont
  {Meerbergen}},\ }\bibfield  {title} {\bibinfo {title} {The quadratic
  eigenvalue problem},\ }\href@noop {} {\bibfield  {journal} {\bibinfo
  {journal} {SIAM Review}\ }\textbf {\bibinfo {volume} {43}},\ \bibinfo {pages}
  {235} (\bibinfo {year} {2001})}\BibitemShut {NoStop}%
\end{thebibliography}
\end{document}